\documentclass[]{emulateapj}
\usepackage{amsmath,graphicx,times,bm,url,xspace,color}

\def\icarus{{Icarus}}

\newcommand{\Eq}[1]{Equation~(\ref{#1})}

\newcommand{\Sec}[1]{Section~\ref{#1}}

\newcommand{\Fig}[1]{Figure~\ref{#1}}
\newcommand{\Figs}[2]{Figures~\ref{#1} and \ref{#2}}

\def\Rm{\mbox{\rm Rm}}

\usepackage{ulem}

\renewcommand\emph[1]{\textit{#1}}

\newcommand\cm{\,\rm cm}

\newcommand\s{\,\rm s}

\newcommand\g{\,\rm g}

\newcommand\K{\,\rm K}
\newcommand\yr{\,\rm yr}
\newcommand\Myr{\,\rm Myr}

\newcommand\au{\,\rm au}
\newcommand\mG{\,\rm mG}

\newcommand\perccm{\,\rm cm^{-3}}

\newcommand\Msun{\,\rm M_\sun}
\newcommand\Me{M_\earth}
\newcommand\Mj{M_{\rm J}}
\newcommand\Mp{M_{\rm p}}
\newcommand\rp{r_{\rm p}}
\newcommand\rH{r_{\rm H}}
\newcommand\cS{c_{\rm s}}

\newcommand\tms{\!\times\!}
\newcommand\cdt{\!\cdot\!}

\newcommand\rr{\mathbf r}

\newcommand\zz{\hat{{\mathbf z}}}

\newcommand\V{\mathbf v}
\newcommand\B{\mathbf B}
\newcommand\J{\mathbf J}

\newcommand{\OO}{\bm{\Omega}}
\newcommand{\dd}{\operatorname{d}}
\newcommand\Rc{R}

\newcommand\visfl{\boldsymbol{\tau}}

\newcommand{\simgt}%
           {\,\hbox{\lower0.35ex\hbox{$\sim$}\llap{\raise0.35ex\hbox{$>$}}}\,}
\newcommand{\simlt}%
           {\,\hbox{\lower0.35ex\hbox{$\sim$}\llap{\raise0.35ex\hbox{$<$}}}\,}

\newcommand\NIII{\textsc{Nirvana-iii}\xspace}

\newcommand\itm{\medskip\noindent\hspace{0.5ex}\textbullet\hspace{1ex}}

\begin{document}

\title{Global hydromagnetic simulations of a planet embedded in a dead zone:\\
gap opening, gas accretion and formation of a protoplanetary jet
}

\author{
  O.~Gressel$^{1,2}$,  R.~P.~Nelson$^{2}$, 
  N.~J.~Turner$^{3}$,  U.~Ziegler$^{4}$
}

\affil{
  $^1$NORDITA, KTH Royal Institute of Technology and Stockholm University, 
      Roslagstullsbacken 23, 106 91 Stockholm,   Sweden\\
  $^2$Astronomy Unit, Queen Mary University of London,
      Mile End Road, London E1 4NS, UK\\ 
  $^3$Jet Propulsion Laboratory, California Institute of
      Technology, Pasadena, CA 91109, USA\\
  $^4$Leibniz-Institut f{\"u}r Astrophysik Potsdam (AIP), 
      An der Sternwarte 16, 14482, Potsdam, Germany
}

\email{OG: gressel@kth.se, RPN: r.p.nelson@qmul.ac.uk,\newline
  NJT: neal.j.turner@jpl.nasa.gov, UZ: uziegler@aip.de}



\begin{abstract}
We present global hydrodynamic and magnetohydrodynamic (MHD)
simulations with mesh refinement of accreting planets embedded in
protoplanetary disks (PPDs). The magnetized disk includes Ohmic
resistivity that depends on the overlying mass column, leading to
turbulent surface layers and a dead zone near the midplane. The main
results are: (i) The accretion flow in the Hill sphere is
intrinsically 3D for hydrodynamic and MHD models. Net inflow toward
the planet is dominated by high latitude flows. A circumplanetary disk
(CPD) forms.  Its midplane flows outward in a pattern whose details
differ between models. (ii) Gap opening magnetically couples and
ignites the dead zone near the planet, leading to stochastic
accretion, a quasi-turbulent flow in the Hill sphere and a CPD whose
structure displays high levels of variability. (iii) Advection of
magnetized gas onto the rotating CPD generates helical fields that
launch magnetocentrifugally driven outflows. During one specific epoch
a highly collimated, one-sided jet is observed. (iv) The CPD's surface
density $\sim 30\g\cm^{-2}$, small enough for significant ionization
and turbulence to develop. (v) The accretion rate onto the planet in
the MHD simulation reaches a steady value $8 \times 10^{-3} \Me
\yr^{-1}$, and is similar in the viscous hydrodynamic runs. Our
results suggest that gas accretion onto a forming giant planet within
a magnetized PPD with dead zone allows rapid growth from Saturnian to
Jovian masses. As well as being relevant for giant planet formation,
these results have important implications for the formation of regular
satellites around gas giant planets.
\end{abstract}

\keywords{magnetohydrodynamics -- methods: numerical -- planets and
  satellites: formation -- protoplanetary disks}


\section{Introduction}
\label{sec:intro}

Gas giant planets are widely believed to form {\it via} core nucleated
accretion, a scenario that begins with the formation of a solid rock
and ice core in a protoplanetary disk (PPD) by the agglomeration of
smaller bodies (planetesimals), and which concludes with the accretion
of a gaseous envelope from the surrounding nebula
\citep{1980PThPh..64..544M}. Although it has been suggested that giant
planets may form through the direct gravitational fragmentation of a
massive protoplanetary disk during its early evolution
\citep{1998ApJ...503..923B}, circumstantial evidence for core
accretion having operated in the Solar System is provided by the
inferred existence of significant cores in Saturn, Uranus and Neptune
\citep[e.g.][]{2004ApJ...609.1170S}. It can further be argued that the
substantial numbers of relatively low-mass super-Earth and
Neptune-like extrasolar planets being discovered
\citep[e.g.][]{2011Natur.470...53L} indicate that core accretion is a
common mode of planetary formation outside of the Solar System.

Detailed one-dimensional models of gas giant planet formation indicate
that envelope accretion occurs in two distinct stages: (i) a
quasi-static contraction phase during which the envelope mass grows
slowly over time scales $\gtrsim 1\Myr$ \citep{1996Icar..124...62P};
(ii) a runaway growth phase during which the envelope accretes
dynamically onto the planet. This later phase normally arises once the
envelope exceeds the core mass, corresponding to a total planet mass
$\gtrsim 35\Me$. Prior to runaway gas accretion the protoplanet
remains embedded in the nebula and the bloated envelope is envisaged
to connect smoothly onto the surrounding disk. During the runaway
phase, however, the planet contracts down to a size $\simeq 3$ Jupiter
radii and gas accretion is expected to occur through a circumplanetary
disk (CPD) that forms by the flow of gas into the planet Hill sphere
\citep{2005A&A...433..247P}. We note that the 1D envelope calculations
indicate that planets with Saturn-like masses ($\sim 0.2-0.6\Mj$) are
likely to experience maximal gas accretion
\citep{1996Icar..124...62P}, as the flow onto the planet is relatively
unimpeded by the compressional heating of the envelope for these
masses. Although these 1D quasi-static calculations cannot determine
the details of the hydrodynamic flow, so some doubt remains about the
actual accretion rate, it is noteworthy that extrasolar planets with
masses similar to Saturn are common\footnote{Out of 720 confirmed
  extrasolar planets listed on \texttt{exoplanets.org}, approximately
  170 have masses in the range $0.2-1\Mj$} even though existing
calculations suggest that once a planet enters the runaway gas
accretion phase it should grow rapidly beyond this mass if in the
presence of a significant gas reservoir. One motivation for performing
the calculations presented in this paper is to address this issue, and
examine whether or not a possible bottle-neck exists that can prevent
rapid growth of planets in this mass range.

The formation of a gas giant planet through gas accretion and the
action of tidal torques leads to the formation of an annular gap
around the vicinity of the planet
\citep{1986ApJ...309..846L,1999ApJ...514..344B,
  1999MNRAS.303..696K,1999ApJ...526.1001L}. This may also herald the
transition of the planet's migration from type~I
\citep{1997Icar..126..261W} during the embedded phase to type~II when
a gap has formed \citep{1986ApJ...309..846L,2000MNRAS.318...18N}.
Material then feeds onto the planet through the gap at the viscous
supply rate. A detailed understanding of this accretion process
necessitates a sophisticated model of the PPD environment in which the
planet is embedded.

Global, multidimensional studies of gas accretion onto embedded
planets began with 2D flat viscous locally isothermal disk models
\citep{1999ApJ...514..344B,1999MNRAS.303..696K,1999ApJ...526.1001L}
and suggested that accretion rates onto Jovian mass planets should be
$\sim 10^{-5}\Mj\yr^{-1}$ for disk models with masses close to the
minimum mass solar nebula \citep{1981PThPS..70...35H} and canonical
values for the disk aspect ratio ($H/r \sim 0.05$) and viscous stress
parameter ($\alpha \sim 10^{-3}$). Models that adopt more realistic
equations of state and/or are 3D find similar accretion rates
\citep{2003ApJ...599..548D,2006A&A...445..747K,2008A&A...478..245P,%
  2009MNRAS.393...49A}. Typically these simulations do not resolve the
flow in the planet Hill sphere very accurately, so although they
indicate the presence of a CPD, the simulations are unable to resolve
the details of its flow. An exception to this are the calculations by
\citet{2009MNRAS.393...49A}, which attempt to simulate the full 3D
radiation-hydrodynamic evolution of the planetary envelope and CPD,
but unfortunately these are hampered by time step restrictions that
prevent long term runs being performed.

Recent 3D hydrodynamic studies of the accretion flow onto giant
planets have been performed that focus on the local evolution in the
vicinity of the Hill sphere
\citep{2008ApJ...685.1220M,2012ApJ...747...47T}. These highly resolved
studies have uncovered a particularly interesting result, namely that
accretion onto the planet occurs not through the midplane region of
the CPD -- where the gas flow appears to be \emph{away} from the
planet on average -- but instead through gas flows that occur at
higher latitudes within the planet Hill sphere. The midplane region of
the CPD does not appear to act as a traditional accretion disk, but
instead is a region where material which enters the Hill sphere with
excess angular momentum is spun-out away from the planet. The most
recent study of \citet{2012ApJ...747...47T} resolves the flow down to
approximately 3\% of the Hill sphere radius, and midplane outflow
appears to occur all the way down to this region. It is unclear how
these results will change when angular momentum transport processes
are included in the evolution of the CPD.

All of the above multidimensional studies have either ignored the disk
viscosity, or have solved the Navier-Stokes equations by adopting an
anomalous value to account for the angular momentum transport arising
from turbulence in the disk, which is believed to be driven by the
magnetorotational instability
\citep[MRI,][]{1998RvMP...70....1B}. Highly resolved studies of gas
accretion onto giant planets embedded in disks that support
MRI-turbulence have not been performed, although low resolution
studies of gap formation and/or gas accretion have been presented by
\citet{2003MNRAS.339..993N}, \citet*{2003ApJ...589..543W},
\citet{2004MNRAS.350..829P}, and \citet{2011ApJ...736...85U} for disks
that support fully developed turbulence. It was observed that magnetic
braking caused an apparent increase in gas accretion in some these
simulations, by removal of angular momentum of material entering the
Hill sphere, but the low resolution and assumption of ideal
magnetohydrodynamics render this observation questionable in terms of
its application to real systems. Although these simulations in some
ways represent a step-up in realism relative to laminar viscous disk
models, the cold dense midplane regions of protoplanetary disks are
believed to host dead zones where the magnetic field and gas decouple
due to the low ionization levels there, with accretion occurring in
the surface layers that are ionized by external sources such as cosmic
rays and stellar X-rays
\citep{1996ApJ...457..355G,1999ApJ...518..848I}.

In this paper, we present the first global simulation of a giant
planet embedded in a magnetized protoplanetary disk with a midplane
dead zone and actively accreting surface layers, where the larger
scale flow features in the planet Hill sphere are resolved.  We adopt
a resistivity prescription in which the ionization structure changes
over time in response to the changing density, temperature and column
of material absorbing the X-rays
\citep{2011MNRAS.415.3291G,2012MNRAS.422.1140G}, and examine gap
formation and accretion onto the protoplanet. We utilize adaptive mesh
refinement to provide good resolution within the planet Hill sphere,
resolving the flow with reasonable accuracy down to a distance from
the planet equal to 5\% of the Hill sphere radius. Similarly resolved
hydrodynamic simulations are also presented for comparison
purposes. The primary aim is to examine the structure of the gap that
forms as the planet accretes gas and exerts tidal torques on the
surrounding protoplanetary disk (including the influence of the
varying ionization fraction as the gap opens), to examine the flow in
the Hill sphere and the rate of gas accretion onto the planet. We find
accretion rates that are consistent with the previous studies
described above, implying that runway gas accretion should form
planets with Jovian masses and above efficiently if planet formation
occurs in a disk with mass similar to the minimum mass
nebula. Enlivening of the gap region into a turbulent state leads to a
highly time-dependent 3D flow within the gap and planet Hill sphere,
with interesting consequences for the dynamics of the circumplanetary
disk.

This paper is organized as follows: In \Sec{sec:methods} we formulate
the equations and lay out the numerical methods used. The utilized
disk model is described in \Sec{sec:model}, where we give initial and
boundary conditions and recapitulate our ionization model. General
results concerning the PPD are presented in \Sec{sec:ppd_results},
followed by \Sec{sec:gap} on the opening of the gap, and \Sec{sec:CPD}
on the circumplanetary disk. Results on the accretion flow onto the
planet are found in \Sec{sec:acc_flow}. We finally summarize our
findings in \Sec{sec:discussion}, where we discuss potential
implications of our results.


\section{Numerical methods}
\label{sec:methods}

We perform hydrodynamic (HD) and magnetohydrodynamic (MHD) simulations
of protoplanetary accretion disks employing a spherical-polar mesh
with adaptive grid refinement around an embedded planetary core. The
planet is modeled \emph{via} the gravitational potential of a softened
point mass, and its position is kept fixed throughout each
simulation. The planet's mass is allowed to grow during the simulation
by accretion of gas from the disk, as detailed below.

\subsection{Numerical scheme} 

The simulations presented in this paper were performed using the
single-fluid MHD code \NIII, which is based on a second-order finite
volume Godunov scheme \citep{2004JCoPh.196..393Z} and employs the
constrained transport (CT) discretization for an intrinsically
divergence-free evolution of the magnetic induction. The code has
recently been extended to orthogonal curvilinear meshes
\citep{2011JCoPh.230.1035Z}, and here we use spherical-polar
coordinates $(r,\theta,\phi)$, denoting spherical radius, co-latitude
and azimuth, respectively. Deviating from the publicly available
version of the code, we here use the upwind reconstruction technique
of \citet{2008JCoPh.227.4123G} to obtain the edge-centered
electromotive force (EMF) needed within the CT update. We have
furthermore generalized the EMF interpolation to curvilinear
coordinates \citep[cf.][for the cylindrical
  case]{2010ApJS..188..290S}. The upwind reconstruction avoids
stability issues present in the original EMF interpolation scheme by
\citet{1999JCoPh.149..270B}, and the relevance of this to the
development of the magneto-rotational instability (MRI) was
demonstrated by \citet{2010A&A...516A..26F}. By default, \NIII
addresses this issue by implementing the two-dimensional Riemann
solver of \citet{2004JCoPh.195...17L}, but we here instead chose the
approach taken by \citeauthor{2008JCoPh.227.4123G}, as this allows us
to easily use the more accurate HLLD approximate Riemann solver of
\citet{2005JCoPh.208..315M}, which provides better numerical accuracy
for low-Mach number flows. The benefits of using HLLD when modeling
MRI turbulence have been demonstrated by \citet{2010arXiv1003.0018B}.

\subsection{Equations solved} 

For our hydrodynamic calculations, we solve the compressible
Navier-Stokes equations subject to a prescribed \emph{enhanced}
viscosity, which we estimate from the turbulent stresses occurring in
the MRI simulation. The hydromagnetic run employs the standard
resistive MHD equations with a position-dependent \emph{molecular}
diffusivity $\eta(\rr,t)$, which we derive self-consistently from a
detailed ionization model (see \Sec{sec:ion}) accounting for
irradiation of the disk surface by ionizing sources. The full set of
equations in a coordinate system co-rotating with the planet
reads\footnote{Note that for clarity, we have suppressed factors of
  the permeability $\mu$ appearing in the equations.}
\begin{eqnarray}
  \partial_t\rho +\nabla\cdt(\rho \V) & = & 0           \,,\nonumber\\
  \partial_t(\rho\V) +\nabla\cdt
          [\rho\mathbf{vv}+p^{\star}-\mathbf{BB}] & = &
          - \rho \nabla\Phi + \rho \mathbf{a}_{\rm i}
          + \nabla\cdt\visfl                            \,,\nonumber\\
  \partial_t e + \nabla\cdt
          [(e + p^{\star})\V - (\V\cdt\B)\B] & = &
          - \rho (\nabla\Phi)\cdt\V 
          + \rho \mathbf{a}_{\rm i}\cdt\V                  \nonumber\\
      & & + \nabla\cdt[\visfl\V\!+\!\eta\B\tms\J ] 
          + \Gamma                                      \,,\nonumber\\
  \partial_t \B -\nabla\tms(\V\tms\B -\eta\J)& = & 0    \,,\nonumber\\
  \nabla\cdt\B & = & 0\,,
\label{eq:mhd}
\end{eqnarray}
with conserved variables $\rho$, $\rho\V$, and the \emph{total} energy
$e= \epsilon+\frac{1}{2}\rho\V^2 +\frac{1}{2}\B^2$, and where
$\epsilon$ denotes the thermal energy density. We use $\J= {\rm
  curl}(\B)$, and introduce the total pressure $p^{\star}=
p+\frac{1}{2}\B^2$. The inertial acceleration due to Coriolis and
centrifugal effects is $\mathbf{a}_{\rm i} = - 2\OO\tms\V -
\OO\tms(\OO\tms \rr)$, where $\OO=\Omega_0\zz$ is the axial vector
representing the angular frequency at the planet radius, and $\zz$ is
the unit vector pointing along the rotation axis of the protoplanetary
disk. The gravitational potential is composed of the central potential
of a solar-mass star (i.e. $M_\star=\Msun$), and the softened
point-mass potential of the planet with mass $\Mp=\Mp(t)$, at the
fixed position $\rr_{\rm p}$, i.e.,
\begin{equation}
  \Phi(\rr,t) = -\frac{GM_\star}{|\rr|}
                -\frac{G\Mp}{|\rr-\rr_{\rm p}|}
                +\frac{\rr\cdt\rr_{\rm p}\,G\Mp}{\rp^3}\,.
                \label{eq:phi}
\end{equation}
The third term appearing in the potential is often referred to as the
`indirect term', and it accounts for the fact that our coordinate
system has its origin at the position of the star, rather than at the
center of mass of the combined system. Note that the gravity of the
disk does not act on any of the bodies, and hence it does note appear
in the potential function. Even with refined meshes we do not yet
resolve the physical radius of the accretion envelope of the planet
core. To avoid the singularity of the assumed point-mass potential,
the second term in \Eq{eq:phi} is softened with a smoothing length
that is fixed at 5\% of the Hill radius corresponding to the initial
planet mass of $100\Me$. This corresponds to half a grid spacing on
the coarse mesh.

For the cases where we evolve an energy equation, we assume
a relation, $p = (\gamma-1)\epsilon$, with $\gamma=7/5$. The
additional heating and cooling term, $\Gamma$, serves the purpose of
relaxing the thermal energy $\epsilon$ towards the initial radial
temperature profile $T_{\rm i}(\Rc)$ on a local dynamical time
scale. The update is implemented operator-split and applies Newtonian
cooling according to the equation
\begin{equation}
  \frac{1}{\rho}\frac{\dd \epsilon}{\dd t} = 
    -\frac{\Omega_{\rm K}(\Rc)}{2\pi}\, 
     \left(\frac{\epsilon}{\rho} - \Theta_{\rm i}(\Rc)\right) \,,
  \label{eq:cooling}
\end{equation}
where $\Omega_{\rm K}(\Rc)$ is the Keplerian rotation frequency as a
function of cylindrical radius, $\Rc$, and the ideal gas relation is
used to compute $\Theta_{\rm i}(\Rc) \equiv T_{\rm i}(\Rc)\,k_{\rm B}
/ (\bar{\mu}m_{\rm u}\,(\gamma-1))$. While such a treatment is a far
cry from a realistic approximation of radiative effects occurring in a
real protoplanetary disk, it adds realism over a purely isothermal
approach. On a practical level, it successfully suppresses the
occurrence of the vertical-shear instability
\citep{2012arXiv1209.2753N} which leads to severe disturbance of the
dead-zone layer \emph{via} the excitation of corrugation waves. Such
disturbances were encountered in inviscid simulations of a locally
isothermal disk with imposed radial temperature profile, but in the
locally isothermal run that we present here the inclusion of viscosity
in the model also suppresses the vertical shear instability.

The viscous stress tensor, $\visfl$, appearing in the momentum
equation is given by $\visfl = \nu\left(\nabla\V +(\nabla\V)^{\top}
-\frac{2}{3}(\nabla\cdt\V)\right)$, with $\nu$ the dynamic viscosity
parameter. Ohmic dissipation enters the induction equation \emph{via}
the $\nabla\tms(\eta\J)$~term. In our explicit time integration
scheme, the numerically allowed time step related to dissipative terms
scales with the square of the grid spacing. This quickly becomes
restrictive when applying mesh refinement, especially in the presence
of high values of the viscosity coefficient $\nu$, and the diffusivity
$\eta$. To circumvent these potentially very restrictive time-step
constraints, we have adopted the super-time-stepping (STS) scheme,
introduced by \citet*{CNM:CNM950}. This concept is based on the idea
of applying a sequence of (first-order accurate) forward-Euler
sub-steps for the dissipation terms. With an appropriate choice of
non-equidistant sub-intervals in time, the scheme gains an advantage
over classical sub-stepping techniques. We here chose the STS
parameter $\nu_{\rm STS}=0.02$, and limit the maximum ratio between
the permissive Courant time-step and the STS time-step to 20. We
remark that the energy equation contains a divergence of a dissipative
flux, $\visfl \V\! + \!\eta\B\tms\J$. Comparing to test solutions
obtained with the unmodified \NIII, we have not found any indications
that the combination of STS with an associated forward-Euler update of
the energy equation leads to spurious results.  We remark that a
future implementation of the \NIII code will make use of the more
accurate second-order scheme recently proposed by
\citet{2012MNRAS.422.2102M}.


\section{Model description}
\label{sec:model}

Our underlying protostellar disk model aims to resemble the
intermediate regions of the early protosolar nebula where giant planet
formation is believed to have taken place. The adopted computational
domain spans a radial extent of $r\in[1,8]\au$. Owing to limited
computational resources, we restrict the azimuthal domain to a quarter
wedge, i.e., $\phi\in[0,\pi/2]$. The latitudinal grid spans a region
$\theta\in[\pi/2-\vartheta,\pi/2+\vartheta]$, with
$\vartheta=4.5\,H/r$, i.e., covering four and a half pressure scale
heights, $H$, on each side of the disk midplane. Unless otherwise
indicated, the base grid resolution is chosen as $N_r \times N_\theta
\times N_\phi = 384\times 96\times 128$ grid points corresponding to
$10.7$ grid points per $H$ in the vertical direction. This coincides
with the requirement to reasonably resolve unstable MRI
modes. Block-adaptive grid refinement (with blocks of size $4\tms
4\tms 4$) is enabled once the planet potential is switched on. Refined
meshes of level $l=1,2,\dots$ have grid spacings of $2^{-l}$ times the
base-level resolution. For reasons of simplicity, the adaptivity is
controlled by a purely geometrical criterion. Approximately-spherical
regions with $r_{\rm s}=1,2,4\,\rH$ around the planet are refined with
$l=3,2,1$ levels, respectively. This implies that the Hill sphere
($r_{\rm s}=\rH$) is always refined at the highest level of $l_{\rm
  max}=3$, resulting in a grid resolution of roughly 5 Jovian radii,
$R_{\rm J}$ in the $r$, and $\theta$ directions, and about $11\,R_{\rm
  J}$ in the $\phi$~direction. Considering the number of grid cells
covering the CPD, we note that a planet of mass $M_{\rm p}=150 \Me$
located at $r_{\rm p}=3.5 \au$ will be surrounded by a CPD with radius
approximately equal to 40\% of the Hill radius, giving $r_{\rm CPD}
\sim 0.074 \au$. The grid-spacings on the most refined level are
($\Delta_r$, $\Delta_{\theta}$, $\Delta_{\phi}$) = (2.28, 2.05,
5.37)$\times 10^{-3} \au$.  In the horizontal plane $r_{\rm CPD}$ is
resolved by approximately 33 cells of length $\Delta_r$ and 14 cells
of length $\Delta_{\phi}$. Assuming a constant aspect ratio of, $H/r
\sim 0.5$, of the CPD, the scale height of the CPD is resolved by
approximately 18 cells at its outer rim. Towards the center the
resolution per scale height drops linearly.  This suggests that the
vertical structure of the CPD should be decently resolved, as should
the circulating flow in the horizontal plane.  Although magnetic
stresses related to larger scale magnetic features in the CPD will be
resolved, we note that development of the MRI on small scales in the
CPD cannot be followed in our simulations.

It is well known that the spherical-polar mesh is ideally suited to
numerically preserve angular momentum within the protoplanetary
disk. However, in the vicinity of the planet the grid is essentially
Cartesian -- in consequence, the angular momentum of the
circumplanetary disk is likely less-well conserved. The associated
numerical truncation error will contribute to the measured accretion
rate onto the planet.

\subsection{Equilibrium disk model} 
\label{sec:disk_model}

We now describe the initial disk model, which is chosen to facilitate
comparison of our results with the existing literature. Although
modeling and observations point to the fact that protoplanetary disks
are moderately flaring\citep{2007prpl.conf..523W,2008A&A...489..633P},
this results in thin disks at small radii, and is suboptimal for
simulating using a spherical-polar grid.  A convenient setting is
obtained when one prescribes a locally-isothermal temperature, $T$,
falling-off with inverse cylindrical radius, i.e., $T(\Rc) = T_0\,
(\Rc/\Rc_0)^{-1}$. It can easily be shown that such a dependence leads
to a constant opening ratio $H/\Rc$ throughout the disk, promoting the
use of a spherical-polar domain. If we further prescribe a power-law
dependence for the disk midplane density, $\rho_{\rm mid}(\Rc) =
\rho_0\, (\Rc/\Rc_0)^{-3/2}$ and assume independent hydrostatic
balance in the vertical and radial direction, respectively, we can
solve for an equilibrium initial solution given by:
\begin{eqnarray}
  \rho(\rr) & = & \rho_0 \left( \frac{\Rc}{\Rc_0} \right)^{-3/2}
                         \exp{\left(\frac{G M_\star}{\cS^2}
                         \left[ \frac{1}{r} - \frac{1}{\Rc} \right] \right)}\,, 
  \label{eqn:rho} \\[4pt]
  \Omega(\rr) & = & \Omega_{\rm K}(\Rc)\,\sqrt{ \frac{\Rc}{r}
                             -\frac{5}{2} \left(\frac{H}{\Rc}\right)^2 }\,,
  \label{eqn:Omega}
\end{eqnarray}
where we have introduced the isothermal sound speed $\cS$, which can
be derived as $\cS^2=c_{\rm s0}^2\, (\Rc/\Rc_0)^{-1}$. We fix the free
parameter $c_{\rm s0}$ according to a value of $h\equiv H/\Rc=0.05$,
noting that $H \equiv c_{\rm s} \Omega_{\rm K}$ where the Keplerian
angular velocity $\Omega_{\rm K}(\Rc) = \sqrt{GM_\star}\,
\Rc^{-3/2}$. Note that, via $r$, the equilibrium rotation profile has
a weak vertical shear. One can also see that in the midplane (i.e. for
$r=\Rc$), the flow is sub-Keplerian to order $(H/\Rc)^2$, which is a
consequence of radial pressure support. We chose the normalization
$\rho_0$ to yield a vertically integrated column density of
$\Sigma=150\g\cm^{-2}$ at the location $\Rc=5\au$, yielding a total
disk mass of $M_{\rm disk}\simeq 3.6\Mj$, i.e., when accounting for
the full azimuthal extent of $2\pi$. In terms of the disk temperature,
we obtain $T=540\K$ at $\Rc=1\au$, and $T=108\K$ at a distance of
$5\au$, comfortably outside of the ice-line where water ice
condenses. The disk is initially threaded by a weak uniform vertical
magnetic field of $B_z=3.6\mG$. This corresponds to a midplane
$\beta_{\rm P}\equiv 2p/B^2$ of $\sim 3\times 10^5$ at the location of
the planet.

\subsection{Boundary conditions} 

To facilitate the long-term evolution of our disk model, we have to
make modifications to the standard boundary conditions (BCs)
implemented in \NIII. With the exception of the azimuthal direction,
where we simply apply periodicity, all BCs are based on the
``outflow'' type, i.e., allowing material to leave the domain, but
preventing inflow. To improve robustness at the inner corners of the
$(r,\theta)$ domain, we impose reflecting boundaries in the
$r$~direction beyond $4\,H$ in $\theta$, and moreover impose Keplerian
rotation for $v_\phi$.

The boundary conditions applied at the upper and lower disk surfaces
for the magnetic field are of the ``perfect conductor'' type, i.e.,
enforcing the normal component to be zero at the boundary, and
applying zero-gradient extrapolation to the parallel field
components. In addition, the BCs have been modified to preserve the
constant vertical field, imposed initially. Because of the vertical
stratification, the hydrodynamic $\theta$ boundaries require special
attention. It is well known that unsplit finite volume schemes make it
hard to exactly preserve a given static equilibrium \citep[see
  e.g.][]{2002ApJS..143..539Z}. This becomes pronounced at the
boundaries because using constant extrapolation of the thermal energy
density under-estimates the pressure gradient. Not balancing gravity
exactly will induce a standing accretion shock in the first grid cell
of the active domain. To obtain a better estimate for the vertical
continuation of the disk profiles outside the active domain, we employ
a second-order Runge-Kutta shooting method to integrate the equations
of hydrostatic equilibrium; thereby reducing the amplitude of spurious
boundary effects significantly.

A similar, but somewhat less dramatic, effect is present in the radial
direction. Taking into account the additional centrifugal force, we
extrapolate the radial equilibrium condition to obtain improved
boundary values; aiding the long-term stability of the disk
evolution. Furthermore, to avoid reflection of spiral density waves,
we have implemented `buffer zones' as described in detail in
\citet{2006MNRAS.370..529D}. This implies that, in narrow annuli
adjacent to the radial domain boundary, we force the density $\rho$,
energy density $\epsilon$, and the velocity $\V$ back to their initial
values. The timescale of this process is proportional to the local
dynamical time scale, and is smoothly tapered-off with increasing
separation from the inner and outer radial boundaries.

\subsection{The ionization model} 
\label{sec:ion}

The central aim of this work is to develop realistic MHD models for
protoplanetary accretion disks containing gaps opened by embedded
planets. This requires that we obtain a self-consistent distribution
of the disk's ionization fraction. For our new global models, we
largely follow the approach taken in the local box models described in
detail in \citet{2011MNRAS.415.3291G,2012MNRAS.422.1140G} -- with the
exception that the dependence on the radial coordinate is now included
explicitly, rather than being evaluated at the reference radius of the
box model.

The disk's ionization balance is heavily affected by the presence of
small dust grains \citep{2000ApJ...543..486S,2006A&A...445..205I}.
The associated short adsorption time scale of free electrons makes it
prohibitive to follow the detailed non-equilibrium chemistry. Instead,
we adopt a simplified approach employing a precomputed table based on
the reaction network used for \texttt{model4} of
\citet{2006A&A...445..205I}.  As before, we consider the contributions
of all the charged species following equations (21)-(31) in
\citet{2007Ap&SS.311...35W}. When deriving the resistivity
$\eta(\rr,t)$, we restrict ourselves to the case of Ohmic resistivity,
which is a reasonable assumption for the denser parts of the PPD
\citep[see, e.g.][]{2007Ap&SS.311...35W} under weak magnetic
fields. Our choice is mainly motivated by reasons of tractability as
the effect of ambipolar diffusion (and Hall EMFs) will greatly
increase the computational complexity of the model. Recent work points
to the potential importance of including these additional non-ideal
MHD effects \citep{2012MNRAS.422.2737W, 2013ApJ...769...76B}, and our
longer term intention is to explore these on a case-by-case basis
rather than presenting a ``kitchen-sink'' simulation from the outset.

For reference, we briefly recapitulate the parameters entering our
model \citep[cf.][]{2011MNRAS.415.3291G}. We presume dust with density
$\rho_{\rm s}=3\g\perccm$ and of a single grain size $d=0.1\mu{\rm
  m}$, which is further assumed to be depleted (by grain growth) to
yield a dust-to-gas mass ratio of $10^{-3}$. Reactions involving
gas-phase ions are represented by Magnesium, which is partly bound-up
in grains and hence taken to be depleted by $10^4$ compared to its
solar abundance. In the reaction network, free electrons are created
according to the ionization rate $\zeta({\bf x},t)$, which we evaluate
based on the external irradiation. The ionization fluxes are
attenuated by the integrated columns of gas to the upper and lower
disk surfaces, respectively. The density integration is complicated by
the requirement to match the integrals at mesh refinement boundaries,
which demands expensive communications. To reduce computational
overheads, we store $\eta(\rr,t)$ as a passive scalar variable. This
allows us to only update $\eta$ from integrated densities every three
computational time-steps, which is short enough to reflect potential
changes of $\eta$ on dynamical time scales.

The underlying ionization prescription used here comprises stellar
X-rays, radionuclides, and energetic protons, and is based on the work
of \citet{2009ApJ...703.2152T}. We closely follow their approach, and
restrict ourselves to stellar X-rays (XRs), and interstellar cosmic
rays (CRs) as the prime ionizing agents. The attenuation of
interstellar CRs within a PPD has been studied by
\citet{2009ApJ...690...69U}, who derived a dependence
\begin{equation}
  \zeta_{\rm CR} = 5\tms10^{-18}\s^{-1}
  \ {\rm e}^{-\Sigma_a/\Sigma_{\rm CR}}\ 
  \left[ 1+\left(\frac{\Sigma_a}{\Sigma_{\rm CR}}\right)^\frac{3}{4}
    \right]^{-\frac{4}{3}},
\end{equation}
with $\Sigma_{\rm CR}=96\g\cm^{-2}$ a typical cosmic ray attenuation
depth \citep[cf.][]{1981PASJ...33..617U}, and with an according
contribution from the second gas column $\Sigma_b$. As in previous
work, we use a simple fit to the Monte-Carlo radiative transfer
calculations of \citet{1999ApJ...518..848I}, and approximate the XR
ionization rate via
\begin{equation}
  \zeta_{\rm XR} = 2.6\tms10^{-15}\s^{-1}\ 
  \left[ {\rm e}^{-\Sigma_a/\Sigma_{\rm X}}
       + {\rm e}^{-\Sigma_b/\Sigma_{\rm X}}
  \right]\,r_{\au}^{-2}\,,
\end{equation}
with $\Sigma_a$ and $\Sigma_b$ being the column densities to the upper
and lower disk surface, respectively, and where $\Sigma_{\rm X} =
8.0\g\cm^{-2}$ is a characteristic absorption depth. We finally
include an ambient ionization due to the decay of short-lived
radionuclides (SR). As in previous models \citep{2012MNRAS.422.1140G},
the related ionization rate is enhanced $10\times$ compared to the
nominal value of $\zeta_{\rm SR}= 3.7\tms10^{-19}\s^{-1}$
\citep[see][]{2009ApJ...703.2152T}. This is done to provide a ceiling
on the resistivity so that the associated time step size does not
become prohibitively small. In conclusion, we remark that when a gap
is formed in the disk, ideally one should include the radial
illumination of the gap edges by the star. For reasons of simplicity,
however, we neglect this effect in our current models.


\begin{figure}
  \includegraphics[width=\columnwidth]{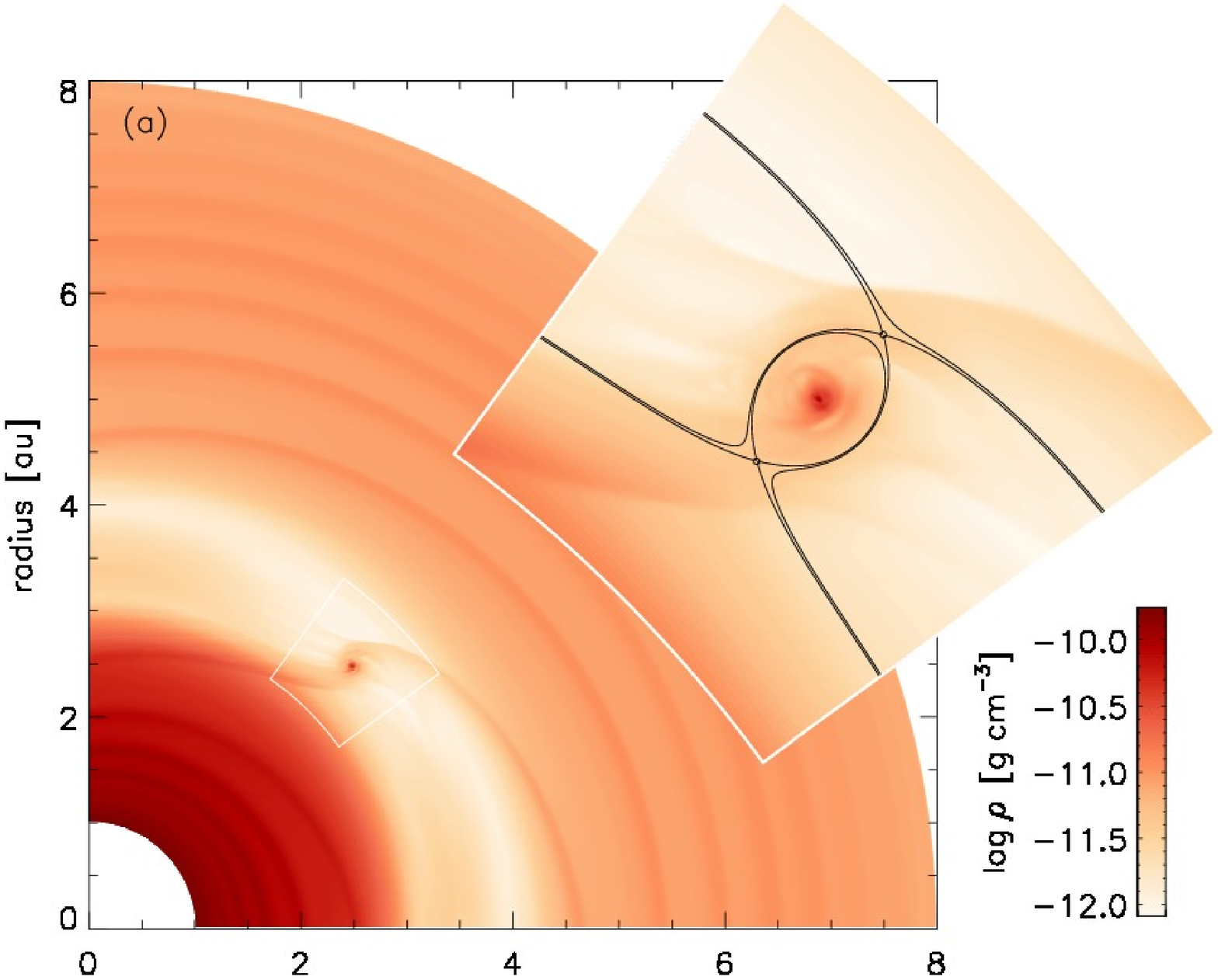}\\[-3ex]
  \includegraphics[width=\columnwidth]{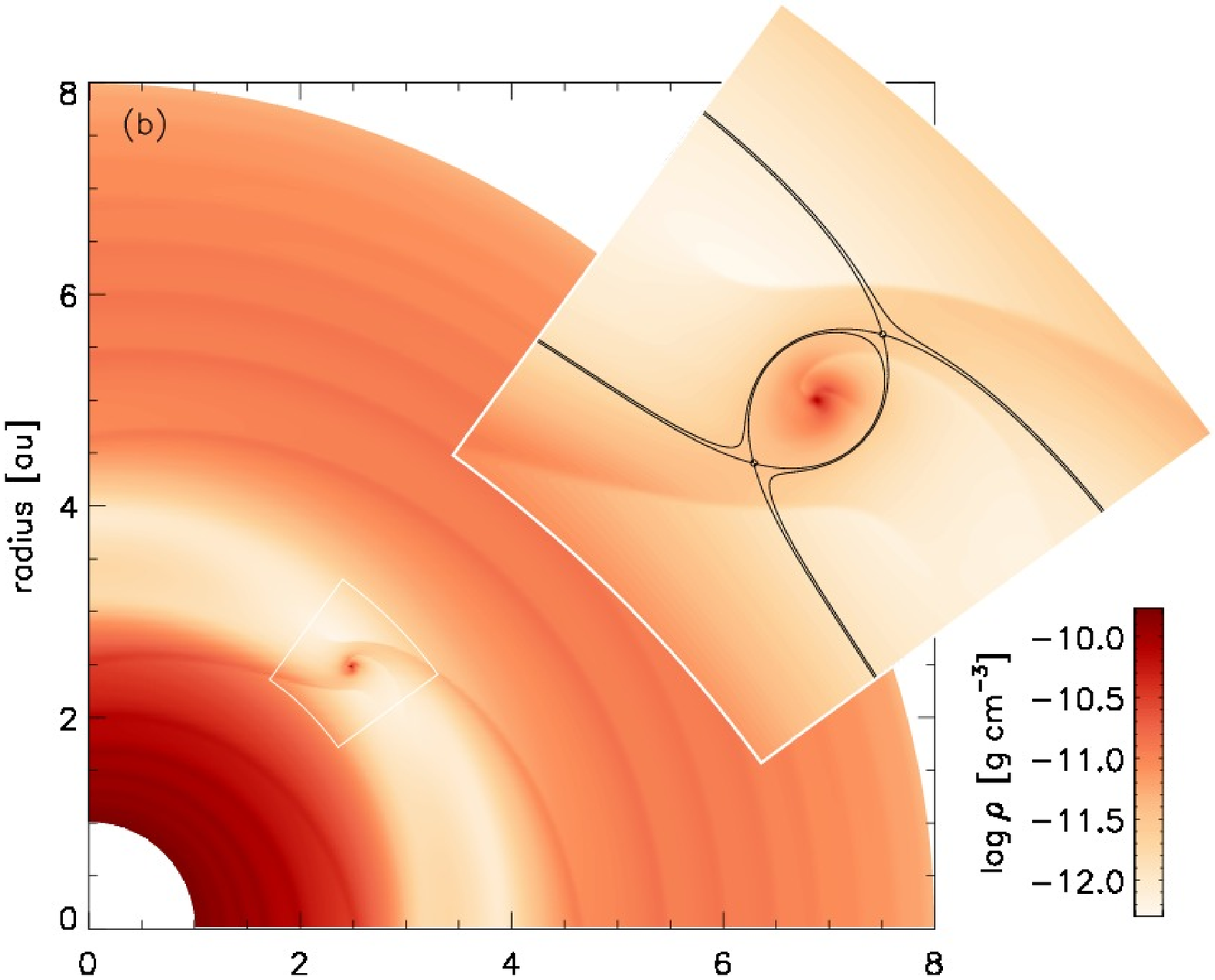}\\[-3ex]
  \includegraphics[width=\columnwidth]{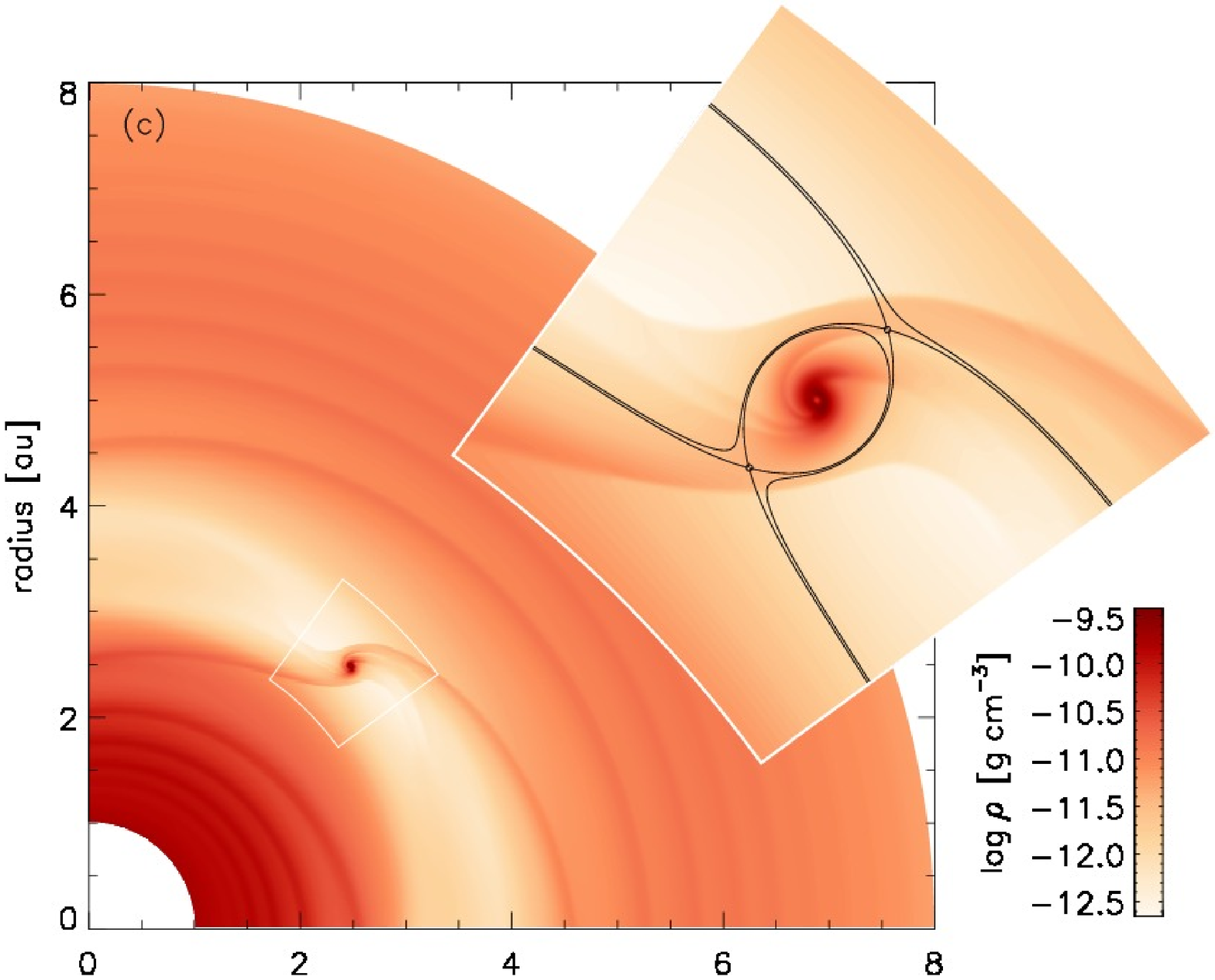}
  \caption{Midplane slices of $\log(\rho)$. From top to bottom: (a)
    MHD run, (b) HD run with cooling, (c) isothermal HD run. Black
    contours indicate the potential surfaces passing through the
    Lagrange points L1 and L2.}\label{fig:midpl_slice}
\end{figure}

\subsection{The accretion sink} 

Even with the aid of adaptive mesh refinement, it remains prohibitive
to resolve the gaseous envelope accreting onto the planetary core.
This is to say that fidelity of resolving small spatial scales is
traded-off against the ability to cover the (comparatively long) time
scale on which gap formation occurs. At the same time, making
meaningful predictions about the envelope of hot accreting gas will
require proper treatment of radiative effects, which are not currently
addressed by our model. Instead focusing on the impact of magnetic
fields, we are hence interested in asking the question whether the
exterior of the accretion region (i.e. the system comprised of the PPD
gap, spiral arms and the CPD) can, in fact, provide the material into
the sphere of influence of the core.

Our accretion sink is similar to that used in \citet{1999MNRAS.303..696K} 
in that a fixed fraction of material is removed from the direct vicinity 
of the planet each time step. Unlike in earlier work, we specify the time 
scale for removal as the local free-fall time (of a test-particle)
\begin{equation}
  \tau_{\rm ff} = \frac{\pi}{2}\sqrt{\frac{r_{\rm acc}^3}{2G\Mp}}\,,
\end{equation}
where $r_{\rm acc}$ is defined as 5\% of the planet's Hill radius --
corresponding roughly to Callisto's semi-major axis. This amounts to a
small fraction of the mass within a sphere of $r_{\rm acc}$ being
removed per time step. To avoid discontinuous behavior in the
accretion flow near the planet, removal is weighted with a
three-dimensional Gaussian kernel. Owing to limited computational
resources, the mass augmented to the planet is artificially enhanced
fourfold to facilitate a speed-up of the gap formation process. In
cases where we evolve an energy equation, the modification to the gas
density is accompanied by a correction in the thermal energy density
as to keep the temperature constant. Within a sphere of $2r_{\rm
  acc}$, we apply additional cooling to avoid the buildup of strong
pressure gradients that would otherwise modify the accretion flow
unphysically and retard the flow of gas toward the sink hole.  In this
region, the temperature is relaxed towards the initial model on a time
scale proportional to the Keplerian angular velocity with respect to
the planet potential, resulting in temperatures similar to those
observed in the radiation-hydrodynamic simulations of
\citet{2006A&A...445..747K}.  To conclude this section, we note that
we do not expect a strong numerical effect from the particular scheme
adopted for the accretion sink \citep[cf. figure~3
  in][]{2008ApJ...685.1220M}.


\section{Protoplanetary disk}
\label{sec:ppd_results}

Previous studies using both laminar and magnetized-turbulent disks
have shown that the presence of an accreting gas giant planet leads to
gap formation, and the formation of a circumplanetary disk surrounding
the planet.

The main goals of this paper are to study the effects of magnetic
fields and time-dependent ionization levels on the evolution of the
gap, the circumplanetary disk, and the gas accretion rate onto the
planet.  To achieve these goals it is necessary to develop a fiducial
hydrodynamic model for comparison purposes. Coming up with a realistic
proxy for a layered turbulent accretion disk is a formidable task in
its own right. Clearly, in the presence of a dead-zone, a
height-dependent $\alpha$ should be used
\citep[e.g.][]{2010A&A...520A..14P}. Ideally, one would obtain an
estimate of the dead-zone structure based on empirical fits derived
from MHD simulations \citep{2011ApJ...742...65O}, but a simple
prescription with $z/H$ is rendered problematic with the presence of a
gap because $\alpha$ then varies spatially and temporally. For lack of
better alternatives, we resort to a standard Shakura-Sunyaev
$\alpha$~viscosity with a constant $\alpha_{\rm SS}=3.25\times
10^{-3}$. This value is derived from globally averaged turbulent
stresses (normalized with the vertically-averaged gas pressure) in the
MHD model, after a quasi-steady state is reached and before the planet
is inserted \citep[also cf.][who use a value of $2\times
  10^{-3}$]{2011ApJ...736...85U}. In comparison to our fiducial box
model \citep{2011MNRAS.415.3291G}, which was located at $5\au$, we
obtain local values (normalized with the midplane gas pressure) of
$\alpha\simeq 3\times 10^{-3}$ in the active layer, which is dominated
by Maxwell stresses, and $\alpha\simeq 3\times 10^{-4}$ in the
dead-zone layer due to residual Reynolds stresses. We note that the
level of turbulence is about a factor of five lower than in the box
simulation, which we attribute to the three-fold weaker net-vertical
field of $B_z=3.6\mG$. For the particular problem of a giant planet
embedded in a disk the approach used in the viscous non-magnetized
disks probably yields reasonable results because the main driver of
evolution on large scales is global angular momentum transport
\citep{2003MNRAS.339..993N}. This argument is supported by our results
which show similar planetary accretion rates for the magnetized and
non-magnetized disk models described later in this paper.

\subsection{Simulation runs} 

We focus on three simulations in this paper, which are based on the
disk model described in \Sec{sec:model}, and all employ three levels
of mesh refinement. Each of the simulations is evolved for roughly 100
planet orbits after the planet is inserted. At the heart of our study
is a full-blown, MHD model based on a \emph{physical}
resistivity. Within this run, due to the shielding of external
ionization, a realistic dead-zone emerges along with self-consistent
MRI-active disk surface layers. Moreover, this MHD model, that we
refer to as ``M1'', is non-isothermal in that an energy equation is
solved based on an adiabatic equation of state and thermal relaxation
towards the initial radial temperature profile (cf. \Eq{eq:cooling}
and \Sec{sec:disk_model}). The influence of numerical resolution and
mesh refinement on the stresses obtained in model M1 are discussed
in the appendix. The same equation of state and thermal relaxation
prescription is adopted for the non-isothermal HD model, ``N1'', 
which includes an \emph{enhanced}
viscosity to model the angular momentum transport associated with the
MRI-induced turbulence that is absent in the HD case. To study the
impact of the equation of state (and ultimately radiative transfer
processes within the PPD and CPD), we furthermore study a second HD
model, ``N2'', which does not solve an energy equation, but instead
adopts a locally isothermal equation of state where the temperature,
$T$, is prescribed as a fixed function of cylindrical radius.

For the magnetized disk run, after the parent disk has reached a
quasi-stationary turbulent state we gradually insert a planet by
smoothly increasing its mass to $100\Me$ over a time scale of roughly
three planet orbits. This mass range corresponds to the rapid
gas-accretion phase of the core accretion scenario
\citep{2005A&A...433..247P}. For the viscous HD runs there is no need
to run the parent disk to steady state prior to inserting the planet,
so the above procedure is followed without relaxing the disk. Once the
planet is fully inserted, the simulations are run until the mass
accretion rate onto the planet approaches a quasi-steady state. To
give an idea of the numerical expense associated with the complexity
of the MHD model, this required 978,600 core hours for simulation M1
alone, amounting to an estimated energy consumption on the order of
$100\,\rm MWh$ equivalent of an emission of about 60 tons of $\rm
CO_2$.


\begin{figure}
  \includegraphics[width=\columnwidth]{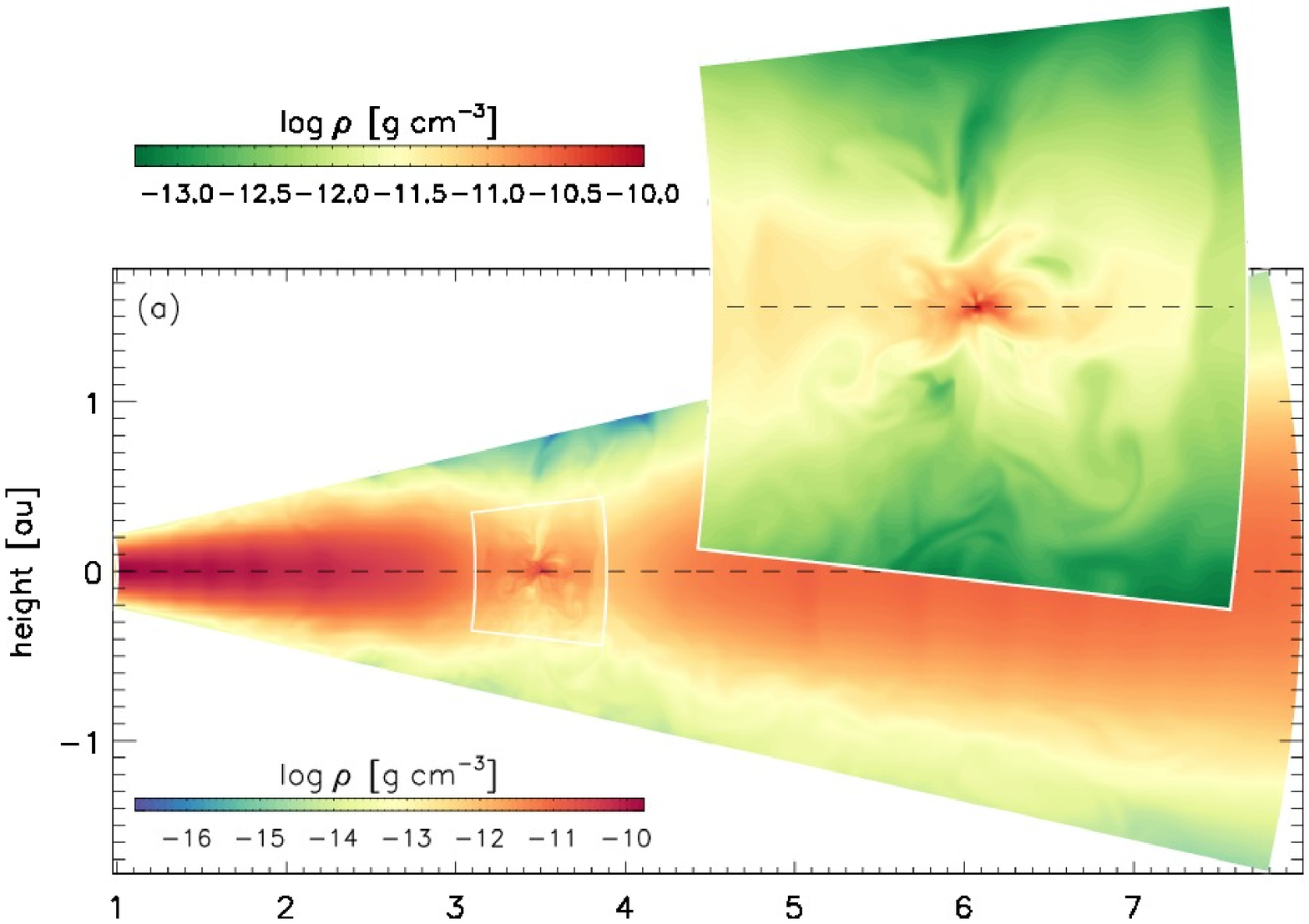}\\[-1.5ex]
  \includegraphics[width=\columnwidth]{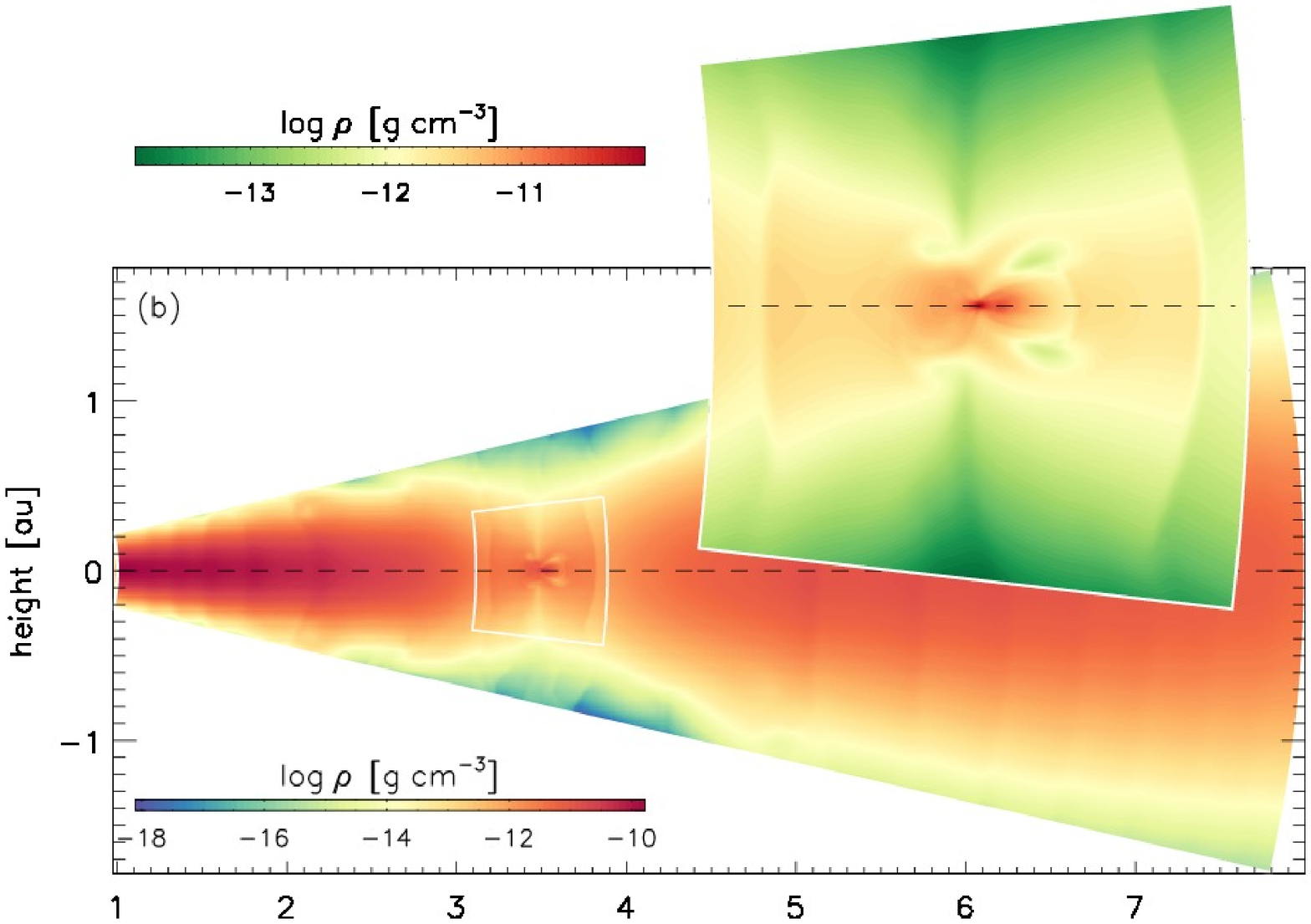}\\[-1.5ex]
  \includegraphics[width=\columnwidth]{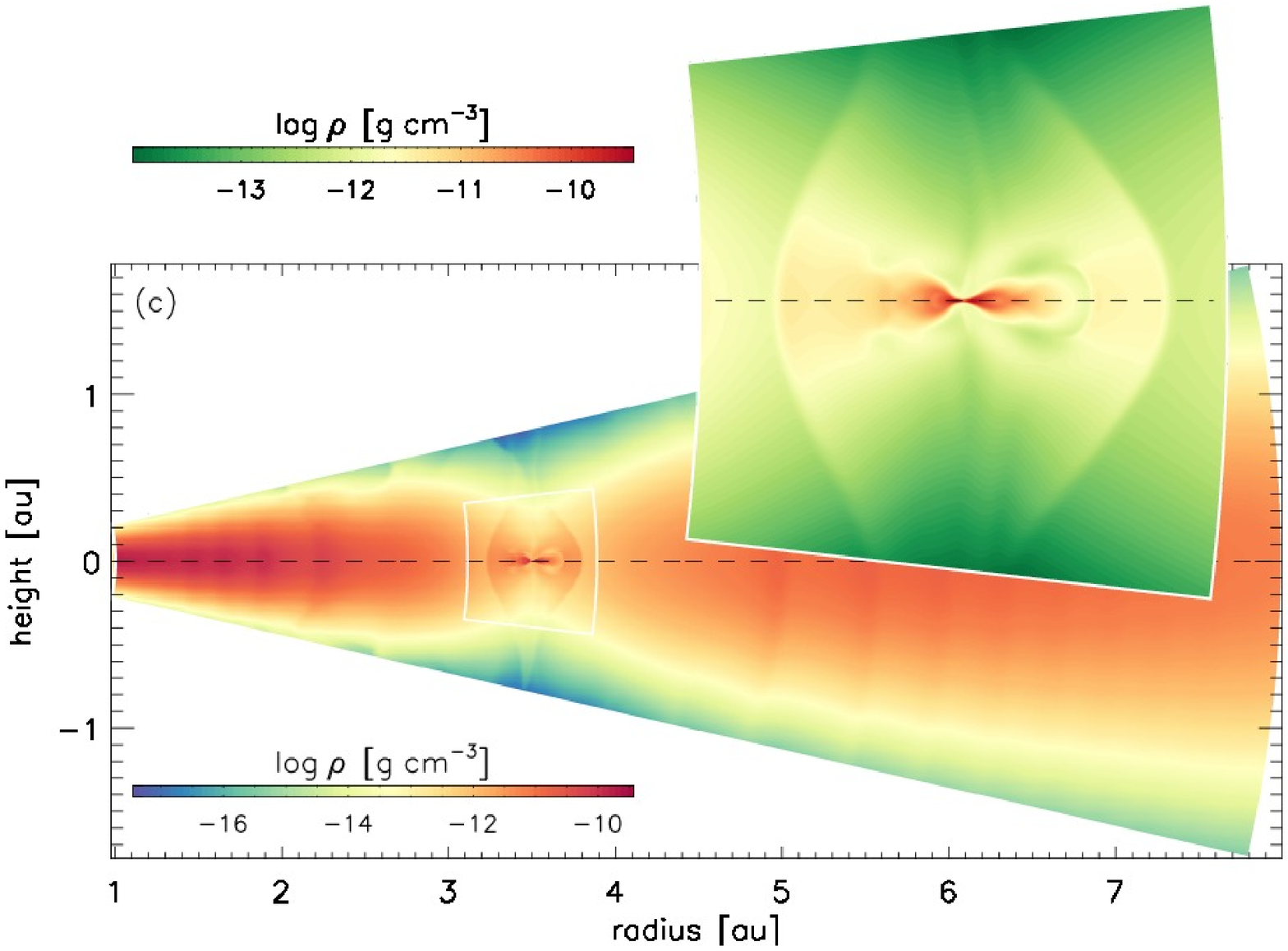}
  \caption{Meridional plane showing $\log(\rho)$ at the planet
    location (note the separate color bars for the insets). From top
    to bottom: (a) MHD run, (b) HD run with cooling, (c) isothermal HD
    run. Under-dense layers above and below the CPD are related to
    pairs of poloidal vortices.}
  \label{fig:polo_slice}
\end{figure}

\subsection{General disk morphology} 
\label{sec:disk_struct}

We begin our analysis by presenting general features of the models. In
Figure~\ref{fig:midpl_slice} we show midplane slices of the gas
density at the end of each simulation. Light colors indicate low
surface density, as in the gap region, and dark colors represent high
density, tracing the spiral wakes induced by the planet. We note that
gap formation in the simulations has not reached completion due to the
high computational cost of these runs, so we see that the gap in the
direct vicinity of the planet is deeper than at azimuthal positions
away from the planet. The material within the co-orbital region
executing horseshoe orbits has yet to either diffuse onto orbits that
cause it to accrete onto the planet or join the rest of the
protoplanetary disk, but eventually this is expected to occur on
longer time scales of about a few 100 planet orbits
\citep{2000MNRAS.318...18N}, or perhaps longer than this in the
presence of dead zone.

The instantaneous cooling in the isothermal model N2 leads to a higher
mass and steeper density profile within the CPD, and to a pronounced
and tightly wound bi-symmetric spiral as observed in earlier work
\citep[e.g.][]{1999ApJ...526.1001L}. In the adiabatic HD model with
prescribed cooling shown in panel (b), the spiral arms are less
tightly-wound because of warmer temperatures in the CPD.  The
temperature in the inner regions of the CPD remains at $T \approx
150\K$ in the isothermal run N2. Even though rapid cooling is applied
in the non-isothermal simulations N2 and M1, temperatures in the inner
regions of the CPDs reach $T \approx 2000\K$ for these runs (similar
to the values $T\approx 1500\K$ obtained by \citet{2006A&A...445..747K} 
in their 3D radiation hydrodynamic runs). The mass of the CPD is
smaller in run N1 and the density profile is shallower (see
\Fig{fig:surface_cmp} below). The flow around the planet is less
regular but still laminar, because each fluid element now has its own
thermal history leading to moderate pressure fluctuations in the CPD.
This changes in the MHD case shown in panel (a), in which the flow
pattern becomes turbulent, but otherwise appears similar to the
equivalent HD simulation.

In their study of giant planets embedded in fully-turbulent disks
without dead-zones, \citet{2003MNRAS.339..993N} comment on the fact
that the spiral wave structure in turbulent protoplanetary disks has a
more washed-out and diffused appearance than observed in laminar
disks. This is clearly not the case here in the midplane because the
dead-zone contains only relatively low-level density perturbations due
to waves propagating from the over-lying active layers. The dominant
perturbations are the spiral wakes excited by the planet. At higher,
more turbulent latitudes in this disk, however, the planet-induced
spirals will have a more diffused appearance. An image of the disk
based on the integrated column density would similarly display a more
diffused rendering than that shown in \Fig{fig:midpl_slice}.

The differences between the isothermal and non-isothermal equation of
state become very apparent when looking at poloidal cuts through the
simulation domain, presented in \Fig{fig:polo_slice}. Compared to both
the HD run N1, panel (b), and the MHD run, panel (a), the CPD is
noticeably thinner for the locally isothermal model, N2, shown in
panel (c). All three cases, however, agree in the asymmetry between
the inner and outer parts of the CPD, with the inner part of the CPD
lying nearest to the central star being vertically thicker than the
component lying further away. This feature is very likely due to the
thermal model that we adopt which assumes that gas lying closer to the
star is warmer than that further out (this is true for both the
locally isothermal and adiabatic disks with thermal relaxation). In a
more realistic model, where the temperature of the CPD is determined
by local heating and cooling processes, we suspect that this in-out
asymmetry will be less pronounced than displayed by our simulations.
We note, however, that in the presence of a gap, a moderately decreasing
temperature at the midplane may be expected as one moves out across the gap,
because reprocessed stellar radiation from the outer gap edge more directly 
illuminates the inner half of the gap \citep{2012ApJ...748...92T}.

The overall structure of the CPD is quite similar in the
non-isothermal HD and MHD cases, indicating that gravity remains the
dominant force (the plasma $\beta_{\rm p}$ -- defined as the ratio of
thermal to magnetic pressure -- typically lies between the values
100-1000 in the CPD region, see \Fig{fig:betap}), but it is also clear
that the circumplanetary region in the MHD run shows considerably more
structure than model N1 due to turbulence. Comparing panels (a) and
(b) reveals an inherent limitation in the ``enhanced'' viscosity
prescription that we applied in the HD case: the resolved flow remains
laminar, and the stochastic nature of the resulting structures
observed in the MHD run cannot be captured in this framework. While
this may be tolerable when studying the secular evolution of a PPD, it
is probably not sufficient for the purpose of studying dynamical
processes occurring in CPDs such as satellite formation where the time
dependent structure of the circumplanetary disk may be crucial.

A low-density funnel along the rotation vector of the PPD can be
observed in all three simulations. In the two HD simulations, this
funnel is always associated with low-angular-momentum material falling
in onto the poles of the planet. In the MHD run this is also usually
the case, but we have also observed the sporadic occurrence of a
(one-sided) magnetically collimated protoplanetary outflow that we
will discuss in more detail later in this paper. The sporadic nature
appears to be related to the fact that the jet can be disrupted by
incoming material accreting through the polar regions.

Taking a global view of the simulation results, we can state that
significant differences arise in the non-magnetized simulations when
moving from an isothermal equation of state to an adiabatic one with
imposed cooling, due to the importance of compressional heating
associated with gas flowing into the planet Hill sphere. Magnetic
fields also play an important role as they transform the
circumplanetary flow from one that is essentially laminar to one that
is turbulent and highly fluctuating, described later in
Sect.~\ref{sec:mhd_cpd}.


\section{Gap formation}
\label{sec:gap}

\begin{figure}
  \includegraphics[width=0.9\columnwidth]{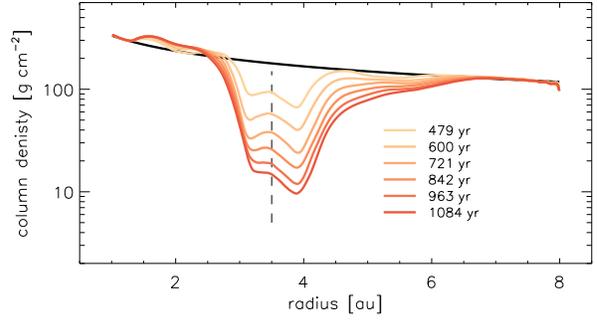}
  \caption{Time sequence of the gap opening for the MHD run,
    M1. Consecutive profiles are for equal time intervals of $\Delta
    t=121\yr$ after the planet is inserted. Gap opening is not
    yet complete at the end of the simulation.}
  \label{fig:gap_evol}
\end{figure} 

As outlined above, we chose an initial planet mass of $100\Me$, as
this mass puts the planet in the rapid growth phase that we are
interested in. It is evident that a combination of tidal torques and
gas accretion onto the planet lead to the formation of a significant
gap within the PPD. Considering the gap opening criteria that the
tidal torque must overcome the effective viscous torque ($q \ge
40/{\cal R}_e$, where $q=M_{\rm p}/M_{\star}$ and ${\cal R}_e$ is the
Reynolds number) and the Hill sphere radius of the planet must exceed
the local scale height\footnote{i.e., $R_{\rm H} > H$, which is
  equivalent to the requirement that the disk response to the planet
  gravity be nonlinear \citep{1993prpl.conf..749L}}, we find that a
$100 \Me$ planet just fails to meet the second criterion, but
satisfies it when its mass grows above $110 \Me$. The first criterion
is fulfilled for planet masses exceeding $60 \Me$ when $\alpha=2
\times 10^{-3}$ and $h=0.05$, so the opening and maintenance of a gap
is expected in the simulations. We note that low mass planets are
known to open modest gaps or surface density depressions in disks due
to shock dissipation of their spiral wakes
\citep{2001ApJ...552..793G,2010ApJ...724..448M}, so the above
gap-opening criteria should not be interpreted too rigidly.

\begin{figure}
    \includegraphics[width=0.9\columnwidth]{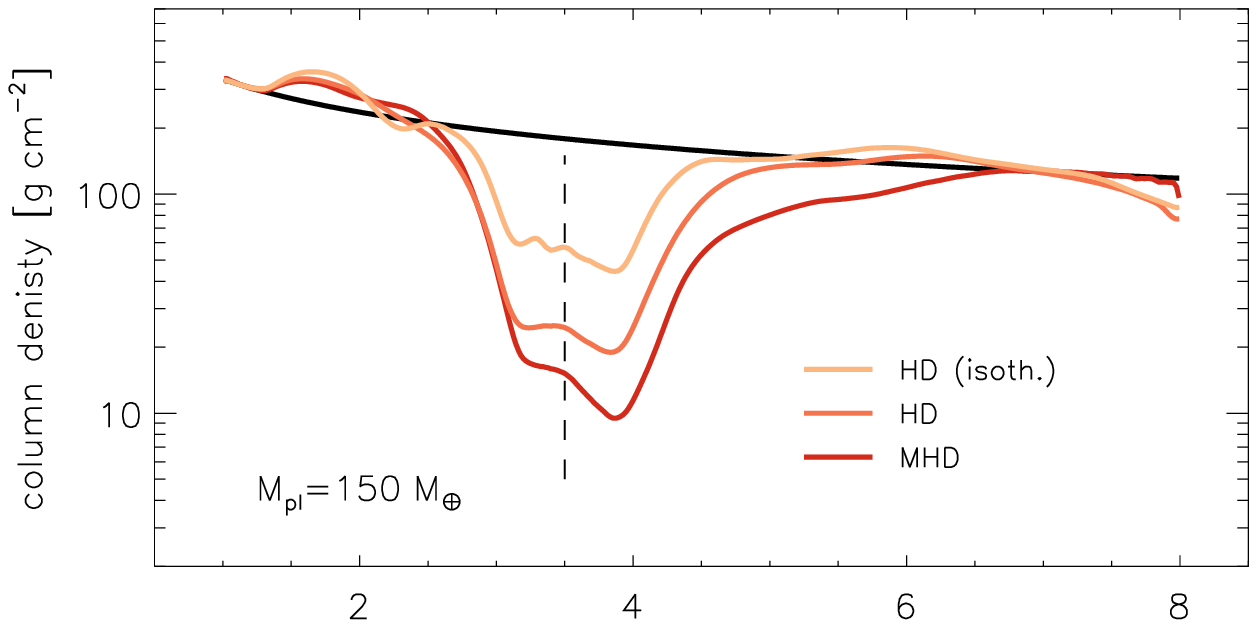}\\[1ex]
    \includegraphics[width=0.9\columnwidth]{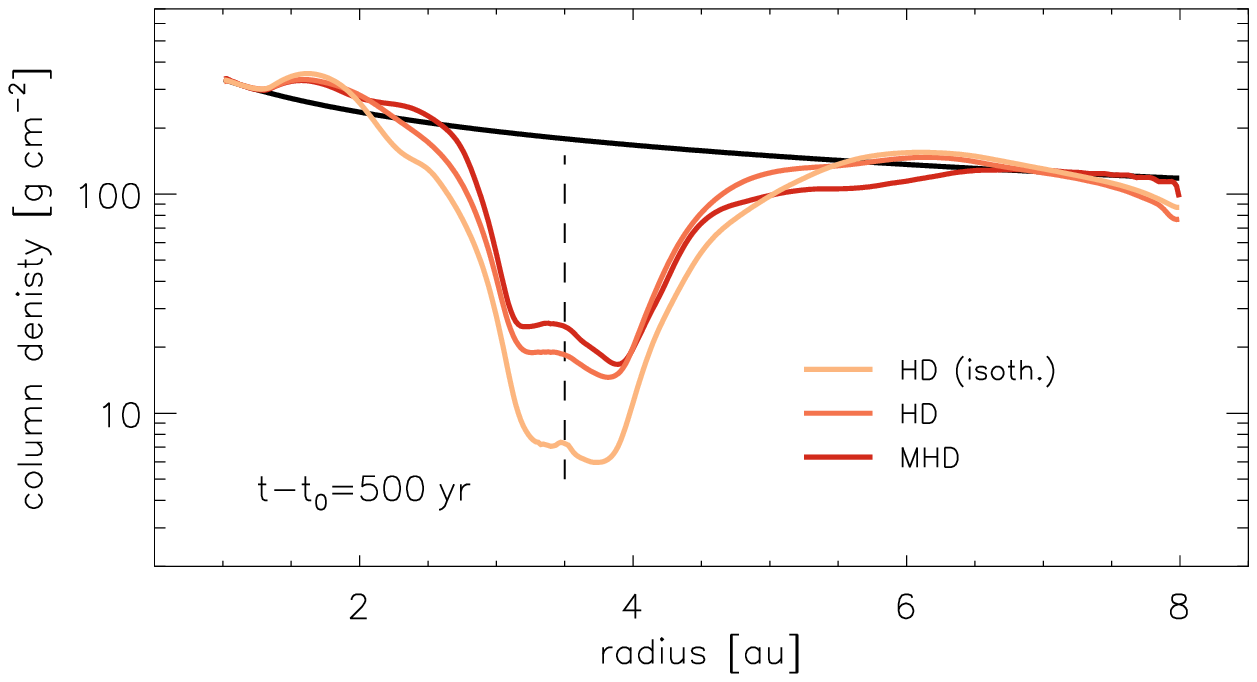}
    \caption{Radial gap profiles (including the mass contained in the
      CPD). \emph{Top:} for constant planet mass of
      $\Mp=150\Me$. \emph{Bottom:} at constant time $t-t_0=500\yr$,
      where the momentary planet masses are 143, 153, and $196\Me$ for
      models M1, N1, and N2, respectively. The solid black line
      indicates the initial profile; the dashed line shows the planet
      position.}
    \label{fig:gap_prof}
  \end{figure}

The gap opening process is illustrated in \Fig{fig:gap_evol}, where we
plot radial profiles of the disk surface density in regular intervals
of $\Delta t=121\yr$, corresponding to roughly 20 orbital times at the
planet location. While the gap has not reached a fully stationary
state, the deepening of the gap is clearly slowing down towards the
end of the simulation. Moreover, there is a trend towards deepening
the outer half of the gap, which is also visible in
\Fig{fig:midpl_slice}. This arises primarily because the disk is
cooler just outside of the planet location compared to the
corresponding region just interior to it. Waves excited at the outer
Lindblad resonances are therefore more nonlinear than their inner
Lindblad resonance counterparts. The density contrast between the edge
and the gap is roughly 20-30, which is a factor of 2-3 deeper than
reported for the local shearing-box models of
\citet{2008ApJ...685.1220M}. The periodic nature of shearing box
simulations may reduce the depth of the gap, an effect that is avoided
in our global simulations. We remark that their simulation had a
similar mass planet ($0.4\Mj\simeq 127\Me$), which was embedded in an
inviscid disk without magnetic fields or turbulence.

In \Fig{fig:gap_prof}, we show a comparison of the gap structure
between the different models. Because of the different accretion rates
(see discussion in Sect.~\ref{sec:acc_rates} below), we perform this 
comparison in two different ways: at constant planet mass of 
$\Mp=150\Me$ (upper panel), and at constant time $t-t_0=500\yr$ after 
the insertion of the planet (lower panel). Gap opening proceeds 
fastest in the isothermal-HD case, N2, presumably due to the more 
efficient mass growth of the planet in this setup. At a given planet mass, 
the model M1 shows the deepest gap, followed by the equivalent HD model, N1.

Comparing the simulation results at a fixed evolution time, our
non-isothermal MHD and HD runs have quite comparable gap structures
(lower panel of \Fig{fig:gap_prof}). There is a noticeable difference
in the gap edges, with the MHD model having a higher density inner
edge, and a lower density outer edge compared to the HD model. While
this may have implications for the detailed torque balance experienced
by the planet, a detailed examination of this issue is probably not
warranted due to the fact that none of the simulations have formed a
steady-state gap. Gap opening has been studied extensively in fully
MRI-active disks. While early studies neglected the vertical
stratification
\citep{2003MNRAS.339..993N,2003ApJ...589..543W,2004MNRAS.350..829P}
this has recently become possible to include
\citep{2011ApJ...736...85U}. These studies generally find that a giant
planet embedded in an MRI-turbulent disk opens a wider gap than one
embedded in an equivalent HD simulation that utilizes the $\alpha$
model for viscosity, and the upper panel of \Fig{fig:gap_prof}
provides some support for this. We caution, however, that direct
comparison with these previously reported trends is difficult because
the planet in this paper has a time dependent mass due to accretion,
unlike the previous studies that used a fixed mass.

\subsection{Consequences of evolving ionization fraction} 
\label{sec:ionize}

Formation of a gap reduces the column density in run M1, increasing
the midplane ionization fraction of disk material there through
increased penetration of X-rays and cosmic rays. In essence we observe
the dead-zone in the vicinity of the planet being ignited into a
turbulent state by gap formation. As we have discussed above, gap
formation does not run to completion in this simulation, so we find
that it is primarily the upper and intermediate layers of the disk in
the gap that share common characteristics with the MRI-active surface
layers of the rest of the PPD. The gap is deepest in the close
vicinity of the planet (i.e. in a wedge of $\phi \simeq\pm\pi/8$), and
the magnetic Reynolds number, $\Rm\equiv \cS H/\eta$, has high values
in the range $10^4$ -- $10^5$ there. At the same time, the Ohmic
Elsasser number, $\Lambda\equiv v_{\rm A}^2/(\Omega\eta)$, with
$v_{\rm A}$ the Alfv{\'e}n speed associated with the vertical field, is
well above unity.  Accordingly, we find this region in particular to
be in a turbulent state all the way down to the midplane. Values of
$\beta_{\rm p}$ are in the typical range for active MRI. As one moves
away from the planet position around the orbit the characteristics of
the flow change. Particularly in the midplane, and away from the
planet, $\Rm$ drops to values as low as $100$, which is generally
insufficient for sustained MRI \citep{2011ApJ...740...18O}. Moreover,
$\Lambda$ becomes smaller than unity; in agreement with linear theory
the flow remains laminar in these regions. At the same time, the
plasma there is only very weakly magnetized with $\beta_{\rm p}\sim
10^5$.

Taking a global view of the flow in the protoplanetary disk and gap
region, we can say that gas accretes through the disk toward the
planet mainly \emph{via} the upper active regions. As gas enters the
gap it is pulled down by the star's gravity and toward the planet by
its gravity. The low density there allows strong coupling between the
gas and the predominantly azimuthal magnetic field such that the field
is advected with the fluid into the gap, where it helps to sustain the
MRI.  Because our simulations do not allow gap formation to run to
completion, gas within the gap region at azimuthal locations away from
the planet remains laminar at the midplane. On longer time scales,
however, we would expect these regions to also become evacuated such
that the MRI could be sustained there also. The end result would be a
protoplanetary disk with active surface layers and laminar midplane
region far from the planet, but a magnetized and turbulent gap in the
planet's vicinity through which material passes as it flows into the
planet Hill sphere.

\subsection{Consequences of accreting from a layered disk} 
\label{sec:gap-accretion}

The feeding of gas into the gap region from the surface layers may
have important consequences for the chemistry and dust content of the
gas that eventually accretes onto the planet, as dust settling and
sequestration in the dead-zone may reduce the heavy element content of
this gas as it accretes through the disk. Possible implications of
this are discussed in Sect.~\ref{sec:discussion}.

\begin{figure}
  \includegraphics[width=\columnwidth]{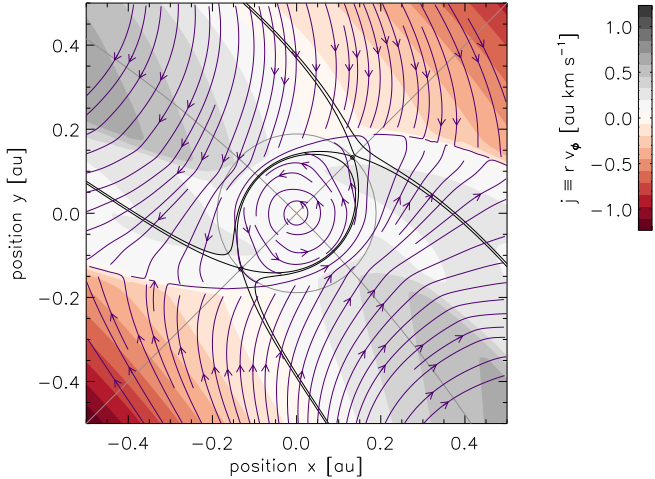}
  \caption{Time-averaged flow field (arrows) and specific angular
    momentum (color coded) for a midplane slice of the isothermal HD
    model. Averages are taken over $\sim 4$ planet orbits after the
    planet has reached a mass of $150\Me$.}
  \label{fig:flow_field_HD_iso}
\end{figure}

An interesting consequence of accretion through a gap in a layered
disk is that the gas flow through the surface layers into the gap
should be able to continue relatively unimpeded by tidal truncation of
the disk even for relatively massive planets, because the magnetic
stresses and associated $\alpha$ values in the upper layers tend to be
very large (i.e. $\alpha > 0.1$). As such, these disk layers will not
satisfy the viscous gap opening criterion discussed above, and should
continue to feed the planet as its mass grows, albeit at a moderate
rate because of the low densities in the upper layers. As we discuss
later in the paper, the accretion rates observed in the viscous
hydrodynamic runs N1 and N2 show a tendency to continuously decrease
as the planet mass increases and the gap deepens. Run M1, however,
appears to reach a steady accretion rate that is higher than observed
in the laminar runs by the end of the simulation (see
\Fig{fig:acc_rate}), apparently for the reason just described.


\section{The circumplanetary disk}
\label{sec:CPD}

We now discuss the features of the flow in our models on scales
appropriate to the planet Hill sphere radius. We begin by describing
the horizontal flow features in the midplane, before examining the
flow pattern in the meridional plane. We discuss and compare the
integrated properties of the circumplanetary disks that arise in all
of our simulations, before focusing on the details of the flow that
arise in the MHD simulation M1.

\subsection{Horizontal flow features in the midplane} 
\label{sec:horizontal}

Within the Hill sphere (where the gravitational force of the planet
dominates), we observe the formation of a circumplanetary disk in all
simulations as a consequence of the approximate conservation of
(specific) angular momentum, $j\equiv r v_\phi$ (with respect to the
planet's position).

\begin{figure}
  \includegraphics[width=\columnwidth]{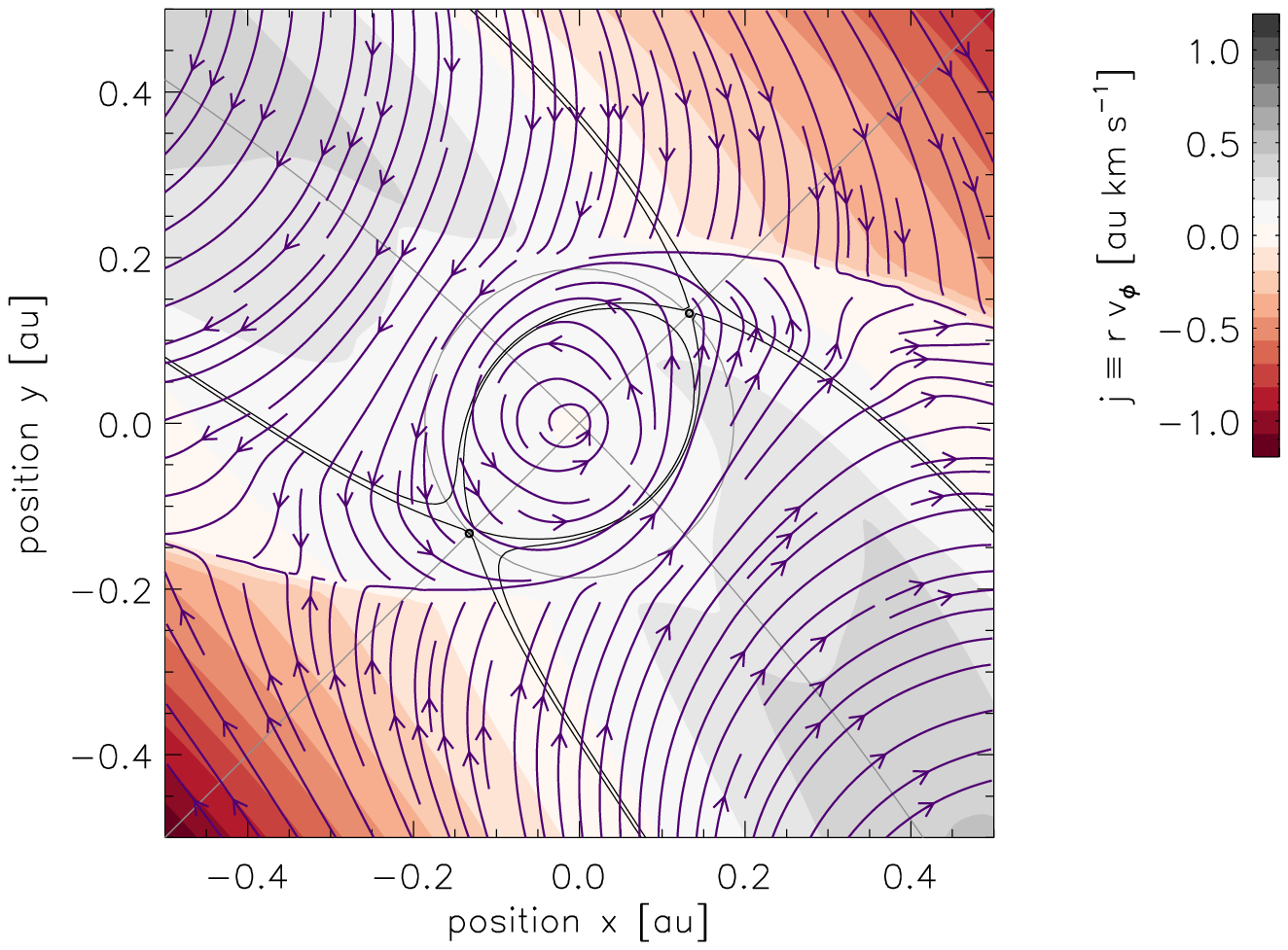}\\[1ex]
  \includegraphics[width=\columnwidth]{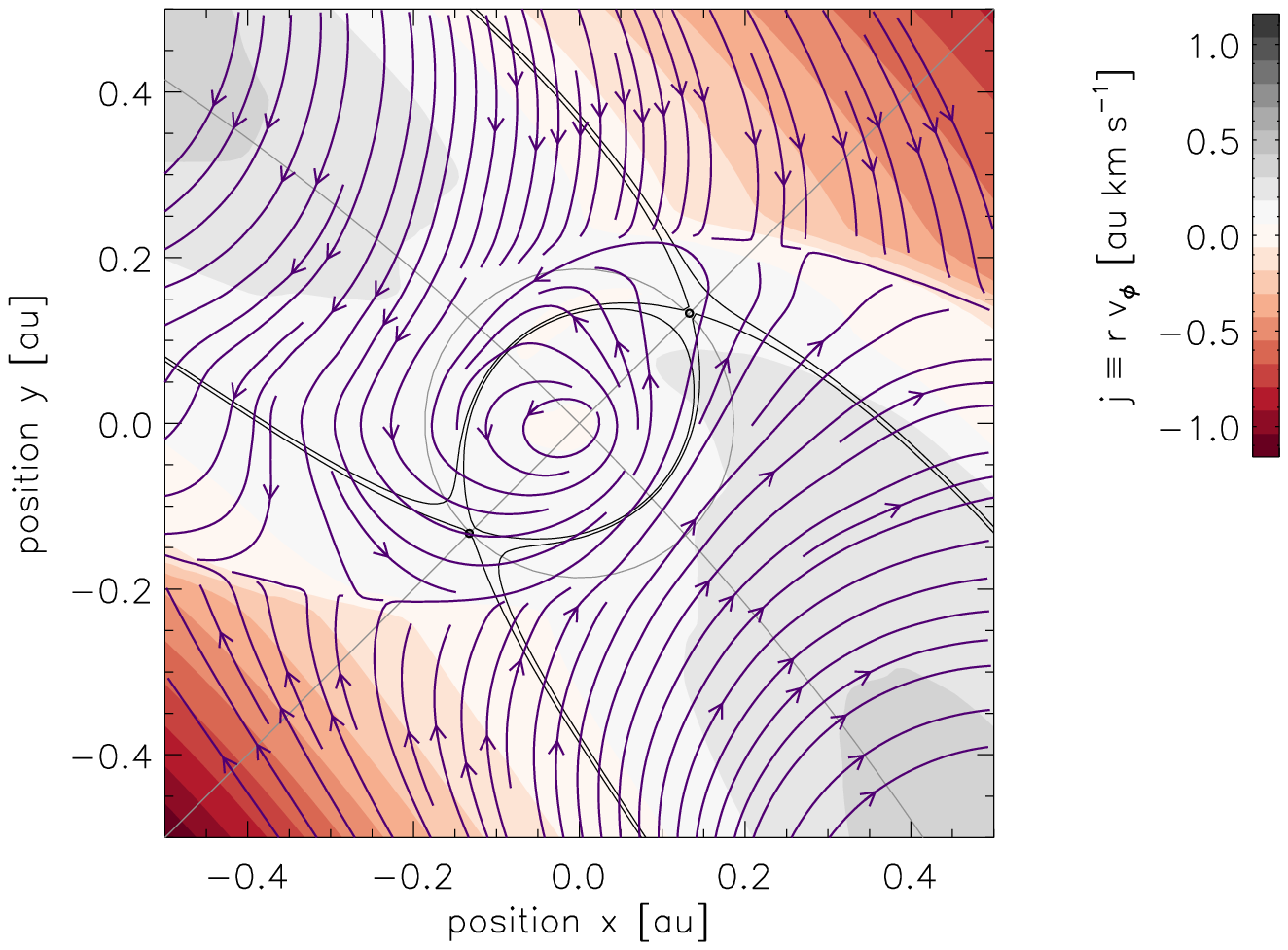}
  \caption{Same as \Fig{fig:flow_field_HD_iso}, but for the
    non-isothermal HD model (top), and the MHD model (bottom panel).}
  \label{fig:flow_field_tau}
\end{figure}

In \Fig{fig:flow_field_HD_iso} we plot the time-averaged specific
angular momentum of the flow region near the planet for the
locally-isothermal run N2. Superimposed streamlines show that the flow
has a high degree of symmetry. Marginally bound trajectories closely
follow the Roche lobe (equipotential surfaces through L1, L2 are shown
as black lines), indicating that the flow is only weakly affected by
pressure forces. The flow within the CPD is nearly circular at this
time, but is observed to become more distorted at late times as
discussed in Sect.~\ref{sec:details}. In comparison, the averaged flow
field appears less symmetric in the simulation with an adiabatic
equation of state and thermal relaxation towards the initial
temperature (upper panel of \Fig{fig:flow_field_tau}). Unlike in the
previous case, the circulating flow appears to extend beyond the Roche
lobe, and shows some level of distortion. This is partly due to the
fact that the upper panel in \Fig{fig:flow_field_tau} corresponds to a
later time when gap formation is more developed; we generally find
that the CPD flow becomes increasingly distorted as the simulations
evolve and gap formation becomes more pronounced. The distortion is
apparently also amplified by the enhanced pressure forces present in
run N1, combined with the in-out asymmetry that the circumplanetary
material displays because of the radial temperature profile imposed on
the gas.

This trend of reduced order in the fluid trajectories is enhanced
substantially when looking at the MHD case, shown in the lower panel
of \Fig{fig:flow_field_tau}: here the flow within the Hill sphere is
significantly distorted and less symmetric, presumably due to the
combined effect of pressure and magnetic forces. Notably, the flow
within the CPD appears to have a non-negligible eccentricity. Magnetic
field lines (not shown) are dominantly azimuthal (with respect to the
PPD) outside the horseshoe region and moderately compressed in the
spiral arms; notably $B_\phi$ has opposite sign in the inner and outer
part of the PPD. In the vicinity of the planet orbit (the horseshoe
region and just beyond), field lines are generally aligned with the
flow, with the exception of the post-spiral shock region where angles
up to $90\degr$ are obtained. Within the CPD, the field lines follow
the winding of the spiral shocks. Overall the field morphology is
consistent with that described by \citet{2003MNRAS.339..993N}.

\subsection{Poloidal and three-dimensional flow features} 
\label{sec:poloidal_flow}

\begin{figure}
  \center\includegraphics[width=\columnwidth]{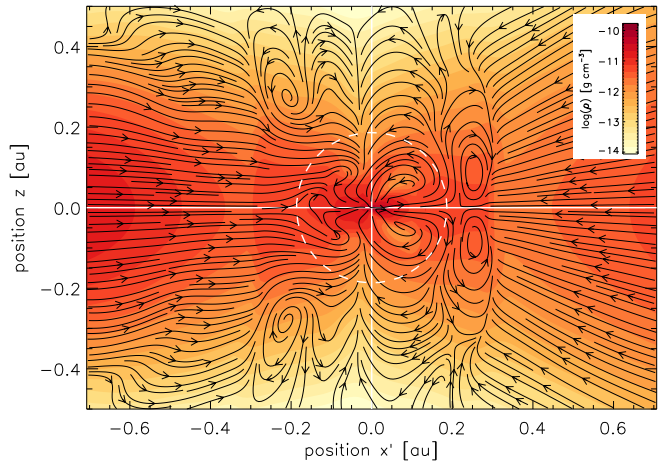}
  \caption{Projected flow field in the vertical plane connecting the
    planet with the star. Small vortical return flows within the Hill
    radius (dotted) are associated with under-dense regions already
    seen in \Fig{fig:polo_slice}. Note the strong vortex pair adjacent
    to the outer spiral arm.}
  \label{fig:flow_polo_HD_tau}
\end{figure}

During the early phase of evolution, prior to pronounced gap
formation, we find that the flow field observed in poloidal slices in
the isothermal run N2 is qualitatively very similar to the highly
symmetric renderings shown in figure~5 of \citet{2008ApJ...685.1220M}
and figure~4 of \citet{2012ApJ...747...47T}. This flow consists of
inflow toward the planet at high latitudes, and outflow toward the L1
and L2 points near the CPD midplane.  During late times, when gap
formation is more developed, we find that the flow symmetry breaks
down and becomes similar to that displayed by the non-isothermal HD
run N1, for which we plot poloidal flow lines near the end of the
simulation in \Fig{fig:flow_polo_HD_tau}. Within the outer half of the
Hill sphere, material is centrifugally spun-out towards the L2 point
near the CPD midplane and then recycled into the vertical accretion
flow along the vertical axis of the planet. Unlike at early times, and
in the (intrinsically symmetric) local simulations of
\citet{2008ApJ...685.1220M}, this is markedly not the case on the side
of the CPD facing the star, where we only observe inflow. This
difference is also clearly seen in
\Figs{fig:mass_flux_HD_iso}{fig:mass_flux_MHD_tau} in
\Sec{sec:details} below.

Outside the Hill sphere, the non-isothermal flow shows substantially
more complex features than the ones seen for an isothermal gas. Most
notably, there is a pair of vertically elongated poloidal vortices
sitting right outside, above and below the L2 point. These vortices
possibly have their origin through baroclinic generation of vorticity
(via the $\nabla P \times \nabla \rho$ term) in the vicinity of the
spiral shock. A similar but weaker (and more widely separated) pair is
seen on the side close to the star. Note that the apparent ending of
stream lines outside the vortices is a projection effect. Near the
spiral shock, the flow abruptly changes direction, pointing
into/out-of the plane shown in \Fig{fig:flow_polo_HD_tau}.

\begin{figure}
  \center\includegraphics[width=0.8\columnwidth]{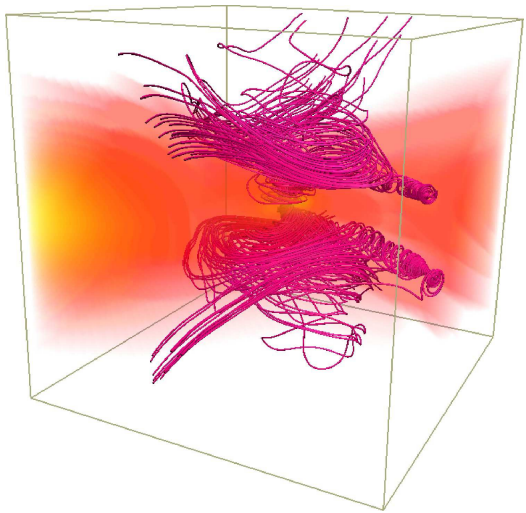}
  \caption{Backward extrapolation of stream lines passing through the
    close vicinity of the planet. Final snapshot from the
    non-isothermal HD model. The size of the cube is $1\au$, and the
    left corner points towards the star. For orientation, the gas
    density is shown. Adjacent to the outer spiral shock, a pair of
    counter-rotating vortices is visible.}
  \label{fig:flow_field_HD_tau}
\end{figure}

A three-dimensional volume rendering of the flow field of model N1 is
attempted in \Fig{fig:flow_field_HD_tau}, where we recognize the outer
vortex pair (the slice from \Fig{fig:flow_polo_HD_tau} passes through
the left-right diagonal here). These flow lines are traced from the
vicinity of the planet in the CPD, and demonstrate that inflow
primarily occurs from high latitudes and not along the midplane of the
CPD. This feature of the accretion flow is discussed in greater detail
in Sect.~\ref{sec:details}.  The flow structure becomes even more
tangled-up in the MHD case, which is not shown here. A detailed
discussion of the intrinsic variability seen in this run is presented
in Sect.~\ref{sec:mhd_cpd} below.

\subsection{Integrated properties of the CPDs} 
\label{sec:CPDs}

\begin{figure}
  \center\includegraphics[width=0.9\columnwidth]{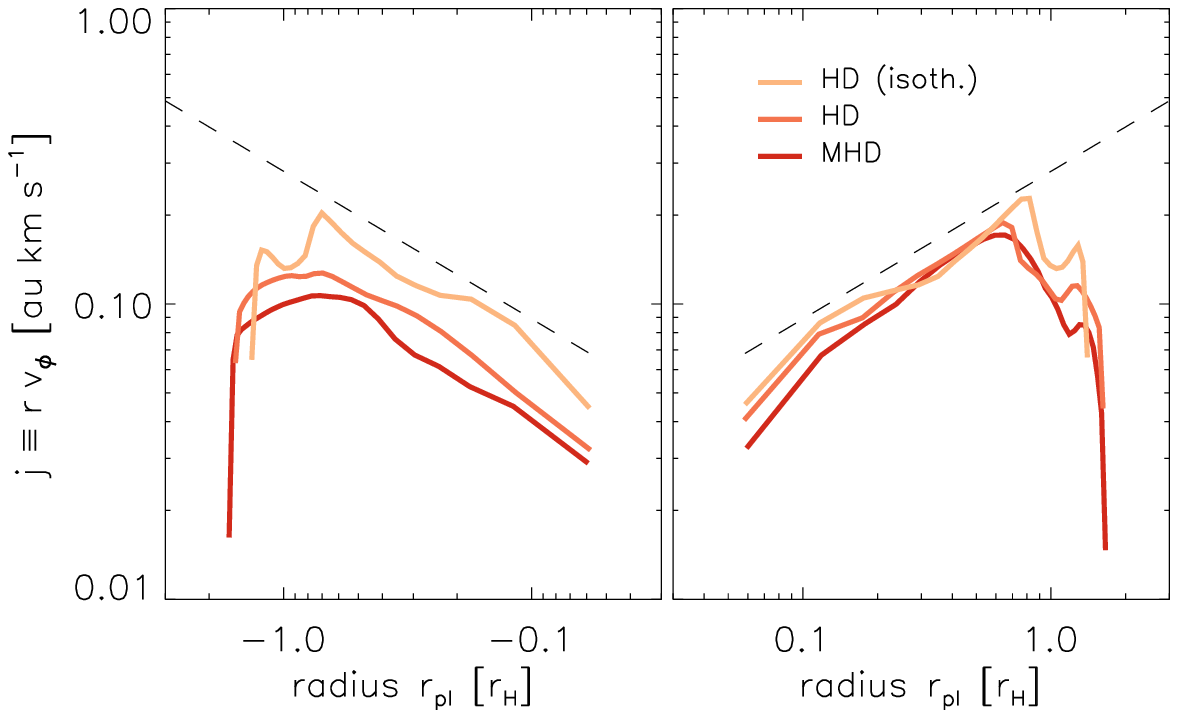}
  \caption{Time-averaged (as in \Fig{fig:flow_field_HD_iso}) specific
    angular momentum profile within the CPD midplane along the ray
    connecting to the star. The left (right) panel shows the half of
    the CPD facing towards (away from) the star. Keplerian rotation is
    indicated by the dashed line.}
  \label{fig:spcf_angm_cmp}
  \center\includegraphics[width=0.9\columnwidth]{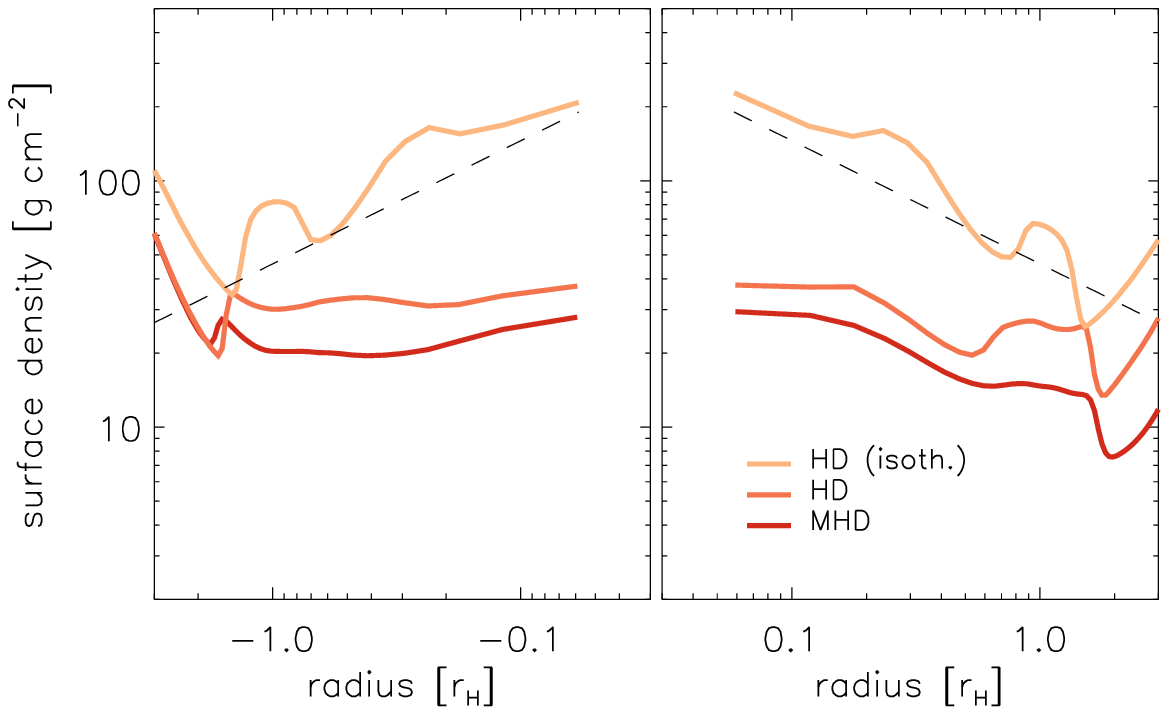}
  \caption{Time-averaged surface density profiles for the three
    cases. Lines represent a radial slice through the CPD
    midplane. The dashed line indicates a surface density profile
    $\sim r^{-0.5}$.}
  \label{fig:surface_cmp}
\end{figure}

In \Fig{fig:spcf_angm_cmp}, we compare time-averaged angular momentum
profiles within the CPD midplane of the three main models.
Differences, due to additional radial pressure support, are more
pronounced for the part of the disk closer to the planet. This is
consistent with the geometrically thicker inner disk seen in the
insets of \Fig{fig:polo_slice}. As expected in the absence of strong
pressure gradients (that might arise because of compressional heating
near the planet), the isothermal model is closest to the Keplerian
rotation profile (dashed line). In the MHD model, the CPD rotates the
slowest, indicating that magnetic forces, at least at some level,
contribute to the overall structure of the disk. Estimating the size
of the CPD according to the radius at which the specific angular
momentum begins to turn, we find the CPD extends to about half the Hill
radius, which is similar to but slightly larger than the CPDs reported
in other studies where values of one-third of the Hill sphere radius
\citep{2009MNRAS.397..657A,2008ApJ...685.1220M,2012ApJ...747...47T} or
$0.4 \rH$ \citep{2011MNRAS.413.1447M} have been estimated.  These
estimates are influenced by the disk temperature because of the role
of pressure forces at the CPD outer edge, and also by the properties
of the inflowing gas such as its angular momentum, so we believe our
results are consistent with those obtained in previous studies.

\begin{figure}
   \center%
   \includegraphics[width=\columnwidth]{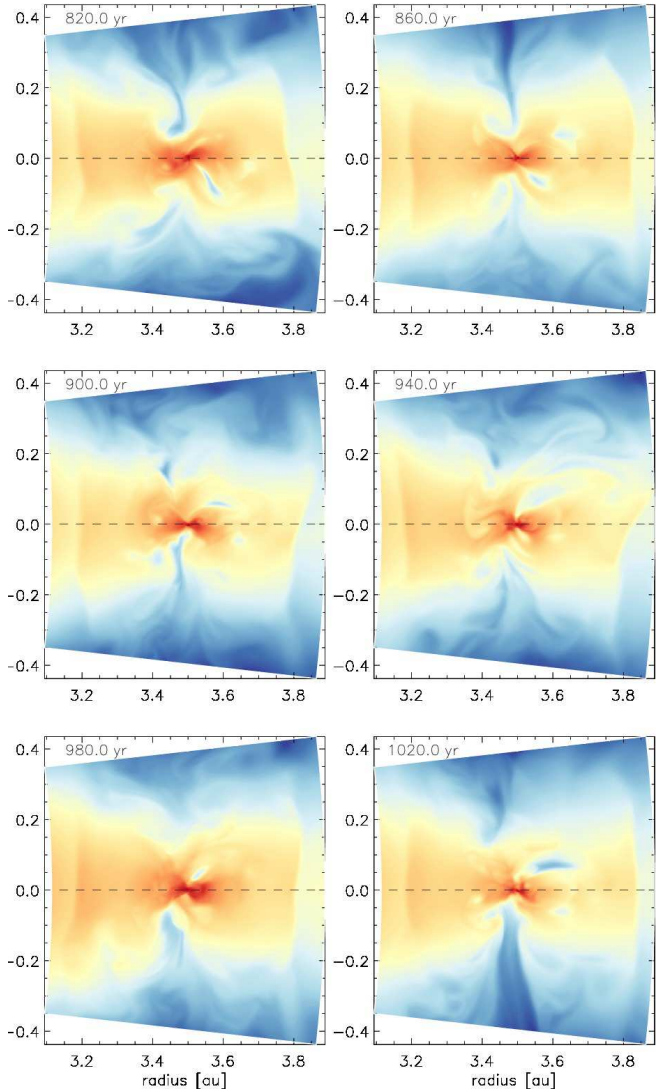}\\[1ex]
   \caption{Time sequence of edge-on slices of the gas density for the
     MHD model illustrating the stochastic, heavily time-varying
     nature of the CPD. While the low-density, tornado-like structures
     (roughly aligned with the rotation axis of the CPD) are
     associated with inflow, the funnel seen in the lower half of the
     last panel is in fact a collimated jet-like outflow
     \citep[see][]{2003A&A...411..623F}.}
   \label{fig:slice_seq_edge_MHD_tau}
   \vspace{0.5ex}
 \end{figure}

The radial surface density profile of the CPDs from all models is
shown in \Fig{fig:surface_cmp}. Unlike the two non-isothermal models,
the isothermal HD model shows a cusp-like inner disk structure, as
well as pronounced spiral features \citep[cf.][]{1999ApJ...526.1001L}
leading to peaks in the surface density profile. We find the profile
is close to a $\sim r^{-0.5}$ dependence, which is shallower than the
$\sim r^{-1.5}$ dependence reported by
\citet{2012ApJ...747...47T}. The reason for this minor discrepancy is
likely to be our adoption of a larger sink hole radius.  When applying
a non-isothermal equation of state, we find lower and nearly constant
surface densities, and much weaker spiral features due to the higher
temperatures. The inclusion of magnetic field leads to a further
reduced disk mass. Generally, the observed surface densities are in
the range of the ``gas-starved'' scenario of
\citet{2002AJ....124.3404C,2006Natur.441..834C}, which is
significantly lower than in the ``minimum mass'' models of
\citet{2003Icar..163..198M}. If we integrate the mass contained within
the Roche region (black lines in
\Figs{fig:flow_field_HD_iso}{fig:flow_field_tau}), we obtain
$0.041\Me$ for model M1, $0.063\Me$ for model N1, and $0.196\Me$ for
model N2, respectively. These numbers should be contrasted with the
the Jovian satellite system mass $\simeq 0.067\Me$, which if augmented to 
solar abundance would imply the presence of $\simeq 7\Me$ of gas.

\subsection{Time-dependent morphology in the MHD case} 
\label{sec:mhd_cpd}

\begin{figure}
   \center%
   \includegraphics[width=\columnwidth]{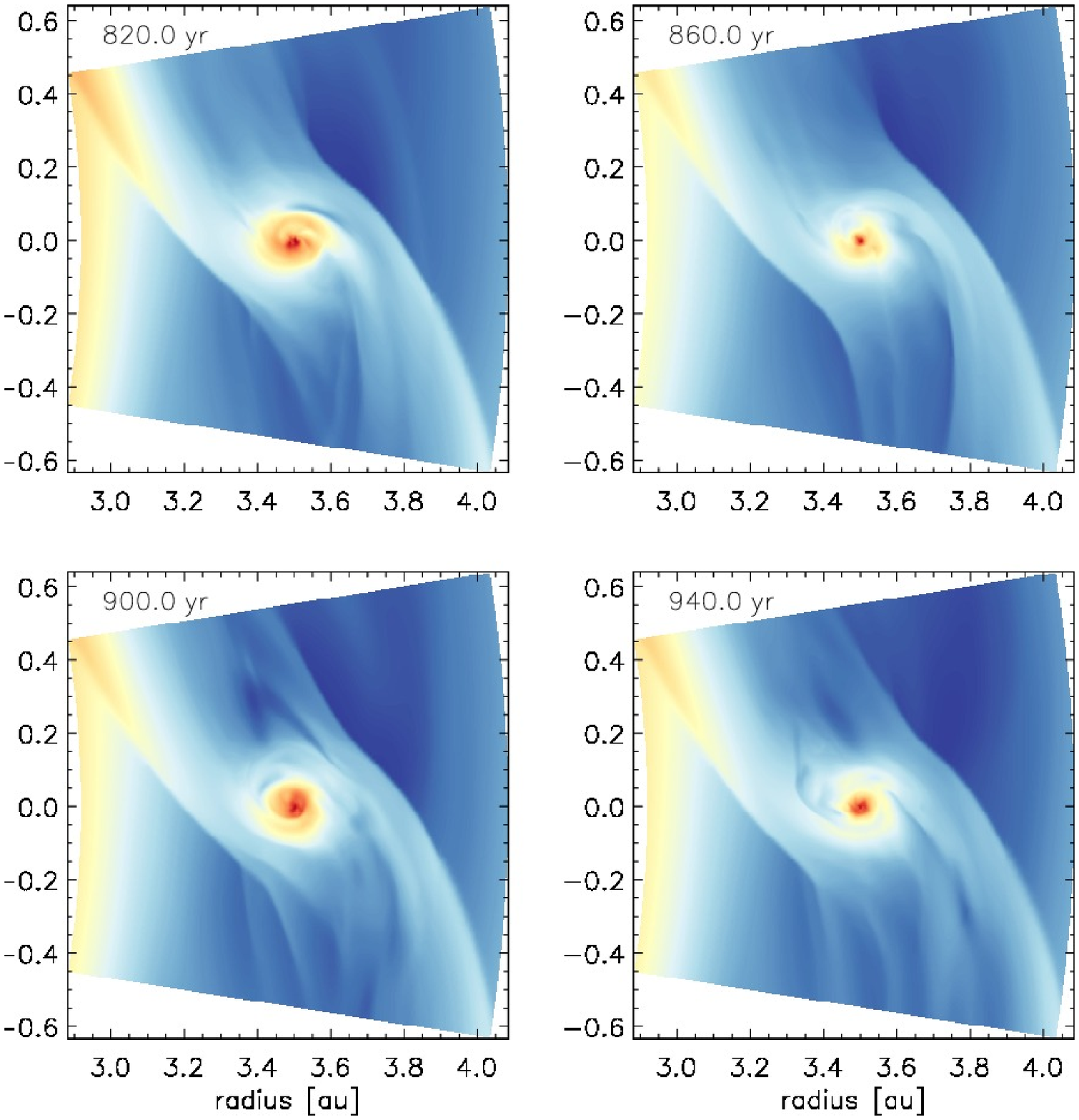}\\[1ex]
   \caption{Similar to \Fig{fig:slice_seq_edge_MHD_tau}, but for
     midplane slices of the gas density.}
   \label{fig:slice_seq_mid_MHD_tau}
\end{figure}

Having inter-compared the gross properties of the CPD regions for our
three runs, we now consider the detailed and time-dependent structure
of the CPD region for the magnetized run M1. We leave discussion about
the geometry of the actual accretion flow onto the planet until
Sect.~\ref{sec:details}. In \Fig{fig:slice_seq_edge_MHD_tau} we
present a series of snapshots showing vertical slices of the density,
demonstrating the substantial time-variability of the accretion flow
in this region because of the turbulence in the gap that feeds
material into the planet Hill sphere. We see low-density patches
within the CPD that are connected with the poloidal vortices that
develop within the flow, described in Sect.~\ref{sec:poloidal_flow}
above. A similar level of temporal variation is seen in the sequence
of midplane slices shown in \Fig{fig:slice_seq_mid_MHD_tau}.

\begin{figure}
   \center\includegraphics[width=\columnwidth]{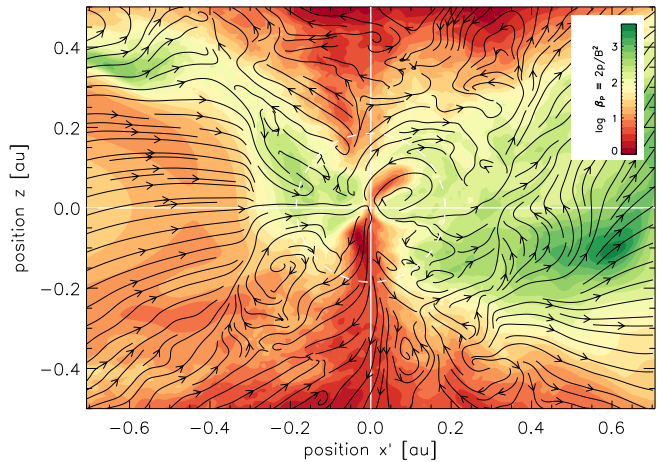}
   \caption{Radial-vertical slice of the logarithm of the plasma
     parameter, $\beta_{\rm P}\equiv 2p/B^2$, and projected magnetic
     field lines. Averages are taken over eight planet orbits at the
     time corresponding to the last panel in
     \Fig{fig:slice_seq_edge_MHD_tau}.}
   \label{fig:betap}
\end{figure}
 
We plot a vertical slice through the circumplanetary disk region
showing the value of the time-averaged plasma $\beta_{\rm p}$ value in
Figure \ref{fig:betap}. The CPD is relatively weakly magnetized, with
$\beta_{\rm p}$ typically in the range $10^2$-$10^3$, while material
falling in through the funnel-shaped regions above and below the
planet are characterized by equipartition-strength magnetic
fields. The field topology in this region can very crudely be
described as follows: the large-scale azimuthal field in the
MRI-active layers is dragged downwards with the accretion flow. Within
the Roche lobe, the cusp of the now V-shaped field lines is twirled-up
by the circular flow around the planet, producing a helical field
topology. One consequence of this field evolution appears to be the
launching of a relatively diffuse, accelerated wind from the conical
regions above and below the circumplanetary disk. More localized
vertical outflows (e.g. note the mushroom-shaped bow shock seen in the
upper half of the fourth panel in \Fig{fig:slice_seq_edge_MHD_tau})
appear sporadically during the simulation but generally do not display
high levels of collimation and are quickly disrupted by the vigorous
MRI turbulence.

\begin{figure}
  \center\includegraphics[width=\columnwidth]{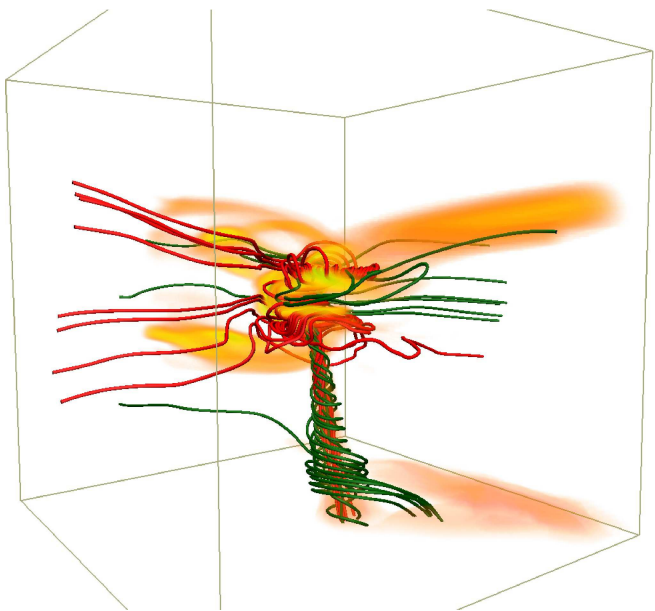}\\[1ex]
  \caption{Volume rendering of the single-sided, cone-shaped outflow
    seen at one instance in time during the MHD run. Selected flow
    lines passing through the vicinity of the planet (red) are shown
    within a cube of $1.5\au$ (with the viewing direction being very
    similar to \Fig{fig:betap} above). Draped around them is a helical
    magnetic field (green lines) leading to collimation of the outflow
    \emph{via} the magnetic hoop stress. Shaded volumes indicate
    regions of strong Ohmic dissipation.}
  \label{fig:jet}
\end{figure}

In contrast to this rather diffuse disk wind just described, the MHD
simulation did produce one episode of sustained, collimated outflow as
shown in the last panel of \Fig{fig:slice_seq_edge_MHD_tau}. This is
more dramatically illustrated by the volume rendering in \Fig{fig:jet}
which shows both the fluid streamlines and magnetic field
topology. This collimated outflow lasted for a period of about four
planet orbits before it was squashed by infalling material.
Unfortunately, the non-aligned mesh geometry and intrinsic variability
make the outflows hard to access quantitatively, so the driving
mechanism cannot be pinned down definitively, but it is clear from the
twisted field wrapped around the outflowing gas that the
protoplanetary jet seen in \Fig{fig:jet} is collimated by magnetic
hoop stresses. Given the importance of magnetic field advection and
rotation in the vicinity of the planet, our suggestion that the
outflow is magnetocentrifugally driven is well-motivated, but
speculative at the present time. It is possible that field winding
generates sufficient magnetic pressure to launch the jet at its base
instead, although this is doubtful because field amplification in the
circumplanetary disk is damped efficiently by Ohmic diffusion (see
volume rendering in \Fig{fig:jet}). A deeper analysis of
protoplanetary jet launching will be presented in a future
publication. We note that CPD jets have been predicted on theoretical
grounds by \citet{1998ApJ...508..707Q} and
\citet{2003A&A...411..623F}, and they have been observed in local
shearing box simulations by \citet{2006ApJ...649L.129M}. The physical
set-up in this latter study was quite different from the one we
consider here, however, because we initiate the simulation with a weak
magnetic field, and strong local fields are built up in the vicinity
of the planet by gas accretion and associated advection of magnetic
field from the surrounding PPD at late times in the simulation. The
study by \citet{2006ApJ...649L.129M}, however, employed an
equipartition strength vertical magnetic field, leading to the
formation of a well-defined bipolar outflow shortly after the
simulation was initiated.  Although interesting as phenomena in their
own right, there is no evidence at present that the launching of a CPD
jet has a significant influence of the growth of the
planet. Examination of the gas accretion rate onto the planet does not
indicate that it is influenced significantly during the time of
strongly collimated jet launching, implying that momentum and energy
injection into the surrounding infalling envelope is not an important
process here. A scenario in which wind or jet launching from a CPD may
be important for determining the accretion rate onto the planet
corresponds to a scaled-down version of the picture presented by
\citet{2013ApJ...769...76B}, where the combined effects of Ohmic
resistivity and ambipolar diffusion cause angular momentum transport
and mass accretion in a protoplanetary disk to arise through the
launching of a magnetized wind. If such a scenario also applies to
accretion onto a planet through a CPD, then wind launching will be
central to determining the accretion rate onto the planet.

\begin{figure}
  \center\includegraphics[width=\columnwidth]{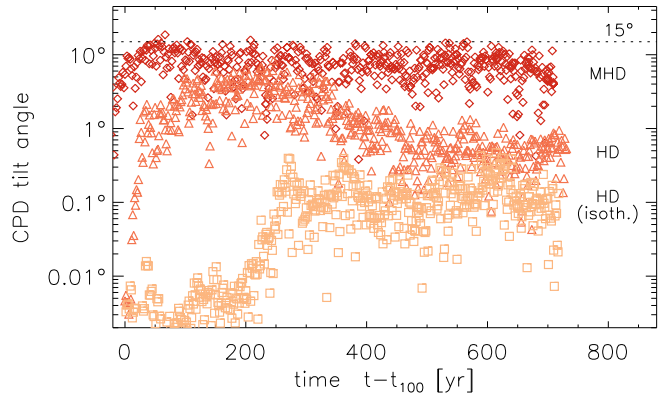}
  \caption{Tilt angle of the mean angular momentum vector (of all the
    material within the Roche lobe) as a function of time $t-t_{100}$,
    i.e., after each planet has reached its initial mass of
    $100\Me$.}
  \label{fig:cpd_tilt}
\end{figure}

The stochastic nature of the accretion flow into the planet Hill
sphere implies that the angular momentum vector of the CPD material
should not maintain a constant orientation. This expectation is
confirmed by \Fig{fig:cpd_tilt} which shows the time evolution of the
angle between the angular momentum vector and the vertical axis
calculated for all material inside the planet Roche lobe for the
various simulations. As expected, we see that both of the HD runs
maintain a reasonably constant direction for the CPD angular momentum
vector after an initial period of relaxation, with the tilt angle
remaining $<1\degr$. The MHD run, however, shows high levels of
variability over the full time span of the run, with tilt angles
reaching up to $15\degr$. Although this measurement applies to all
material in the Roche lobe, and not just the material in the
well-defined CPD confined to less than half the Hill radius, it seems
very likely that the inner part of the CPD close to the planet will
experience substantial disturbance that causes it to tilt and
precess. In particular, stochastic accretion of material with
different specific angular momenta will cause local warping on orbital
time scales that may excite bending waves that propagate through the
circumplanetary disk
\citep[e.g.][]{1995ARA&A..33..505P,1996MNRAS.282..597L}, with
interesting consequences for its dynamical evolution and the formation
of regular satellite systems in the inner regions. Indeed, we
speculate that when all other things are equal, a higher mass planet
will host a circumplanetary disk that is subject to reduced levels of
perturbation in the inner satellite forming regions compared to a
lower mass planet, by virtue of the strength of the local
gravitational field relative to the external perturbing forces that
are independent of the planet mass. This may be of relevance when
comparing the regular satellite systems of Jupiter and Saturn, where
the Jovian system clearly is more substantial than Saturn's.

At late times during the MHD simulation, the component of the CPD that
we model is itself marginally unstable to the MRI -- i.e., according
to the Elsasser number, $\Lambda \sim 0.2-20$, based on Ohmic
diffusivity alone. We would predict this to be the case from the
surface density values displayed earlier in \Fig{fig:surface_cmp}. The
penetration depth of X-rays is $\sim 9\g\cm^{-2}$ and for cosmic rays
it is $\sim 90\g\cm^{-2}$. Allowing for partial attenuation of these
ionizing sources by overlying material in the gap region, we would
predict that the circumplanetary disk is MRI-active based on the
measured column densities \citep[but see][for a more detailed
  assessment]{2013arXiv1306.2276T}. The magnetic Reynolds number is
rather low ($\Rm\simeq 100-1000$), which implies that a relatively
strong vertical net flux will be required to facilitate sustained
MRI. Build-up of strong fields through advection into the Hill sphere,
and subsequent amplification of the toroidal component by differential
rotation, is opposed by Ohmic diffusion, so we might reasonably expect
the disk to attain a marginal state of MRI.

Given the stochastic nature of the accretion flow and the associated
tilting and warping of the disk, it is hard to assess whether the CPD
in our MHD simulations is in fact laminar or MRI-turbulent. Within the
restrictions of our diffusivity model, and based on the associated
dimensionless numbers, the latter possibility can certainly not be
excluded. If we, on the other hand, estimate the vertical wavelength
of the fastest growing (ideal) MRI mode, we find a typical value of
$\lambda_{\rm MRI} \simeq 5\times 10^{-3}\au$ -- approximately
coinciding with the Nyquist frequency of our finest grid (and hence
falling short of the \emph{minimal} resolution requirement by roughly
a factor of five). In view of this issue, and regarding a possible
existence of a disk-wind and/or magneto-centrifugal jet, even better
resolved mesh-refined models are certainly called-for.


\section{The accretion flow onto the planet}
\label{sec:acc_flow}

As has been pointed out in various previous studies
\citep[e.g.][]{2006A&A...445..747K,2008ApJ...685.1220M,2013ApJ...767...63S},
the accretion flow onto the planet in a laminar disk is genuinely
three-dimensional. This is also confirmed by all our models which
develop complex flow structures in both the vertical and horizontal
directions, with inflow toward the planet arising primarily from high
latitudes. In agreement with previous high-resolution studies of the
accretion flow onto a giant planet from laminar, inviscid, isothermal
disks \citep{2008ApJ...685.1220M, 2012ApJ...747...47T}, we also find
that material flows \emph{away} from the planet near the midplane of
the CPD and toward the planet from higher latitudes in our locally
isothermal run.  Outflow away from the planet near the midplane of the
CPD is also observed in the two non-isothermal runs, although the
situation in these runs is less clear-cut.

\subsection{Mass accretion rates} 
\label{sec:acc_rates}

\begin{figure}
  \includegraphics[height=0.495\columnwidth]{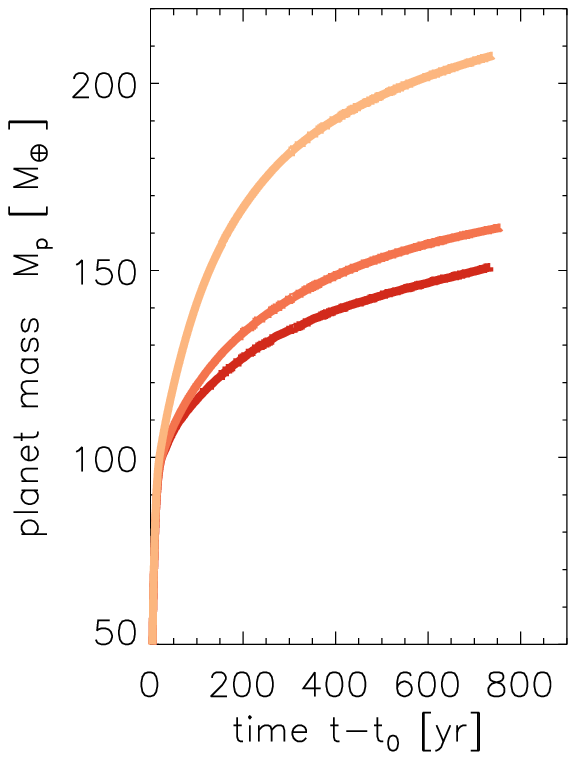}\hfill
  \includegraphics[height=0.495\columnwidth]{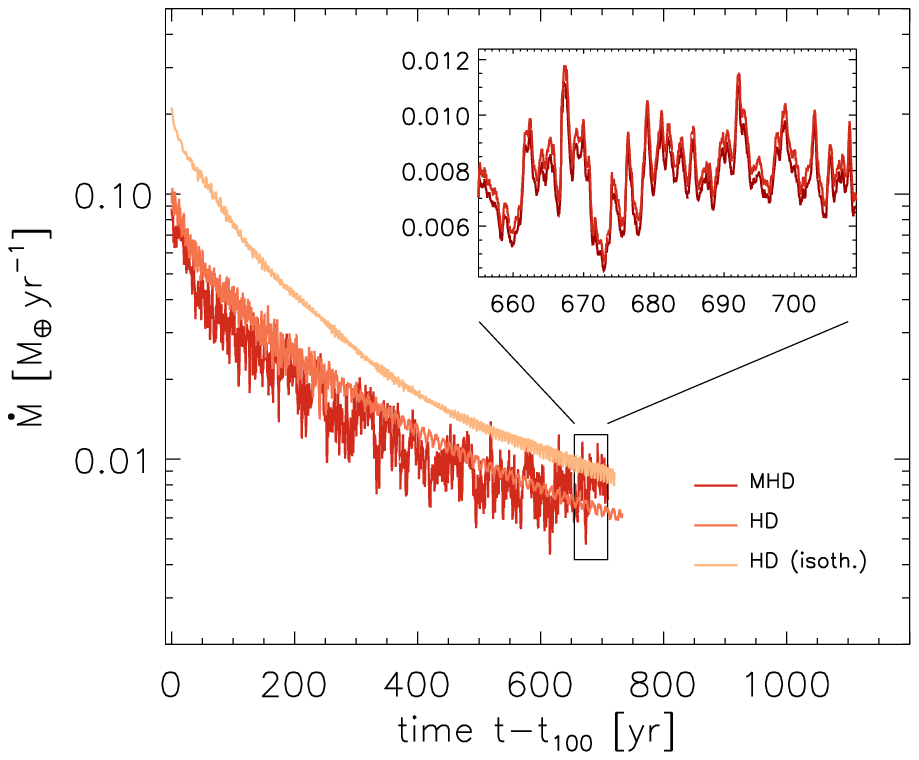}
  \caption{\emph{Left:} Evolution of the planet mass after insertion
    ($t_0$). \emph{Right:} Mass accretion rates after $100\Me$ are
    reached ($t_{100}$). The inset additionally shows (for the MHD
    case) the mass accretion rate into a sphere with $2\,r_{\rm acc}$,
    demonstrating that the flow near the sink is essentially
    ballistic.}
  \label{fig:acc_rate}
  \vspace{1ex}
\end{figure}

\begin{figure*}
  \begin{center}
    \includegraphics[width=1.7\columnwidth]{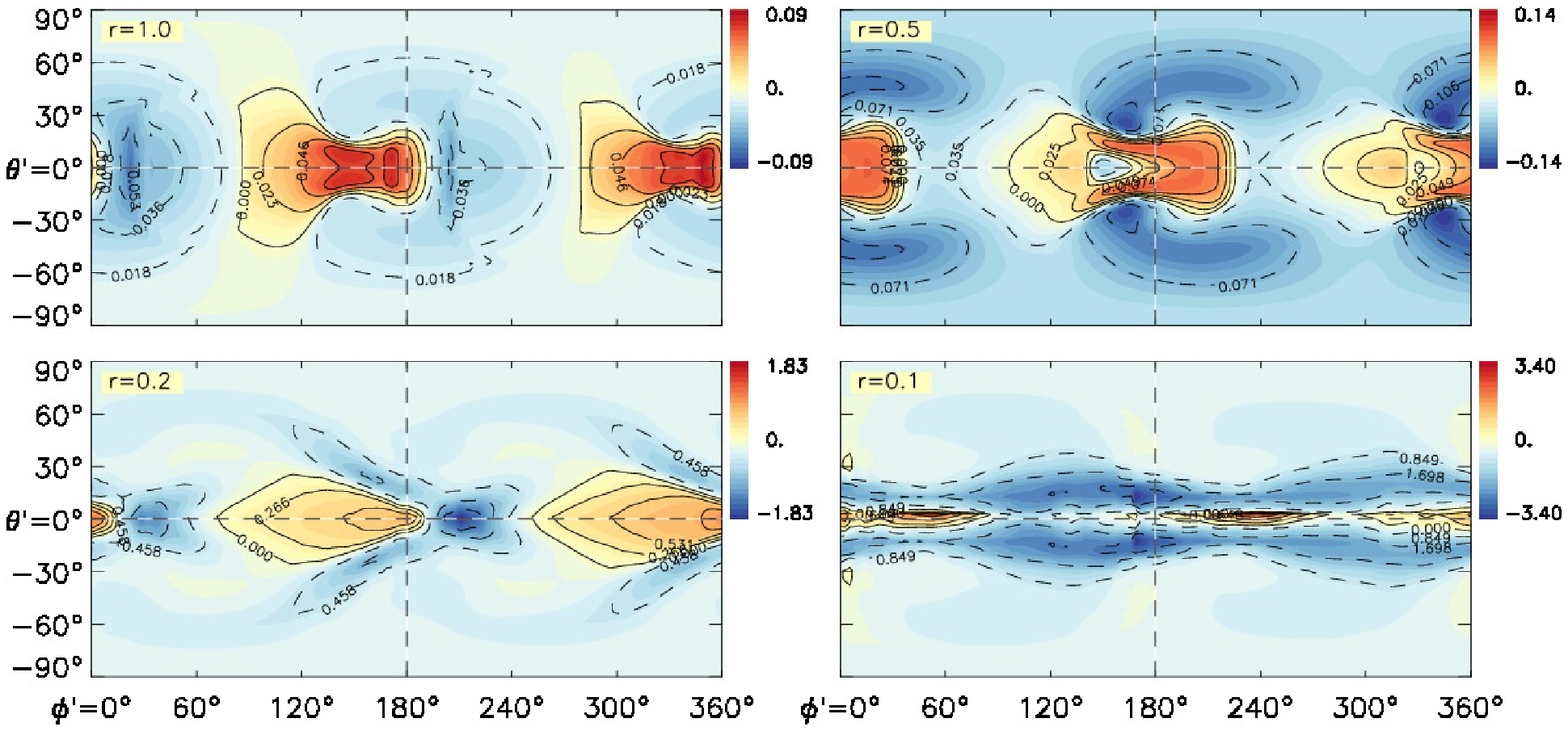}\\[5ex]
    \includegraphics[width=1.7\columnwidth]{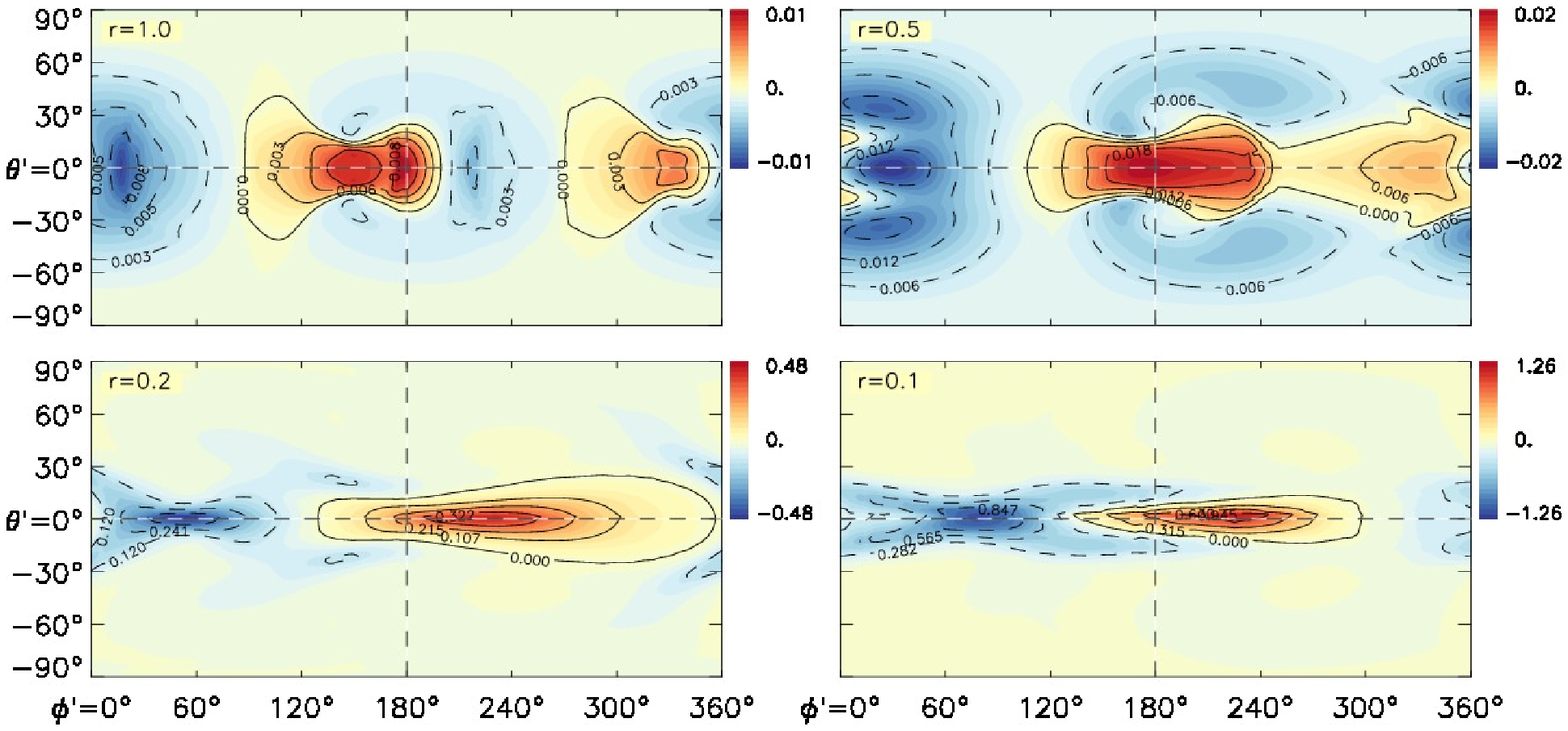}
  \end{center}
  \caption{Angular distribution of mass flux ($\rho\,v_r$) into
    spherical shells around the planet (with $r=1$, 0.5, 0.2, and 0.1
    Hill radii) for the isothermal HD case at early times (top four
    panels), and at the end of the simulation (bottom four
    panels). The coordinates $(\theta',\phi')=(0,0)$, and $(0,180)$,
    correspond to the sub-solar and anti-solar points,
    respectively. Inflow is represented by negative values.}
  \label{fig:mass_flux_HD_iso}
\end{figure*}

Before discussing the complex geometry of the accretion flow onto the
planet we consider the total accretion rates that are produced by the
simulations. The left panel of \Fig{fig:acc_rate} shows the evolving
planet mass for the three models.\footnote{Note that we artificially
  enhance the mass accumulated by the planet by a factor of four to
  speed-up tidally-induced gap opening. The reported accretion rates,
  however, reflect the actual gas flow onto the planet and do not
  include this factor of four enhancement.} The mass accretion rate
into the sink particle is plotted in the right panel of the same
figure. The inset shows a magnification of the late accretion phase in
model M1. The additional (darker) line shows the mass flux through a
spherical surface with twice the radius compared to the actual
``sink'' region, within which gas is removed from the domain. We
remind the reader that the constant sink hole radius equals 5\% of the
Hill sphere radius of the initial planet mass $100 \Me$. The minimal
offset between the curves demonstrates that the mass flux toward the
planet is essentially constant near the accretion sink, indicating
that the flow of gas into the sink hole is not retarded by any
processes occurring in this region. For the magnetized model, after
$500\yr$, we reach a quasi-stationary state, for which we infer a mass
accretion rate of $(8.0 \pm 1.4)\times 10^{-3}\Me\yr^{-1}$. In other
words, a Saturn-mass planet can grow to become a Jovian mass within
$\sim 25,000\yr$, a small fraction of the expected disk life-time. In
comparison, the total Maxwell stress measured within the CPD is
inferred as $\alpha_{\rm M}\simeq 0.01$, translating into an estimated
accretion rate of roughly $10^{-3}\Me\yr^{-1}$, indicating that
probably only a fraction of the mass is actually delivered \emph{via}
viscous transport within the disk.

In the other two models, the accretion rate still drops, presumably
since gap formation has not completed yet, and there is no MRI-active
surface layer with large stresses replenishing material into the gap
region.  It is evident, however, that contraction of gas onto the
planet is more efficient in the locally-isothermal model which lacks
compressional heating. Even though we apply cooling on the local
dynamical time scale to material deep in the planet Hill sphere in the
non-isothermal models, compressional heating influences the dynamics
significantly through the build-up of a central pressure gradient,
with consequent reduction in the accretion rate. As described earlier
in Sect.~\ref{sec:disk_struct}, the temperature in the non-isothermal
runs reaches values of $T \simeq 2000\K$ in the vicinity of the planet
despite the rapid cooling applied there, whereas in the locally
isothermal run it remains fixed at $T \simeq 150\K$. It is clear that
in addition to consideration of MHD processes, an \emph{accurate}
model of planetary accretion also needs to account for the influence
of radiation transport on the infalling material, even during the
runaway growth phase of giant planet formation. Nonetheless, the
accretion rates between all models vary by less than a factor of two
at the end of the simulation when a quasi-steady state has been
reached, indicating that qualitatively the results are in agreement
regarding rapid gas accretion onto a growing Saturn-mass planet. We
conclude that the adoption of two different equations of state induces
only a moderate change in the medium-term accretion rates, in
agreement with previous analyzes
\citep[e.g.][]{2009MNRAS.393...49A,2010MNRAS.405.1227M}.  In
particular, we note that \citeauthor{2010MNRAS.405.1227M} quote
accretion rates of (6--18) $\times 10^{-3}\Me\yr^{-1}$ in their
discussion, in excellent agreement with our results.

Unlike in the viscous hydrodynamic runs, the accretion in model M1 is
stochastic, showing variation by a factor of two over time scales
$\approx 5\yr$. In terms of the average accretion rate, however, the
non-isothermal HD and MHD runs produce very similar values. The MHD
run produces a slightly smaller accretion rate during early phases,
when magnetic pressure effects within the Hill sphere seem to
moderately impede accretion, but at later times when a significant gap
has formed the accretion rate for this run levels-off at a higher
value than observed for the HD run. As discussed above, this latter
effect seems to be due to the large magnetic stresses operating in the
MRI-active regions near the PPD surfaces that are not present in a
viscous model with constant $\alpha$. What is clear from these
simulations is that magnetic effects within a semi-realistic
protostellar disk model do not provide a barrier to planetary gas
accretion on scales larger than 5\% of the Hill sphere radius. As far
as we can tell from our simulations, giant planets can accrete
substantial gaseous envelopes within $\sim 3 \times 10^4\yr$, in broad
agreement with earlier studies of non-magnetized disks.

\begin{figure*}
  \begin{center}
    \includegraphics[width=1.7\columnwidth]{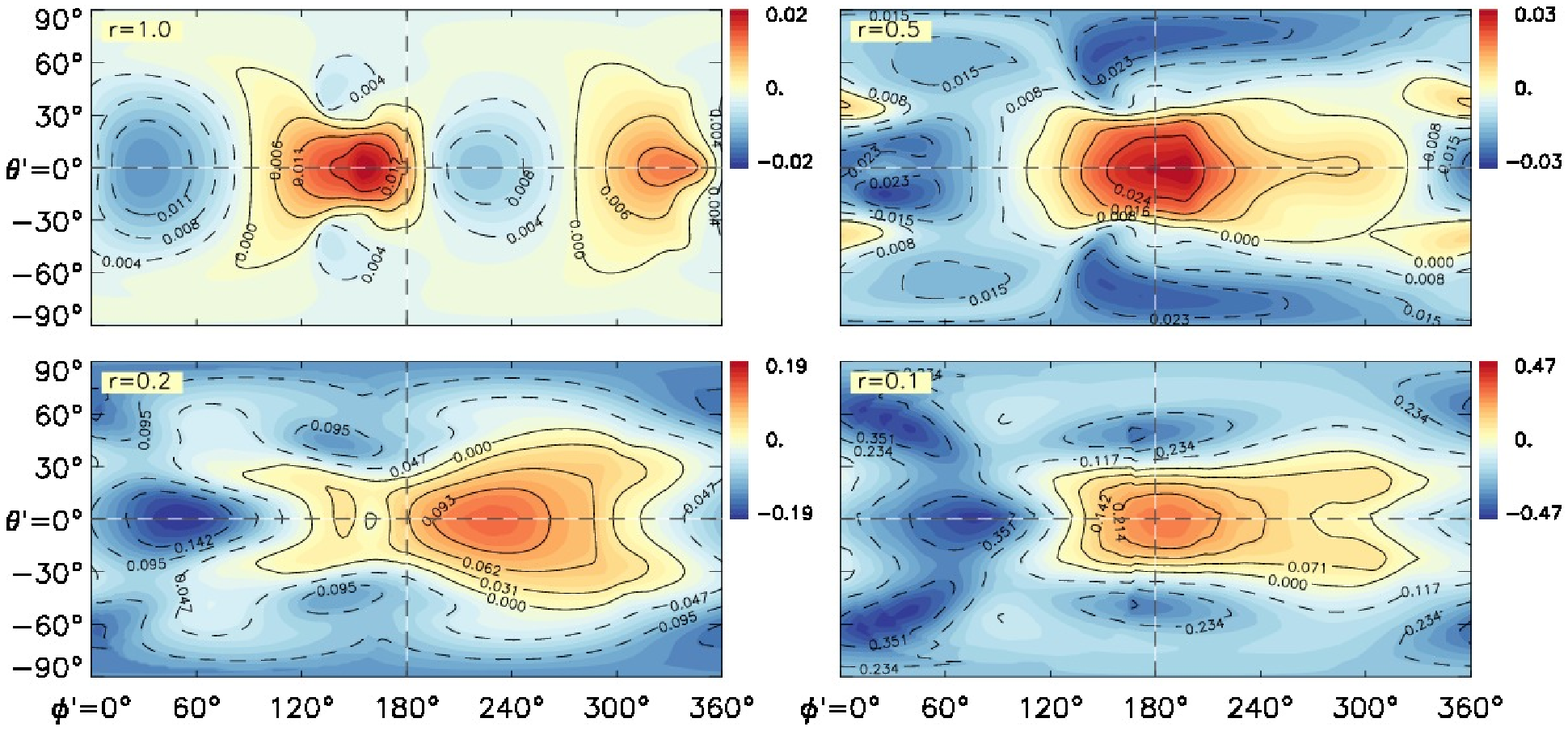}\\[5ex]
    \includegraphics[width=1.7\columnwidth]{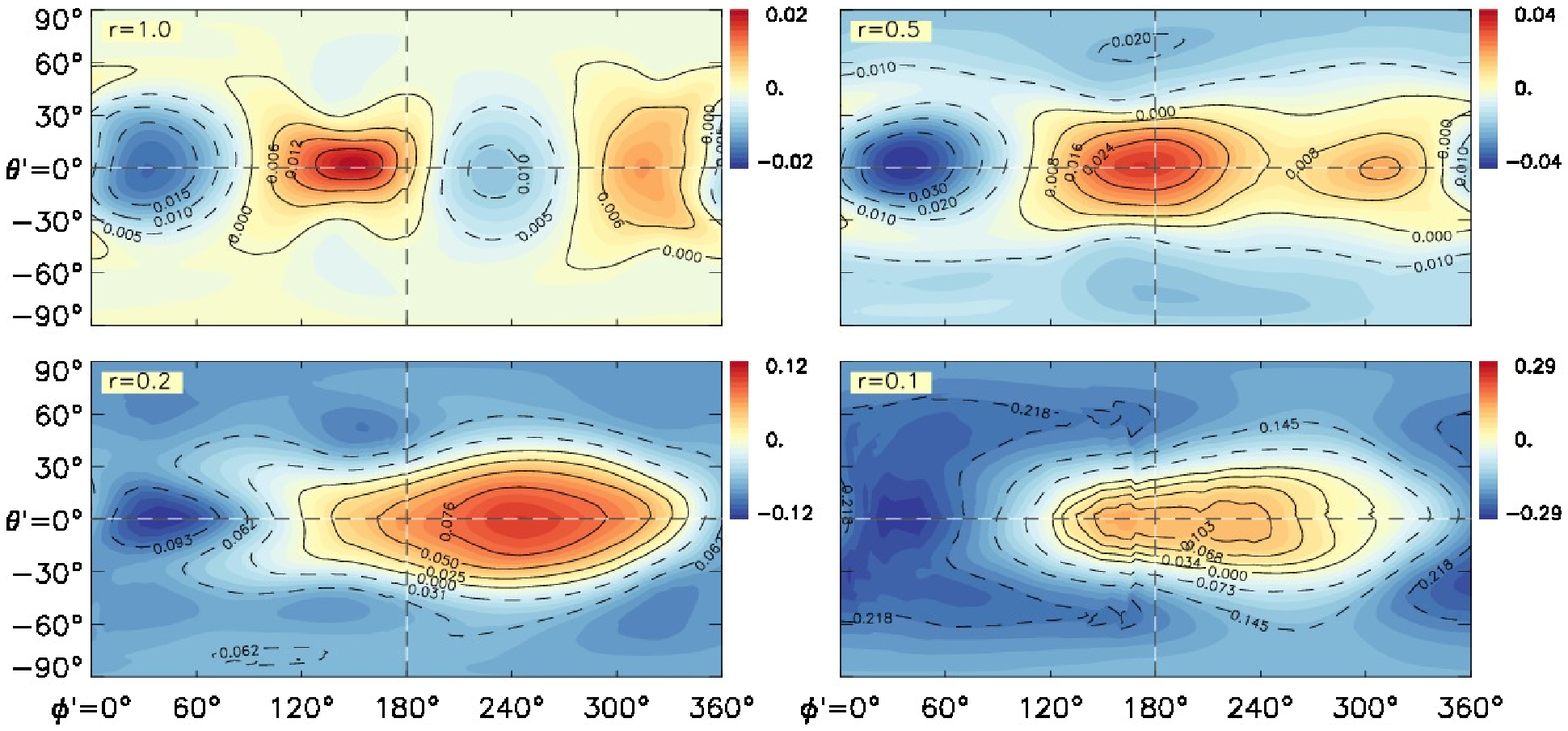}
  \end{center}
  \caption{Same as \Fig{fig:mass_flux_HD_iso}, but for the HD model,
    N1, (top four panels), and the MHD model (bottom four
    panels). Owing to the chaotic nature of the flow, time averages
    are taken over approximately 50 planet orbits to obtain a better
    picture of the long-term effective flow pattern.}
  \label{fig:mass_flux_MHD_tau}
\end{figure*}

\subsection{Details of the accretion flow geometry} 
\label{sec:details}

Given the complex flow structure in the planet Hill sphere reported
here and in previous studies, it is interesting to look at the
spherical distribution of the mass flux toward the planet and to compare
it with the laminar, inviscid, isothermal models of \citet{2012ApJ...747...47T}.
This is done in \Fig{fig:mass_flux_HD_iso}, where we look at the mass
flux through shells centered around the planet arising in run N1
\citep[cf. figure~5 in][note however the different set of radii, owing to 
  the lower
  resolution in our global model compared with their shearing
  box]{2012ApJ...747...47T}. Mass fluxes are averaged over
approximately one planet orbit, and our convention is such that
outflow corresponds to positive values (red), and accretion is
indicated by negative values (blue). The top four panels display
the mass flux onto the planet after $t \simeq 20$ orbits when its 
mass equals $150 \Me$. These are comparable with \citet{2012ApJ...747...47T}
because their model ran for $\simeq 25$ planet orbits,
leading only to moderate gap formation.
The bottom four panels correspond the end of the simulation
when the planet mass moderately exceeds $200 \Me$ and gap formation
is more pronounced.

Considering the early evolution first, we see that the first panel, at
$r=\rH$, matches well with the one from \citet{2012ApJ...747...47T}.
Material flows into the Hill sphere at two diametrically opposite
longitudes $\sim 20\degr$ in the counter-clockwise direction (when
viewed from above) from the L1 and L2 points, respectively. The
pronounced $m=2$ structure indicates that material also leaves the
Hill sphere close to the Lagrange points. The region of inflow seems
to be rather loosely defined while the corresponding outflow is more
focused, due to the influence of the spiral wake induced by the
planet. We remark that the spherical $r=\rH$ surface is somewhat
misleading in that the true shape of the Roche lobe is closer to a
rugby ball than to a football (this is evident in
\Fig{fig:flow_field_HD_iso}). We conclude that the flow pattern seen
in the first panel of \Fig{fig:mass_flux_HD_iso} hence reflects the
horseshoe orbits entering and leaving the spherical surface at
$r=\rH$, which is of little significance when considering accretion
onto the planet.  For the remaining three panels we find a
qualitatively similar distribution as in
\citet{2012ApJ...747...47T}. At all radii the mass fluxes show a clear
$m=2$ structure related to the spiral features observed in the CPD,
again in agreement with \citet{2012ApJ...747...47T} .  At $r=0.5\rH$
inflow occurs at high latitudes and outflow arises around the midplane
region at latitudes between $\theta' \simeq \pm 30 \degr$. On scales
of $r=0.1\rH$ there is a noticeable flattening of the disk toward the
midplane because the near-constant temperature leads to a flaring CPD
structure.

The lower four panels show that the detailed flow geometry toward the
planet evolves over time, such that the distinct $m=2$ feature
observed at all radii during early evolution is lost. At $r=0.5 \rH$
we observe a superposition of $m=2$ and $m=1$ features, and at smaller
radii $m=1$ dominates. This indicates that the inner CPD develops an
elongated/elliptical structure, confirmed by examination of
streamlines in the Hill sphere. The reason for this change in
structure is not clear, but we conjecture that more pronounced gap
opening at late times may modify the symmetry of the accretion flow
onto the CPD arising from the inner and outer disk. This is an effect
that can only arise in a global disk model and is not expected in
local shearing box simulations.  At late times the CPD midplane
continues to display outflow away from the planet, and inflow remains
confined to overlying latitudes above and below the
midplane. \citet{2012ApJ...747...47T} report that the midplane region
continues to display outflow on scales as small as $r=0.03$. Taking
this result at face value suggests that further simulations are
required that probe even deeper into the Hill sphere to examine at
which radius the midplane region of the circumplanetary disk behaves
as a classical accretion disk, instead of transporting mass away from
the planet.

The situation described above changes only moderately when we consider
the two non-isothermal runs, for which the geometry of the mass flux
is shown in \Fig{fig:mass_flux_MHD_tau}. We note that the isothermal
model N2 showed only modest temporal variation over short time
intervals, so the plots in \Fig{fig:mass_flux_HD_iso} show averages
taken over one orbital period. As discussed already, both of the
non-isothermal models N1 and M1 showed higher levels of temporal
variability, and \Fig{fig:mass_flux_MHD_tau} displays averages taken
over the last 50 orbits of the simulations.

When comparing the last four panels of \Fig{fig:mass_flux_HD_iso} with
\Fig{fig:mass_flux_MHD_tau} we see that the main features of the flow
in and out of the Hill sphere at $r=\rH$ are similar in all runs.  The
basic $m=1$ symmetry of the mass flux at smaller radii is also
similar. The main differences arise because of the flattening of the
CPD in model N2 which is not replicated in the hotter CPDs in runs N1
and M1.

Focusing first on the non-magnetized simulation N1, we note that
inflow is observed at essentially all longitudes for high latitudes
$|\theta'| \gtrsim 30 \degr$. This is true at all radii from
$r=0.5\rH$ down to $r=0.1 \rH$, as seen in
\Fig{fig:mass_flux_MHD_tau}.  We remark that this is consistent with
the flow field shown in \Fig{fig:flow_polo_HD_tau}. Considering the
flow at lower latitudes in the equatorial region, we see that for
$|\theta'| \lesssim 30 \degr$ there lies a contiguous band of inflow
at all radii $r \le 0.5 \rH$ that lies between $\phi' \approx 330
\degr$ and $\phi' \approx 90 \degr$, straddling the substellar point.
Moving round to longitudes lying between $100 \lesssim \phi' \lesssim
330 \degr$ we see a well-defined band of outflow near the
equator. These bands of inflow and outflow are also consistent with
the flow field shown in \Fig{fig:flow_polo_HD_tau}, which shows inflow
through the hemisphere facing the star, and outflow near the midplane
through the hemisphere facing away from the star.  Summarizing, we can
say that inflow occurs from high latitudes, causing gas to fall toward
the planet onto a circumplanetary disk whose half opening angle is
$\simeq 30 \degr$. The persistent bands of in- and outflow at all
radii indicate that the flow within the CPD is decidedly non-circular,
and instead exhibits an elongated or elliptical topology, similar to
that observed in run N2 at late times.

Moving on to the magnetized run, we see that the mass flux geometry
shown in the last four panels of \Fig{fig:mass_flux_MHD_tau} are very
similar to those described for run N1. Examination of individual
snapshots during the run show significant time dependence on orbital
time scales, but the average accretion flow in the magnetized disk
looks similar to that in the viscous model.

\begin{figure}
  \includegraphics[width=0.95\columnwidth]{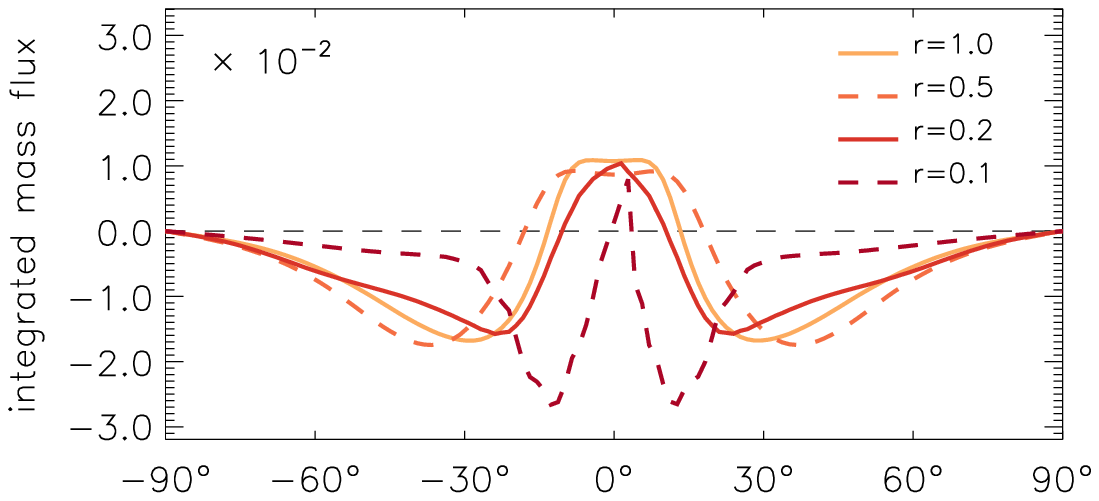}\\[2ex]
  \includegraphics[width=0.95\columnwidth]{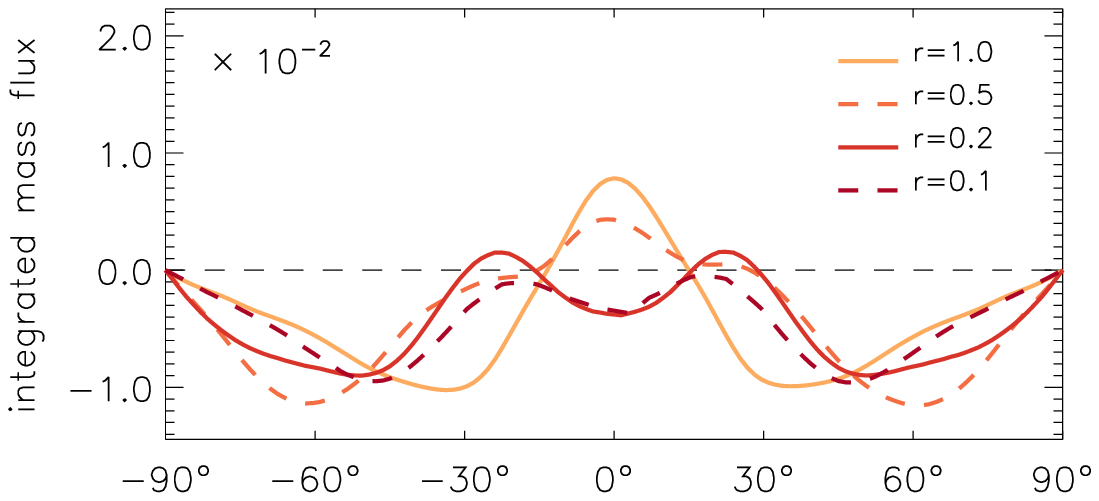}\\[2ex]
  \includegraphics[width=0.95\columnwidth]{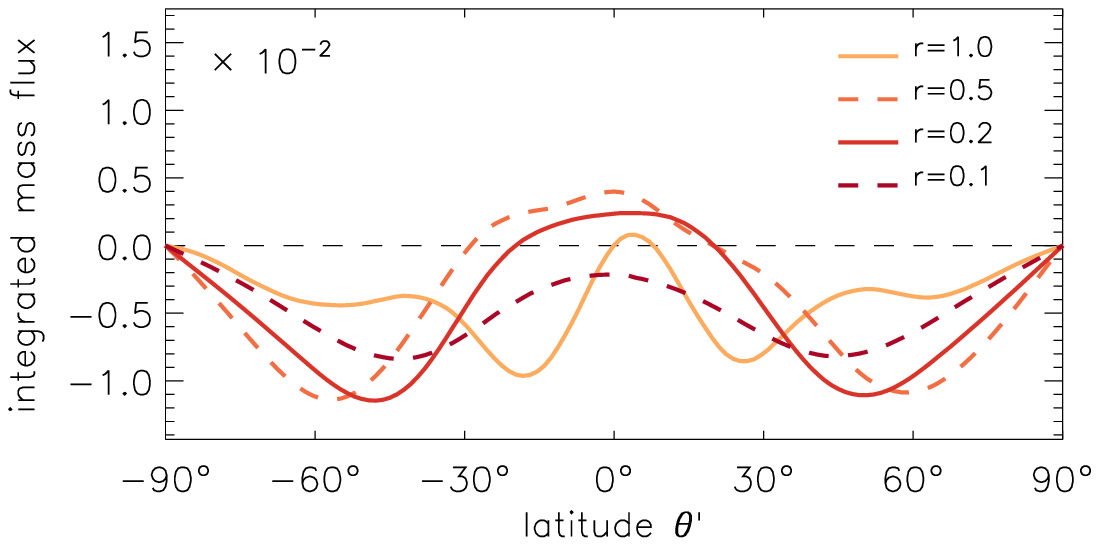}
  \caption{Azimuthally integrated mass fluxes (in code units) as a
    function of latitude for the isothermal HD model (top), the
    non-isothermal HD model (middle), and the non-isothermal MHD model
    (bottom). For low latitudes, and $r\ge 0.5 \rH$, material is
    actually \emph{expelled} (positive values).}
  \label{fig:mass_flux2}
  \medskip
\end{figure}

We now consider the flow rate toward the planet as a function of
latitude only by azimuthally-averaging the mass fluxes displayed in
Figures~\ref{fig:mass_flux_HD_iso} and \ref{fig:mass_flux_MHD_tau}.
These averaged mass fluxes are displayed in \Fig{fig:mass_flux2}.  The
top panel shows the isothermal model N2 at the end of the simulation,
and this compares very well with previous results \citep[cf. figure 6
  in][]{2012ApJ...747...47T}.  In particular, we see that mass flow
toward the planet at all radii occurs from high latitudes $|\theta'|
\ge 30 \degr$, with the midplane region showing net outflow away from
the planet at these radii. It is interesting to note that the
inclusion of viscosity and angular momentum transport in the CPD in
our run N2 leads to a net flow profile that agrees qualitatively with
the inviscid model presented by \citet{2012ApJ...747...47T}.  The
non-isothermal HD case N1 is shown in the middle panel of
\Fig{fig:mass_flux2}, and agrees reasonably well with the isothermal
run at radii $r=1$ and $0.5 \rH$, where outflow near midplane and
inflow at high latitudes are observed.  At $r=0.2 \rH$ we see that the
mass flux near the midplane is of small magnitude, and alternates
between inflow and outflow.  Net inflow at this radius is dominated by
contributions from high latitudes. At $r=0.1 \rH$ inflow occurs at all
latitudes, with a modest contribution from the midplane region and a
dominant contribution from higher latitudes. The results from model M1
shown in the bottom panel are broadly consistent with model N1, and
again display the important inflow contribution from high latitudes.
We again note that the nature of the accretion flow is not
dramatically altered by the inclusion of magnetic fields compared with
the viscous disk model.


\section{Summary and Conclusions}
\label{sec:discussion}

In this paper we have presented the results of global 3D hydrodynamic
and magnetohydrodynamic simulations of accreting planets embedded in
protoplanetary disks. The MHD simulation utilized a detailed
ionization model from which Ohmic resistivity was calculated, leading
to a disk with active surface layers that sustain MRI turbulence, and
midplane regions that host a dead-zone where MRI turbulence is damped
by resistivity. The hydrodynamic simulations adopted the
$\alpha$~model for anomalous viscosity, calibrated to give the same
volume averaged stress as the magnetized disk, and were computed for
the purpose of comparison with the turbulent disk simulation. One of
the hydrodynamic simulations adopted a locally isothermal equation of
state. The MHD run and the other hydrodynamic simulation both adopted
an adiabatic equation of state combined with thermal relaxation, where
the temperature in the disk was continuously forced back toward its
initial value on the local orbital time scale. All simulations used
three levels of adaptive mesh refinement to resolve the region inside
the planet Hill sphere. The accretion flow in these models is followed
down to a radius equal to 5\% of the Hill sphere. Interior to this a
sink hole accretes inflowing gas, mimicking accretion onto the planet.

The main aims of this work are: (i) to examine gap formation in a
layered protoplanetary disk and the influence of the changing
ionization fraction of material near the planet as the gap opens, (ii)
to examine the physical state and dynamical evolution of material that
enters the planet Hill sphere, (iii) to measure the rate at which gas
accretes onto a planet of mass $100 \Me$, placing it firmly in the
regime of runaway gas accretion during giant planet formation.  The
main results may be summarized as follows:

\itm All simulations lead to gap formation in the vicinity of the
planet. Accretion of gas into the Hill sphere leads to the formation
of a rotationally supported circumplanetary disk.  Significant
accretion occurs onto the planet in all runs.

\itm The locally isothermal hydrodynamic simulation produces results
very similar to those described previously by
\citet{2008ApJ...685.1220M} and \citet{2012ApJ...747...47T}. Infalling
material leads to strong inflow toward the planet from \emph{high
  latitudes}. Outflow away from the planet occurs in the midplane
regions of the CPD for all radii down to $0.1 \rH$, this being the
limit of where we can measure the flow geometry.

\itm Adoption of an adiabatic equation of state with thermal
relaxation in a viscous hydrodynamic model leads to a warmer and
thicker circumplanetary disk. Even in the presence of rapid cooling
compressional heating increases the temperature to $T\simeq 2000\K$ in
the inner CPD. The increased pressure support reduces the accretion
rate onto the planet during early evolution compared to the isothermal
model. At late times, when higher angular momentum material accretes
into the Hill sphere, the accretion rates converge.

\itm In agreement with the locally isothermal run, both the MHD and
hydrodynamic simulations with thermal relaxation demonstrate that
accretion toward the planet occurs from high latitudes. The midplane
region of the CPD displays net outflow away from the planet down to
radii $r \ge 0.2 \rH$.

\itm Gap opening in the MHD simulation leads to ignition of the
dead-zone into a turbulent state as X-rays and cosmic rays penetrate
the gap region. The global structure of the disk is one in which
accretion occurs in the active surface layers far from the planet,
with the midplane region there being largely inert. Near the planet
there is deep gap where the ionization fraction allows development of
MRI-turbulence.

\itm Enlivening the gap in the MHD simulation causes the accreting
planet to be embedded in a turbulent environment.  Accretion onto the
planet becomes stochastic, and the flow in the Hill sphere displays
significant temporal variability.  The CPD has a surface density $\sim
30\g\cm^{-2}$ between its outer edge at $\sim 0.5$ Hill radii and the
inner boundary at 5\% of the Hill radius. This is low enough to be
sufficiently ionized by X-rays and cosmic rays to sustain the MRI. Our
measurement of the Elsasser number there suggests that the
circumplanetary disk should indeed be MRI active, at least in its
outer regions.

\itm Gas accretion into the gap and planet Hill sphere occurs largely
from the active surface layers of the surrounding protoplanetary
disk. As gas enters the gap it is pulled toward the midplane by the
star and planet gravity, dragging the (largely azimuthal) magnetic
field with it. Gas and magnetic field lines that enter the Hill sphere
join the rotating circumplanetary disk, leading to the generation of
helical magnetic fields. These launch what appear to be sporadic
magneto-centrifugally driven outflows. These are generally found to be
loosely collimated, but we have observed at least one extended time
interval during which a highly collimated jet was launched from the
CPD region. The protoplanetary jet does not influence strongly the gas
accretion rate onto the planet.

\itm Stochastic accretion into the Hill sphere causes the direction of
the angular momentum vector of material there to vary significantly.
Our model suggests that CPDs display substantial time variability in
their tilt angle, amounting to changes $\sim 10 \degr$ on orbital time
scales. Although we cannot model the inner regions of the CPD
explicitly, we speculate that oscillations in global tilt angle will
allow propagation of bending waves into the inner satellite forming
region, creating a source of disturbance that may influence the
formation of regular satellite systems. In this scenario, it seems
likely that satellite building blocks orbiting within the inner CPD
will develop mutually inclined orbits if they are strongly coupled to
the gaseous CPD, leading to slower growth of the satellites.

\itm The accretion rate onto the planet in the MHD simulation reaches
a steady value of ${\dot M} = 8 \times 10^{-3}\Me\yr^{-1}$, large
enough for a Saturn-mass planet to become a Jovian planet in $\sim 3
\times 10^4\yr$, much shorter than the expected disk life-time. This
steady state is reached $\sim 100$ orbits after insertion of the
planet. The accretion rates in the viscous, hydrodynamic simulations
continue to fall below this value, and do not reach steady values by
the end of the runs. This is a consequence of the accretion flow in
the magnetized disk being confined to the active layers where the
magnetic stresses are large. This flow is impervious to tidal
truncation and gap formation by the planet, even if its mass becomes
quite large, because the effective viscous stress there is large
(i.e. $\alpha \simgt 0.1$), leading to saturation of the mass
accretion rate at a higher value than obtained in a viscous model with
a constant $\alpha$ value.

\subsection{Implications} 

The above results have a number of implications for planet and
satellite formation.  The feeding of gas into the gap region from the
surface layers of the PPD may have important consequences for the
chemistry and dust content of the gas that eventually accretes onto
the planet, as dust settling and sequestration in the dead-zone may
reduce the heavy element content of this gas as it accretes through
the disk.  One implication is that gas accreted through a gap in a
disk with active surface layers and a midplane dead-zone will have
lower opacity.  Models of giant planet formation
\citep[e.g.][]{1996Icar..124...62P,2005A&A...433..247P,2010Icar..209..616M}
show that the upper-envelope opacity is crucial in determining the
envelope accretion time scale. A low mass planet (i.e. 5-$10 \Me$)
deeply embedded in a disk without a gap will probably not experience a
significant reduction in opacity of the accreted gas as it originates
from the local midplane during the early phases of envelope accretion.
If placed in a region of the disk where the scale-height is much
smaller, however, such as in the inner few tenths of an ${\rm au}$
where $h \lesssim 0.02$ \citep[depending on the viscous dissipation
  rate, see e.g.][]{2012ApJ...757...50D} gap formation is expected
even for planets with masses $< 10 \Me$ and these may then accrete
metal-poor, low opacity gas, reducing the core mass required to
build-up a substantial gaseous envelope. This may provide an
explanation for the low mass and low density planets that have been
discovered by the Kepler mission such as ``Kepler 11e''
\citep{2011Natur.470...53L}. One potential caveat is that most
protoplanetary disk models have a transition where the dead-zone
disappears and the disk becomes fully active at a distance of a few
tenths of an ${\rm au}$ from the central star
\citep{1996ApJ...457..355G,2006A&A...445..205I}. We note that a
planetary core accreting gas from this disk region will still accrete
metal-poor gas, because gas arriving at the inner regions will have
accreted through the disk surface layers from further out in the disk
where dust settling will have reduced the heavy element content of
this gas. Examination of this idea will require development of
detailed planetary envelope accretion models in which the usual outer
boundary condition that matches the envelope onto the background
protoplanetary nebula model will need to be replaced with one that
assumes the envelope has a quasi-free surface, perhaps surrounded by a
thick circumplanetary disk. Such calculation were presented by
\citet{2005A&A...433..247P} for giant planets undergoing runaway gas
accretion, but have not been performed for planets in the earlier
phase of quasi-static gas settling. Alternatively this scenario could
be explored using 3D radiation-hydrodynamic simulations similar to
those presented by \citet{2009MNRAS.397..657A,2009MNRAS.393...49A}.

The comments above concerning the heavy element content of accreted
gas also have implications for the formation of satellite systems.
\citet{2012ApJ...747...47T} have noted already that delivery of gas
deep into the Hill sphere from high latitudes may lead to accretion of
low-metallicity gas onto the inner CPD if grain growth and settling
have occurred in the surrounding PPD. The fact that low-metallicity
gas is delivered into the gap largely from the active surface layers
simply reinforces this point.

The temperatures that we obtained for the inner CPDs in the models
with applied cooling should not be taken too seriously given our crude
thermal model. In spite of this, it is noteworthy that inner
temperatures obtained $T \simeq 2000\K$ are similar to those reported
by \citet{2006A&A...445..747K} in their radiation-hydrodynamic
simulations. Temperatures in excess of $T \simeq 1500\K$ are large
enough to vaporise refractory materials in addition to volatiles such
as water ice, so taken at face value our results and those of
\citet{2006A&A...445..747K} indicate that satellite building material
is unlikely to be transported to the vicinity of the planet by this
early-stage accretion flow. Achieving temperatures low enough to
support condensation of ices for building icy satellites clearly
requires a phase of evolution in which material is delivered to the
inner regions of CPDs at much lower rates because of the dependence of
the temperature on the local gas accretion rate.

\subsection{Concluding remarks} 

Although the model we have presented of a planet accreting gas from a
layered protoplanetary disk represents a significant step forward in
terms of complexity, realism and numerical resolution, there are
numerous omissions to the physical model that need to be addressed in
future work before we can be confident that gas accretion rates onto
forming giant planets are fully understood. The non-ideal MHD model
that we have adopted neglects the potentially important effects of
ambipolar diffusion and Hall EMFs, both of which may significantly
modify the results we have presented
\citep{2012MNRAS.422.2737W,2013ApJ...769...76B}. In a future paper we
will be particularly interested in examining the influence of
ambipolar diffusion in the gap region. The low densities there may
allow this effect to quench the turbulent nature of the flow in the
gap, and this will have significant consequences for the time
dependence of the flow in the Hill sphere.

Related issues that need to be explored are the transport of stellar
X-rays into the gap region (as this is crucial for determining the
ionisation levels close to the giant planet), and the level of cosmic
ray ionization experienced by the PPD. For X-ray ionization, we have
taken a simple approach and adopted results from the model of
\citet{1999ApJ...518..848I} in which X-rays are scattered vertically
toward the disk midplane from the overlying disk atmosphere, with
attenuation depending simply on the column density. This model,
however, does not account for the effects of gap opening and a more
accurate approach will require transfer calculations of X-rays being
scattered into the gap.  A further refinement would be to include the
effects of X-ray flares.  These lead to both substantial increases in
the X-ray flux, and also to a substantial hardening of the spectra
\citep{2007A&A...468..477A,2007A&A...468..485F,2008ApJ...688..437G}.
Given that hard X-rays are able to penetrate more deeply into the disk
this may influence the ionization of material in the vicinity of the
planet. Cosmic ray ionization rates experienced at PPD surfaces, and
the attenuating influences of magnetized stellar winds and disk winds,
were examined recently by \citet{2013ApJ...772....5C}. By
extrapolating from the solar wind modulation of cosmic rays in the
heliosphere, with its observed dependence on sunspot coverage, these
authors conclude that the stellar wind can reduce cosmic ray
ionization rates at PPD surfaces by many orders of magnitude compared
to the value adopted in this paper. The disk wind and its associated
ordered field, by contrast, has little influence because of the
compensating effects of magnetic funneling and mirroring. An issue
that remains to be addressed is the interaction between the stellar
wind and the disk wind.  If it carries sufficient momentum, the disk
wind may retard the equatorial flow of the stellar wind and confine it
to polar regions, allowing significant penetration of cosmic rays to
the disk surfaces.  If the disk wind is ineffective in this regard,
however, then it seems likely that cosmic ray ionization rates will
indeed be low. The effect of this on our results will be to simply
reduce the depths of the active layers in the PPD but not remove them
altogether because the stellar X-rays penetrate to column densities of
$\sim 10\g\cm^{-2}$. The enlivening of the gap region will still arise
because the column densities there become low enough for significant
penetration by X-rays.

A proper treatment of the gas thermodynamics is also needed. This is
illustrated by the fact that during the time before a steady state has
been achieved, our locally isothermal model displays a higher gas
accretion rate than in the models with an adiabatic equation of state
and local cooling. In other words, thermodynamics play a role in
regulating the accretion rate prior to gap formation while the planet
is deeply embedded in the PPD. Our simulations suggest that if giant
planets form in protoplanetary disks with properties similar to
minimum mass solar nebula models \citep{1981PThPS..70...35H}, then
these disks are very capable of supplying large amounts of gas to
forming planets that can allow them to grow to super-Jovian
masses. The fact that massive planets are relatively rare suggests
either that the available gas reservoir during gas accretion is
smaller than in a minimum mass disk, or that accretion into the Hill
sphere and onto the planet experiences a bottle-neck. Our simulations
suggest that compressive heating and inefficient cooling in the Hill
sphere can contribute to slowing the accretion, although the
radiation-hydrodynamic simulations of \citet{2009MNRAS.393...49A}
indicate that rapid runaway gas accretion still occurs when radiative
processes are accounted for. Inefficient angular momentum and mass
transport through the inner regions of the circumplanetary disk, on
scales smaller than we have been able to model may also play a role
here, as considered for example in the disk models of
\citet{2013MNRAS.tmp.1252M}.

The possibility, however, that mass can build-up in the inner
circumplanetary disk to levels that allow gravitational instabilities
to develop, which then transport angular momentum efficiently, may
signify that any model that allows efficient supply of mass into the
Hill sphere must also lead inevitably to substantial accretion onto
the planet. If MRI turbulence does not transport mass and angular
momentum efficiently, then gravitational instabilities may develop to
do the job instead, leading to the same net result of an efficiently
accreting circumplanetary disk. This favors a picture of giant planet
formation in which the final mass is determined by the flow of gas
from circumstellar to circumplanetary disk, rather than peculiarities
associated with the flow within the planet's Hill sphere.
Demonstration of this basic idea, however, will require substantial
further work.

\acknowledgments 

We thank the anonymous referee for useful comments that led to an
improvement of this paper. Part of the research was carried out at the
Jet Propulsion Laboratory, California Institute of Technology, under a
contract with the National Aeronautics and Space Administration. The
simulations presented in this paper were run on the QMUL HPC
facility. Three-dimensional imagery produced by \textsc{vapor}
\citep[\texttt{www.vapor.ucar.edu}]{2007NJPh....9..301C}, a product of
the Computational Information Systems Laboratory at the National
Center for Atmospheric Research.


\appendix 

It is clear that insufficient numerical resolution will lead to
erroneous results being obtained in terms of stresses and accretion
rates. For example, \citet{2006A&A...457..343F} presented one of the
first systematic studies of MRI-driven turbulence in vertically
stratified global disk models and found that approximately 6 cells
were required to resolve the fastest growing MRI modes in order for
nonlinear MRI-turbulence to be sustained in disks containing weak
net-toroidal magnetic fields. In recent work the influence of
numerical resolution in producing robust results has been emphasized
strongly \citep{2013ApJ...772..102H}, where it has been suggested that
between 20-30 cells per unstable MRI wavelength are required to obtain
reliable results in disk models that sustain MRI turbulence throughout
(i.e. no dead zone present). 

\begin{figure}
\begin{center}
  \includegraphics[width=0.48\columnwidth]{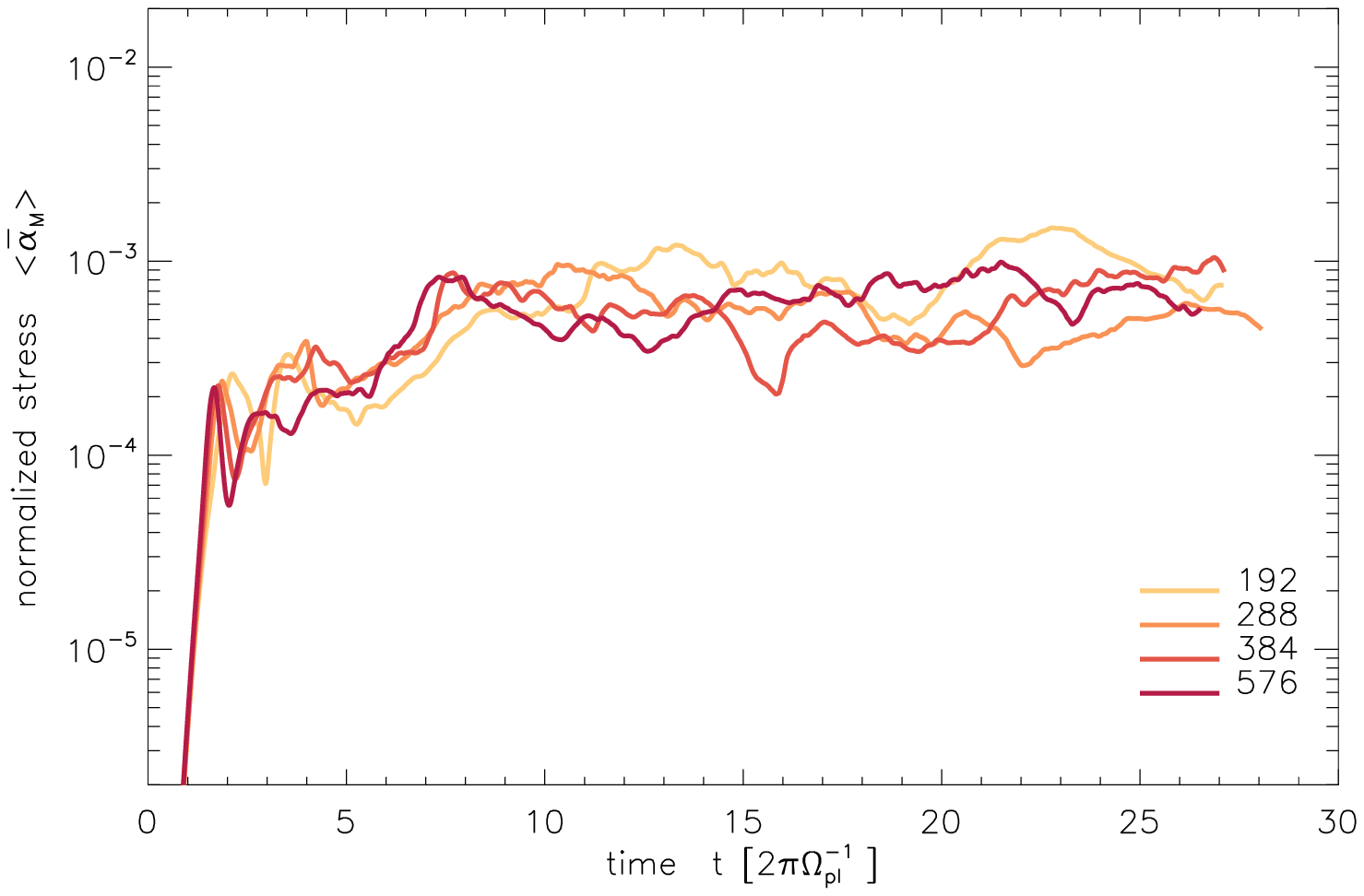}\hspace{4ex}
  \includegraphics[width=0.48\columnwidth]{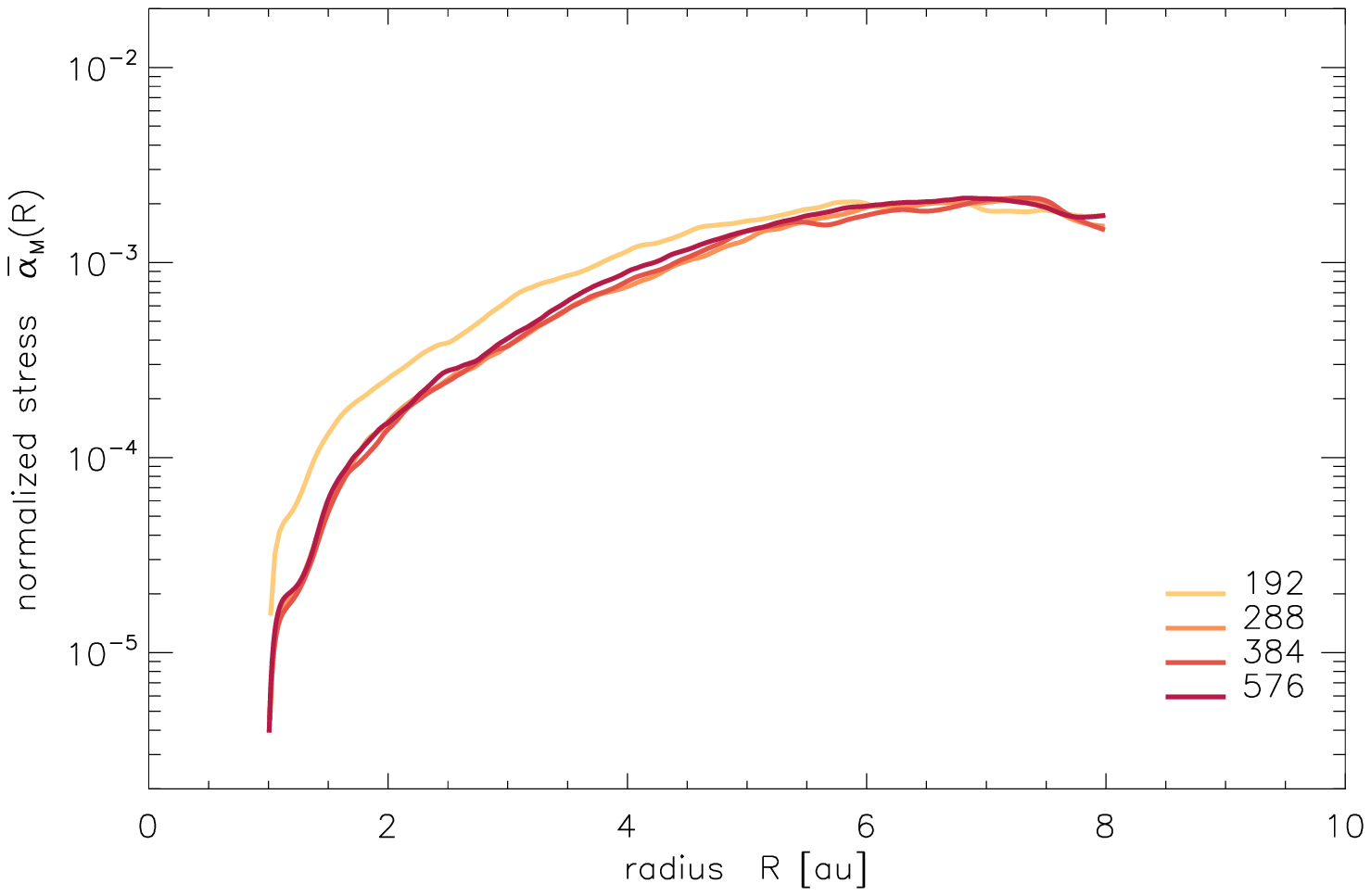}
  \caption{Resolution study of the unperturbed disk model in the
    absence of mesh refinement. Numbers are for the adopted number of
    radial grid cells; the resolution in the vertical and azimuthal
    directions are scaled accordingly. \emph{Left:} Time series of the
    volume averaged Maxwell stress evaluated in the radial range $3
    \le r \le 4\au$.  \emph{Right:} Radial profile of the volume- and
    time-averaged Maxwell stress.  }
  \label{fig:appendix}
  \medskip
 \end{center}
\end{figure}

The basic disk model, M1, presented in the main body of this paper has
a significantly lower resolution than this, and so we have computed a
suite of test calculations to examine the influence of changing the
numerical resolution in the simulations. These calculations did not
use mesh refinement. The size of the azimuthal domain used was $\pi/4$
instead of $\pi/2$, but adopting the same grid size ratio
$\Delta_r:\Delta_{\phi}$ as in the original simulation.  The time
evolution of the Maxwell stress \citep[as defined in Equation~(13)
  of][]{2006A&A...457..343F} evaluated between the radii $3 \le r \le
4\au$ for the different models is shown in the left-hand panel of
\Fig{fig:appendix}. The number shown in the inset is the number of
radial grid cells, $N_r$, where $N_r=384$ in model M1.  When changing
the radial resolution we also changed the resolution in the other
dimensions by the same factor. We see that there is little evidence of
a systematic trend in the results as a function of numerical
resolution. This point is further supported by the right-hand panel of
the same figure, where we show radial profiles of the
temporally-averaged Maxwell stresses. Apart from the low-resolution
model with $192$ grid cells in radius, all models produce nearly
indistinguishable results.

The reasons for this are two-fold. First, we adopt a net-vertical
magnetic field in our simulations that is essentially conserved
throughout the runs. This has the effect of reducing the influence of
the numerical resolution on the results because we do not rely on any
sort of disk dynamo to maintain the large-scale vertical
field. Second, the presence of the dead zone means that only a
relatively narrow region of the disk above the midplane is susceptible
to growth of the MRI -- there the fields are generally quite strong
(compared to the gas pressure), leading to MRI developing on
comparatively larger scales. By contrast, the resolution requirements
are more severe in a disk where turbulence can also develop near the
midplane because the value of $\beta_{\rm P} \equiv P_{\rm gas}/P_{\rm
  mag}$ will be larger there.  The low density and pressure in the
upper disk layers means that $\beta_{\rm P}$ is fairly small (between
20--30), and this allows the MRI to be quite well resolved even for
weak fields. For ideal MHD, the wavelength of the fastest growing mode
can be approximated by $\lambda_{\rm MRI}\equiv 2\pi\,v_{\rm
  A}/\Omega$. Using the relation $\cS=H\Omega$, and noting that
$\beta_{\rm P}=\cS^2/v_{\rm A}^2$, we can express this as
$\lambda_{\rm MRI}= 2\pi\,H/\sqrt{\beta_{\rm P}}$. With this relation,
we estimate that in our case $\lambda_{\rm MRI}$ is on the order of
the pressure scale height itself and thus adequately resolved.

Another issue of potential importance is the role of mesh refinement
in determining the evolution of the MRI in the disk and the strength
of the turbulent stresses. In a disk without a dead zone, where MRI
turbulence can develop throughout the body of the disk, we would
expect the region in which mesh-refinement is applied to show enhanced
activity compared to the surrounding disk regions. This is because
shorter wavelength MRI modes will be resolved in this region,
potentially boosting the strength of the turbulence there. In our
model, however, the mesh refinement is applied in a spherical region
centered around the planet near the midplane. The refined region
therefore lies almost entirely within the dead zone and has
essentially no effect on the turbulence in the active zones near the
disk surfaces.  We have tested this by running a simulation with
refinement applied around the nominal planet location, but with the
planet's gravity switched off.  When the planet gravity is switched
on, however, gap formation ensues and we might be concerned that
developing turbulence in the annular region centered on the planet
semi-major axis will be affected by the refinement. We note, however,
that the refinement is applied in a spherical region around the
planet, and therefore does not extend far in azimuth. As such, we
conclude that the refinement has little influence on the turbulence in
the surrounding protoplanetary disk, and is only effective in the
region where the planet gravity dominates, allowing small scale fluid
and magnetic structures to be resolved there, as intended.



\begin{thebibliography}{81}
\expandafter\ifx\csname natexlab\endcsname\relax\def\natexlab#1{#1}\fi

\bibitem[{Alexiades {et~al.}(1996)Alexiades, Amiez, \& Gremaud}]{CNM:CNM950}
Alexiades, V., Amiez, G., \& Gremaud, P.-A. 1996, Communications in Numerical
  Methods in Engineering, 12, 31

\bibitem[{{Arzner} {et~al.}(2007){Arzner}, {G{\"u}del}, {Briggs}, {Telleschi},
  \& {Audard}}]{2007A&A...468..477A}
{Arzner}, K., {G{\"u}del}, M., {Briggs}, K., {Telleschi}, A., \& {Audard}, M.
  2007, \aap, 468, 477

\bibitem[{{Ayliffe} \& {Bate}(2009{\natexlab{a}})}]{2009MNRAS.397..657A}
{Ayliffe}, B.~A., \& {Bate}, M.~R. 2009{\natexlab{a}}, \mnras, 397, 657

\bibitem[{{Ayliffe} \& {Bate}(2009{\natexlab{b}})}]{2009MNRAS.393...49A}
---. 2009{\natexlab{b}}, \mnras, 393, 49

\bibitem[{{Bai} \& {Stone}(2013)}]{2013ApJ...769...76B}
{Bai}, X.-N., \& {Stone}, J.~M. 2013, \apj, 769, 76

\bibitem[{{Balbus} \& {Hawley}(1998)}]{1998RvMP...70....1B}
{Balbus}, S.~A., \& {Hawley}, J.~F. 1998, \rm RvMP, 70, 1

\bibitem[{{Balsara} \& {Meyer}(2010)}]{2010arXiv1003.0018B}
{Balsara}, D.~S., \& {Meyer}, C. 2010, (astro-ph:1003.0018)

\bibitem[{{Balsara} \& {Spicer}(1999)}]{1999JCoPh.149..270B}
{Balsara}, D.~S., \& {Spicer}, D.~S. 1999, \rm JCoPh, 149, 270

\bibitem[{{Boss}(1998)}]{1998ApJ...503..923B}
{Boss}, A.~P. 1998, \apj, 503, 923

\bibitem[{{Bryden} {et~al.}(1999){Bryden}, {Chen}, {Lin}, {Nelson}, \&
  {Papaloizou}}]{1999ApJ...514..344B}
{Bryden}, G., {Chen}, X., {Lin}, D.~N.~C., {Nelson}, R.~P., \& {Papaloizou},
  J.~C.~B. 1999, \apj, 514, 344

\bibitem[{{Canup} \& {Ward}(2002)}]{2002AJ....124.3404C}
{Canup}, R.~M., \& {Ward}, W.~R. 2002, \aj, 124, 3404

\bibitem[{{Canup} \& {Ward}(2006)}]{2006Natur.441..834C}
---. 2006, \nat, 441, 834

\bibitem[Cleeves et al.(2013)]{2013ApJ...772....5C} Cleeves, L.~I., Adams, 
F.~C., \& Bergin, E.~A.\ 2013, \apj, 772, 5 

\bibitem[{{Clyne} {et~al.}(2007){Clyne}, {Mininni}, {Norton}, \&
  {Rast}}]{2007NJPh....9..301C}
{Clyne}, J., {Mininni}, P., {Norton}, A., \& {Rast}, M. 2007, New Journal of
  Physics, 9, 301

\bibitem[{{D'Angelo} {et~al.}(2003){D'Angelo}, {Henning}, \&
  {Kley}}]{2003ApJ...599..548D}
{D'Angelo}, G., {Henning}, T., \& {Kley}, W. 2003, \apj, 599, 548

\bibitem[{{D'Angelo} \& {Marzari}(2012)}]{2012ApJ...757...50D}
{D'Angelo}, G., \& {Marzari}, F. 2012, \apj, 757, 50

\bibitem[{{de Val-Borro} {et~al.}(2006){de Val-Borro}, {Edgar}, {Artymowicz},
  {Ciecielag}, {Cresswell}, {D'Angelo}, {Delgado-Donate}, {Dirksen}, {Fromang},
  {Gawryszczak}, {Klahr}, {Kley}, {Lyra}, {Masset}, {Mellema}, {Nelson},
  {Paardekooper}, {Peplinski}, {Pierens}, {Plewa}, {Rice}, {Sch{\"a}fer}, \&
  {Speith}}]{2006MNRAS.370..529D}
{de Val-Borro}, M., {Edgar}, R.~G., {Artymowicz}, P., {et~al.} 2006, \mnras,
  370, 529

\bibitem[{{Fendt}(2003)}]{2003A&A...411..623F}
{Fendt}, C. 2003, \aap, 411, 623

\bibitem[{{Flock} {et~al.}(2010){Flock}, {Dzyurkevich}, {Klahr}, \&
  {Mignone}}]{2010A&A...516A..26F}
{Flock}, M., {Dzyurkevich}, N., {Klahr}, H., \& {Mignone}, A. 2010, \aap, 516,
  A26+

\bibitem[{{Franciosini} {et~al.}(2007){Franciosini}, {Pillitteri}, {Stelzer},
  {Micela}, {Briggs}, {Scelsi}, {Telleschi}, {Audard}, {Palla}, \&
  {G{\"u}del}}]{2007A&A...468..485F}
{Franciosini}, E., {Pillitteri}, I., {Stelzer}, B., {et~al.} 2007, \aap, 468,
  485

\bibitem[{{Fromang} \& {Nelson}(2006)}]{2006A&A...457..343F}
{Fromang}, S., \& {Nelson}, R.~P. 2006, \aap, 457, 343

\bibitem[{{Gammie}(1996)}]{1996ApJ...457..355G}
{Gammie}, C.~F. 1996, \apj, 457, 355

\bibitem[{{Gardiner} \& {Stone}(2008)}]{2008JCoPh.227.4123G}
{Gardiner}, T.~A., \& {Stone}, J.~M. 2008, \rm JCoPh, 227, 4123

\bibitem[{{Getman} {et~al.}(2008){Getman}, {Feigelson}, {Micela}, {Jardine},
  {Gregory}, \& {Garmire}}]{2008ApJ...688..437G}
{Getman}, K.~V., {Feigelson}, E.~D., {Micela}, G., {et~al.} 2008, \apj, 688,
  437

\bibitem[{{Goodman} \& {Rafikov}(2001)}]{2001ApJ...552..793G}
{Goodman}, J., \& {Rafikov}, R.~R. 2001, \apj, 552, 793

\bibitem[{{Gressel} {et~al.}(2011){Gressel}, {Nelson}, \&
  {Turner}}]{2011MNRAS.415.3291G}
{Gressel}, O., {Nelson}, R.~P., \& {Turner}, N.~J. 2011, \mnras, 415, 3291

\bibitem[{{Gressel} {et~al.}(2012){Gressel}, {Nelson}, \&
  {Turner}}]{2012MNRAS.422.1140G}
---. 2012, \mnras, 422, 1140

\bibitem[{{Hawley} {et~al.}(2013){Hawley}, {Richers}, {Guan}, \&
  {Krolik}}]{2013ApJ...772..102H}
{Hawley}, J.~F., {Richers}, S.~A., {Guan}, X., \& {Krolik}, J.~H. 2013, \apj,
  772, 102

\bibitem[{{Hayashi}(1981)}]{1981PThPS..70...35H}
{Hayashi}, C. 1981, \rm Progress of Th. Phys. Suppl., 70, 35

\bibitem[{{Igea} \& {Glassgold}(1999)}]{1999ApJ...518..848I}
{Igea}, J., \& {Glassgold}, A.~E. 1999, \apj, 518, 848

\bibitem[{{Ilgner} \& {Nelson}(2006)}]{2006A&A...445..205I}
{Ilgner}, M., \& {Nelson}, R.~P. 2006, \aap, 445, 205

\bibitem[{{Klahr} \& {Kley}(2006)}]{2006A&A...445..747K}
{Klahr}, H., \& {Kley}, W. 2006, \aap, 445, 747

\bibitem[{{Kley}(1999)}]{1999MNRAS.303..696K}
{Kley}, W. 1999, \mnras, 303, 696

\bibitem[{{Larwood} {et~al.}(1996){Larwood}, {Nelson}, {Papaloizou}, \&
  {Terquem}}]{1996MNRAS.282..597L}
{Larwood}, J.~D., {Nelson}, R.~P., {Papaloizou}, J.~C.~B., \& {Terquem}, C.
  1996, \mnras, 282, 597

\bibitem[{{Lin} \& {Papaloizou}(1986)}]{1986ApJ...309..846L}
{Lin}, D.~N.~C., \& {Papaloizou}, J. 1986, \apj, 309, 846

\bibitem[{{Lin} \& {Papaloizou}(1993)}]{1993prpl.conf..749L}
{Lin}, D.~N.~C., \& {Papaloizou}, J.~C.~B. 1993, in Protostars and Planets III,
  ed. E.~H. {Levy} \& J.~I. {Lunine}, 749--835

\bibitem[{{Lissauer} {et~al.}(2011){Lissauer}, {Fabrycky}, {Ford}, {Borucki},
  {Fressin}, {Marcy}, {Orosz}, {Rowe}, {Torres}, {Welsh}, {Batalha}, {Bryson},
  {Buchhave}, {Caldwell}, \& {Carter}}]{2011Natur.470...53L}
{Lissauer}, J.~J., {Fabrycky}, D.~C., {Ford}, E.~B., {et~al.} 2011, \nat, 470,
  53

\bibitem[{{Londrillo} \& {del Zanna}(2004)}]{2004JCoPh.195...17L}
{Londrillo}, P., \& {del Zanna}, L. 2004, \rm JCP, 195, 17

\bibitem[{{Lubow} {et~al.}(1999){Lubow}, {Seibert}, \&
  {Artymowicz}}]{1999ApJ...526.1001L}
{Lubow}, S.~H., {Seibert}, M., \& {Artymowicz}, P. 1999, \apj, 526, 1001

\bibitem[{{Machida} {et~al.}(2006){Machida}, {Inutsuka}, \&
  {Matsumoto}}]{2006ApJ...649L.129M}
{Machida}, M.~N., {Inutsuka}, S.-i., \& {Matsumoto}, T. 2006, \apjl, 649, L129

\bibitem[{{Machida} {et~al.}(2008){Machida}, {Kokubo}, {Inutsuka}, \&
  {Matsumoto}}]{2008ApJ...685.1220M}
{Machida}, M.~N., {Kokubo}, E., {Inutsuka}, S.-i., \& {Matsumoto}, T. 2008,
  \apj, 685, 1220

\bibitem[{{Machida} {et~al.}(2010){Machida}, {Kokubo}, {Inutsuka}, \&
  {Matsumoto}}]{2010MNRAS.405.1227M}
{Machida}, M.~N., {Kokubo}, E., {Inutsuka}, S.-I., \& {Matsumoto}, T. 2010,
  \mnras, 405, 1227

\bibitem[{{Martin} \& {Lubow}(2011)}]{2011MNRAS.413.1447M}
{Martin}, R.~G., \& {Lubow}, S.~H. 2011, \mnras, 413, 1447

\bibitem[{{Martin} \& {Lubow}(2013)}]{2013MNRAS.tmp.1252M}
---. 2013, \mnras

\bibitem[{{Meyer} {et~al.}(2012){Meyer}, {Balsara}, \&
  {Aslam}}]{2012MNRAS.422.2102M}
{Meyer}, C.~D., {Balsara}, D.~S., \& {Aslam}, T.~D. 2012, \mnras, 422, 2102

\bibitem[{{Miyoshi} \& {Kusano}(2005)}]{2005JCoPh.208..315M}
{Miyoshi}, T., \& {Kusano}, K. 2005, \rm JCoPh, 208, 315

\bibitem[{{Mizuno}(1980)}]{1980PThPh..64..544M}
{Mizuno}, H. 1980, Progress of Theoretical Physics, 64, 544

\bibitem[{{Mosqueira} \& {Estrada}(2003)}]{2003Icar..163..198M}
{Mosqueira}, I., \& {Estrada}, P.~R. 2003, \icarus, 163, 198

\bibitem[{{Movshovitz} {et~al.}(2010){Movshovitz}, {Bodenheimer}, {Podolak}, \&
  {Lissauer}}]{2010Icar..209..616M}
{Movshovitz}, N., {Bodenheimer}, P., {Podolak}, M., \& {Lissauer}, J.~J. 2010,
  \icarus, 209, 616

\bibitem[{{Muto} {et~al.}(2010){Muto}, {Suzuki}, \&
  {Inutsuka}}]{2010ApJ...724..448M}
{Muto}, T., {Suzuki}, T.~K., \& {Inutsuka}, S.-i. 2010, \apj, 724, 448

\bibitem[{{Nelson} {et~al.}(2012){Nelson}, {Gressel}, \&
  {Umurhan}}]{2012arXiv1209.2753N}
{Nelson}, R.~P., {Gressel}, O., \& {Umurhan}, O.~M. 2012, astro-ph: 1209.2753

\bibitem[{{Nelson} \& {Papaloizou}(2003)}]{2003MNRAS.339..993N}
{Nelson}, R.~P., \& {Papaloizou}, J.~C.~B. 2003, \mnras, 339, 993

\bibitem[{{Nelson} {et~al.}(2000){Nelson}, {Papaloizou}, {Masset}, \&
  {Kley}}]{2000MNRAS.318...18N}
{Nelson}, R.~P., {Papaloizou}, J.~C.~B., {Masset}, F., \& {Kley}, W. 2000,
  \mnras, 318, 18

\bibitem[{{Oishi} \& {Mac Low}(2011)}]{2011ApJ...740...18O}
{Oishi}, J.~S., \& {Mac Low}, M.-M. 2011, \apj, 740, 18

\bibitem[{{Okuzumi} \& {Hirose}(2011)}]{2011ApJ...742...65O}
{Okuzumi}, S., \& {Hirose}, S. 2011, \apj, 742, 65

\bibitem[{{Paardekooper} \& {Mellema}(2008)}]{2008A&A...478..245P}
{Paardekooper}, S.-J., \& {Mellema}, G. 2008, \aap, 478, 245

\bibitem[{{Papaloizou} \& {Lin}(1995)}]{1995ARA&A..33..505P}
{Papaloizou}, J.~C.~B., \& {Lin}, D.~N.~C. 1995, \araa, 33, 505

\bibitem[{{Papaloizou} \& {Nelson}(2005)}]{2005A&A...433..247P}
{Papaloizou}, J.~C.~B., \& {Nelson}, R.~P. 2005, \aap, 433, 247

\bibitem[{{Papaloizou} {et~al.}(2004){Papaloizou}, {Nelson}, \&
  {Snellgrove}}]{2004MNRAS.350..829P}
{Papaloizou}, J.~C.~B., {Nelson}, R.~P., \& {Snellgrove}, M.~D. 2004, \mnras,
  350, 829

\bibitem[{{Pierens} \& {Nelson}(2010)}]{2010A&A...520A..14P}
{Pierens}, A., \& {Nelson}, R.~P. 2010, \aap, 520, A14

\bibitem[{{Pinte} {et~al.}(2008){Pinte}, {Padgett}, {M{\'e}nard},
  {Stapelfeldt}, {Schneider}, {Olofsson}, {Pani{\'c}}, {Augereau},
  {Duch{\^e}ne}, {Krist}, {Pontoppidan}, {Perrin}, {Grady}, {Kessler-Silacci},
  {van Dishoeck}, {Lommen}, {Silverstone}, {Hines}, {Wolf}, {Blake}, {Henning},
  \& {Stecklum}}]{2008A&A...489..633P}
{Pinte}, C., {Padgett}, D.~L., {M{\'e}nard}, F., {et~al.} 2008, \aap, 489, 633

\bibitem[{{Pollack} {et~al.}(1996){Pollack}, {Hubickyj}, {Bodenheimer},
  {Lissauer}, {Podolak}, \& {Greenzweig}}]{1996Icar..124...62P}
{Pollack}, J.~B., {Hubickyj}, O., {Bodenheimer}, P., {et~al.} 1996, \icarus,
  124, 62

\bibitem[{{Quillen} \& {Trilling}(1998)}]{1998ApJ...508..707Q}
{Quillen}, A.~C., \& {Trilling}, D.~E. 1998, \apj, 508, 707

\bibitem[{{Sano} {et~al.}(2000){Sano}, {Miyama}, {Umebayashi}, \&
  {Nakano}}]{2000ApJ...543..486S}
{Sano}, T., {Miyama}, S.~M., {Umebayashi}, T., \& {Nakano}, T. 2000, \apj, 543,
  486

\bibitem[{{Saumon} \& {Guillot}(2004)}]{2004ApJ...609.1170S}
{Saumon}, D., \& {Guillot}, T. 2004, \apj, 609, 1170

\bibitem[{{Shabram} \& {Boley}(2013)}]{2013ApJ...767...63S}
{Shabram}, M., \& {Boley}, A.~C. 2013, \apj, 767, 63

\bibitem[{{Skinner} \& {Ostriker}(2010)}]{2010ApJS..188..290S}
{Skinner}, M.~A., \& {Ostriker}, E.~C. 2010, \apjs, 188, 290

\bibitem[{{Tanigawa} {et~al.}(2012){Tanigawa}, {Ohtsuki}, \&
  {Machida}}]{2012ApJ...747...47T}
{Tanigawa}, T., {Ohtsuki}, K., \& {Machida}, M.~N. 2012, \apj, 747, 47

\bibitem[{{Turner} {et~al.}(2012){Turner}, {Choukroun}, {Castillo-Rogez}, \&
  {Bryden}}]{2012ApJ...748...92T}
{Turner}, N.~J., {Choukroun}, M., {Castillo-Rogez}, J., \& {Bryden}, G. 2012,
  \apj, 748, 92

\bibitem[{{Turner} \& {Drake}(2009)}]{2009ApJ...703.2152T}
{Turner}, N.~J., \& {Drake}, J.~F. 2009, \apj, 703, 2152

\bibitem[{{Turner} {et~al.}(2013){Turner}, {Lee}, \&
  {Sano}}]{2013arXiv1306.2276T}
{Turner}, N.~J., {Lee}, M.~H., \& {Sano}, T. 2013, arXiv astro-ph:1306.2276

\bibitem[{{Umebayashi} \& {Nakano}(1981)}]{1981PASJ...33..617U}
{Umebayashi}, T., \& {Nakano}, T. 1981, \pasj, 33, 617

\bibitem[{{Umebayashi} \& {Nakano}(2009)}]{2009ApJ...690...69U}
---. 2009, \apj, 690, 69

\bibitem[{{Uribe} {et~al.}(2011){Uribe}, {Klahr}, {Flock}, \&
  {Henning}}]{2011ApJ...736...85U}
{Uribe}, A.~L., {Klahr}, H., {Flock}, M., \& {Henning}, T. 2011, \apj, 736, 85

\bibitem[{{Ward}(1997)}]{1997Icar..126..261W}
{Ward}, W.~R. 1997, \icarus, 126, 261

\bibitem[{{Wardle}(2007)}]{2007Ap&SS.311...35W}
{Wardle}, M. 2007, \apss, 311, 35

\bibitem[{{Wardle} \& {Salmeron}(2012)}]{2012MNRAS.422.2737W}
{Wardle}, M., \& {Salmeron}, R. 2012, \mnras, 422, 2737

\bibitem[{{Watson} {et~al.}(2007){Watson}, {Stapelfeldt}, {Wood}, \&
  {M{\'e}nard}}]{2007prpl.conf..523W}
{Watson}, A.~M., {Stapelfeldt}, K.~R., {Wood}, K., \& {M{\'e}nard}, F. 2007,
  Protostars and Planets V, 523

\bibitem[{{Winters} {et~al.}(2003){Winters}, {Balbus}, \&
  {Hawley}}]{2003ApJ...589..543W}
{Winters}, W.~F., {Balbus}, S.~A., \& {Hawley}, J.~F. 2003, \apj, 589, 543

\bibitem[{{Ziegler}(2004)}]{2004JCoPh.196..393Z}
{Ziegler}, U. 2004, \rm JCoPh, 196, 393

\bibitem[{{Ziegler}(2011)}]{2011JCoPh.230.1035Z}
---. 2011, \rm JCoPh, 230, 1035

\bibitem[{{Zingale} {et~al.}(2002){Zingale}, {Dursi}, {ZuHone}, {Calder},
  {Fryxell}, {Plewa}, {Truran}, {Caceres}, {Olson}, {Ricker}, {Riley},
  {Rosner}, {Siegel}, {Timmes}, \& {Vladimirova}}]{2002ApJS..143..539Z}
{Zingale}, M., {Dursi}, L.~J., {ZuHone}, J., {et~al.} 2002, \apjs, 143, 539

\end{thebibliography}
\end{document}